\documentclass[preprint,aps,tightenlines,showpacs]{revtex4}
\usepackage[dvips]{graphicx}
\usepackage{dcolumn}
\usepackage{bm}

\newcommand{\slas}[1]{\not\! #1}
\newcommand{\beq}{\begin{equation}}
\newcommand{\eeq}{\end{equation}}

\begin{document}
\title{Dynamical Coupled-channel Model of Meson Production \\

Reactions in the Nucleon Resonance Region }
                                                                                
%%%%%%%%%%%%%%%%%%%% Authors %%%%%%%%%%%%%%%%%%%%%
                                                                                
\author{
A. Matsuyama$^a$, T. Sato$^b$, and
T.-S. H. Lee$^c $ }

%%%%%%%%%%%%%%%%%%%% Addresses %%%%%%%%%%%%%%%%%%%%%
                                                                                
\affiliation{
$^a$ Department of Physics, Shizuoka University, Shizuoka 422-8529, Japan \\
$^b$
Department of Physics, Osaka University, Toyonaka, Osaka 560-0043,
 Japan\\
$^c$
Physics Division, Argonne National Laboratory, Argonne,
Illinois 60439}

\begin{abstract}
A dynamical coupled-channel model is  presented for
investigating the nucleon resonances ($N^*$)
in the meson production reactions induced by pions and photons.
Our objective is to extract the $N^*$ parameters
and to investigate the meson production
reaction mechanisms for mapping out the
quark-gluon substructure of $N^*$ from the data.
The model is based on an
energy-independent Hamiltonian which is
derived from a set of Lagrangians by using a unitary transformation method.
The constructed model Hamiltonian consists of (a) 
$\Gamma_V$ for describing the vertex interactions
 $N^*\leftrightarrow MB, \pi\pi N$ 
with $MB=\gamma N, \pi N, \eta N, \pi\Delta, \rho N, \sigma N$,
 and $\rho\leftrightarrow \pi\pi$ and
$\sigma \leftrightarrow \pi\pi$, (b) $v_{22}$ for the non-resonant
$MB \rightarrow M'B'$ and $\pi\pi \rightarrow \pi\pi$ 
interactions,
(3) $v_{MB,\pi\pi N}$ for the 
non-resonant $MB \rightarrow \pi\pi N$ transitions,
and (4) $v_{\pi\pi N,\pi\pi N}$ for the
non-resonant $\pi\pi N \rightarrow \pi\pi N$
interactions.  By applying the projection operator techniques,
we derive a set of coupled-channel equations
which satisfy the unitarity conditions within
the channel space spanned by the considered two-particle  
$MB$ states and the three-particle $\pi\pi N$ state.
The resulting amplitudes are written as a sum of non-resonant and
resonant amplitudes such that the meson cloud effects on the
$N^*$ decay can be explicitly calculated for interpreting the
extracted $N^*$ parameters in terms of hadron structure calculations.
We present and explain in detail a numerical method based on
a spline-function expansion
for solving the resulting
coupled-channel equations which contain logarithmically divergent
one-particle-exchange driving terms $Z^{(E)}_{MB,M'B'}$  resulted from the
$\pi\pi N$ unitarity cut.  This
method is convenient,  and perhaps more practical and accurate
than the commonly employed methods of
contour rotation/deformation, for
calculating the two-pion production observables.
For completeness in explaining our numerical procedures, we also present
explicitly the formula for efficient calculations of a very large number
of partial-wave matrix elements which are the input to the coupled-channel
equations.  Results for two pion photo-production
are presented to illustrate the dynamical consequence of
the one-particle-exchange driving term $Z^{(E)}_{MB,M'B'}$ of 
the coupled-channel equations.
We show that this mechanism, which contains the
effects due to $\pi\pi N$ unitarity cut,
 can generate rapidly varying structure in the
reaction amplitudes associated with the unstable particle channels 
$\pi \Delta$, $\rho N$, and $\sigma N$, in agreement with the analysis
of Aaron and Amado [Phys. Rev. {\bf D13}, 2581 (1976)].
It also has large effects in determining the two-pion production
cross sections.  Our results
indicate that cautions must be taken to interpret the
$N^*$ parameters extracted from using models 
which do not include $\pi\pi N$ cut effects.
Strategies for performing a complete
dynamical coupled-channel 
analysis of all of available data of meson photo-production and
electro-production are discussed.

\end{abstract}
%\pacs{12.15.Ji, 13.60.Le, 25.30.-c}
                                                                                
\maketitle

\section{Introduction}

With the very intense experimental efforts at Jefferson Laboratory (JLab),
Mainz, Bonn, GRAAL, and Spring-8, extensive data of photo-production
and electro-production of $\pi$, $\eta$, $K$, $\omega$, $\phi$, and
two pions have now become available\cite{lee-review}. 
Many approaches
have been developed accordingly to investigate how 
the excitations of  nucleon resonances ($N^*$) can be identified from
these data. The objective is to extract the $N^*$ parameters for
investigating the dynamical structure of Quantum Chromodynamics (QCD) in
the non-perturbative region. The outstanding questions which can be
addressed are, for
example, how the spontaneously broken chiral symmetry is realized, 
and how the constituent quarks emerge as effective degrees of freedom
and how they are  confined.
In this work, we are similarly motivated and have developed a dynamical 
coupled-channel model for analyzing these data.

The $\pi N$ and $\gamma N$ reaction data in the $N^*$ region are most
often analyzed by using two different kinds of approaches.
The first kind is to apply the models which are  mainly
the continuations and/or extensions of the earlier works.
These include the analyses by using 
the Virginia Polytechnic Institute-George Washington 
University (VPI-GWU)
Model (SAID)\cite{said}, the Carnegie-Mellon-Berkeley (CMB) model\cite{cmb},
and the Kent State University (KSU) model\cite{ksu}.  Apart from
imposing the unitarity condition, these models
are very phenomenological in treating the reaction mechanisms.  
In particular, they  
assume that the non-resonant amplitudes, which are often comparable to or
even much larger
than the resonant amplitudes, can be parameterized in terms of
separable or polynomial forms in fitting the data. Furthermore, their
isobar model parameterizations do
not fully account for the analytical properties due to the $\pi\pi N$ unitarity
condition, as discussed, for example, by Aaron and Amado\cite{aa-76}.
We will address this important question later in this paper.

The second kind of  analyses  account for 
the reasonably understood meson-exchange mechanisms. For numerical
simplicity in solving the scattering equations, they however  neglect 
the off-shell multiple-scattering dynamics
which determines the meson-baryon scattering wavefunctions in the
short range region where we want to map out the 
quark-gluon substructure of $N^*$.
The $\pi\pi N$ unitarity condition is also  not satisfied
rigorously  in these analyses.  The most well-developed along this line
are the Unitary Isobar Models (UIM) developed by
the Mainz group (MAID)\cite{maid} and the
Jlab-Yeveran collaboration\cite{jlab-yeve}, K-matrix
coupled-channel models developed by the Giessen group\cite{giessen}
and KVI group\cite{kvi}, and the
JLab-Moscow State University (MSU) model of two-pion production.
More details of these approaches
have been reviewed recently in Ref.\cite{lee-review}. 

As we have learned recently in the
$\Delta$ region, the results from the approaches described above
are useful, but
certainly not sufficient for making real progress 
in understanding the structure of $N^*$ states. For example, the empirical
values of $N$-$\Delta$ transitions extracted by using
SAID and MAID are understood within the constituent quark model only when
the very  large pion cloud effects are identified in
the analyses based on dynamical models\cite{sl-1,sl-2,ky}.
The essence of a dynamical model is to separate the reaction mechanisms
from the internal structure of hadrons in interpreting the data.
To make similar progress in investigating the higher mass $N^*$,
it is highly desirable to extend such a dynamical approach to
analyze the meson production data up to the energy with
invariant mass $W \sim 2$ GeV.
This is the objective of this work.
Our goal is  not only to extract the resonance parameters, but also
to interpret them in terms of the current hadron structure
calculations. The achievable goal at the
present time  is to test the
predictions from various QCD-based models of baryon structure. 
It is also important to make  
connections with Lattice QCD calculations.
The Lattice QCD calculations are now being carried out\cite{dina} to
give a deeper understanding of the $N$-$\Delta$ transition. 
A systematic Lattice QCD program on $N^*$ is also under 
development\cite{richards}.

\begin{figure}
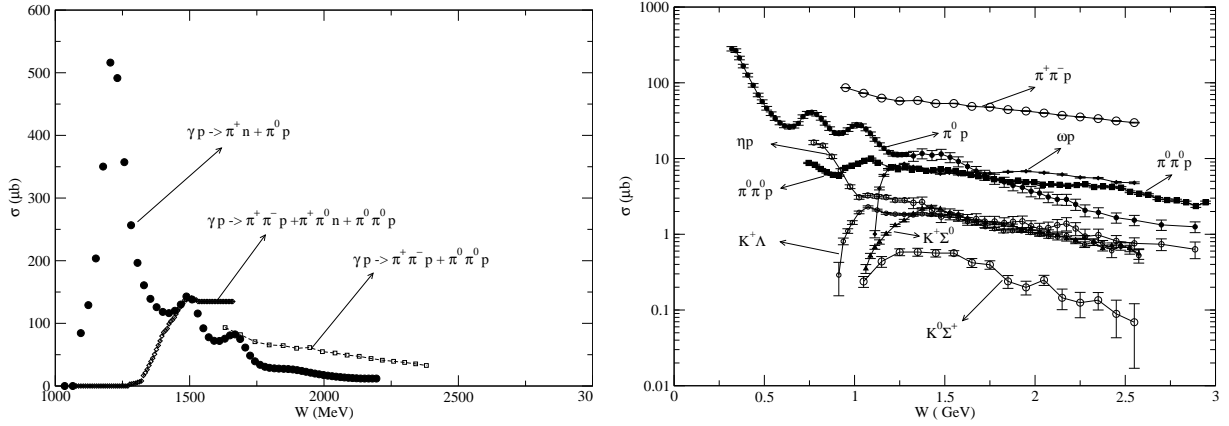

\centering
\includegraphics[width=8cm,angle=-0]{gp_tot-1pi-2pi.eps}
\includegraphics[width=8cm,angle=-0]{ky-etan-omen-1pi-2pi.eps}
\caption{The total cross section data of meson production
in $\gamma p$ reaction.
 Left:  $1-\pi$ and
$2-\pi$ production  are compared. Right:
KY ( $K^+\Lambda$,
$K^+\Sigma^0$, $K^0\Sigma^+$), $\eta p$, and $\omega p$ production
are compared with some of the $1-\pi$ and $2-\pi$ production}
\label{fig:totcross}
\end{figure}

The main challenge of developing dynamical
reaction models of  meson production reactions
in the $N^*$ region 
can be seen in Fig.\ref{fig:totcross}. We see that 
two-pion photo-production cross sections shown in the left-hand-side
become larger than the one-pion photo-production
as the $\gamma p$ invariant mass exceeds $W \sim 1.4 $ GeV.
In the right-hand-side, KY ( $K^+\Lambda$,
$K^+\Sigma^0$, $K^0\Sigma^+$), $\eta p$, and $\omega p$ production
cross sections are a factor of about 10 weaker than the dominant
$\pi^+\pi^-p$ production.
From the unitarity condition, we have for any single meson production
process $\gamma N \rightarrow M B$ with $M B=\pi N, \eta N, \omega N,
  K \Lambda, K\Sigma$
\begin{eqnarray}
i(T_{\gamma N,MB} - T^*_{ MB,\gamma N}) &=& \sum_{M'B'}
T_{\gamma N, M'B'}\rho_{M'B'} T^*_{M B, M'B'} \nonumber \\ 
& & + T_{\gamma N, \pi\pi N} \rho_{\pi\pi N}T^*_{M B, \pi\pi N} \,,
\label{eq:unitarity-0}
\end{eqnarray}
where $\rho_\alpha$ denotes an appropriate phase space factor
for the channel $\alpha$. The large two-pion production cross sections
seen in Fig.\ref{fig:totcross} indicate that the second term in the
right-hand-side of
Eq.(\ref{eq:unitarity-0}) is significant and hence the
single meson production reactions above the $\Delta$ region
must be influenced strongly by the coupling with the two-pion channels.
Similarly, the two-pion production $\gamma N \rightarrow \pi\pi N$ is
also influenced by the transition to two-body $MB$ channel
\begin{eqnarray}
i(T_{\gamma N,\pi\pi N} - T^*_{ \pi\pi N,\gamma N}) &=& \sum_{M'B'}
T_{\gamma N, M'B'}\rho_{M'B'} T^*_{\pi\pi N, M'B'} \nonumber \\
& & + T_{\gamma N, \pi\pi N} \rho_{\pi\pi N}T^*_{\pi\pi N, \pi\pi N} \,.
\label{eq:unitarity-1}
\end{eqnarray}
Clearly, a  sound dynamical reaction model must be able to
describe the two pion production and to account for the above unitarity
conditions.

The development of meson-baryon reaction
 models including two-pion production channel
has a long history.
It was already recognized in 1960's, as discussed
by Blankenbecler and Sugar\cite{bs}, that
the  dispersion-relation approach, which has been very successful in
analyzing the data of $\pi N$ elastic scattering\cite{hohler}
 and $\gamma N \rightarrow \pi N$ reactions\cite{cgln,donnachie},
can not be used to analyze the data of two-pion production. The reason is
that apart from the $\pi N$ and $\pi\pi N$ unitarity cuts,
it is rather impossible to
even guess the analytic structure of two-pion production amplitudes.
Furthermore, the dispersion relation models are difficult to solve
because of their bi-linear structure which is the price of only
dealing with the on-shell amplitudes. 

Ideally, one would like to find alternatives to
analyze the $\pi N$ and $\gamma N$ reaction data completely within the
framework of relativistic quantum field theory. The Bethe-Salpeter (BS) equation
has been taken historically as the starting point of such
an ambitious approach.
The complications involved in solving the BS equation have been known for
long time. For example, its singularity structure and the
associated numerical problems were very well discussed
in Refs.\cite{tjon-1,tjon-2, afnan-1}.
The BS equation contains serious singularities arising from the pinching of
the integration over the time component.
In addition to the two-body unitarity cut, it has a selected
set of n-body unitarity cuts, as explained in great detail in
Ref.\cite{afnan-1}.
Considerable
numerical efforts are already  needed to solve the Ladder BS equation for
$\pi N$ elastic scattering, as can be seen in the work of
Lahiff and Afnan\cite{afnan-2}.
Using the Wick rotation, they can
solve the Ladder BS equation below two-pion production threshold
with very restricted choices of
form factors. It is not clear how to extend their
work to higher energies.

The first main progress in finding 
an alternative to the dispersion-relation approach 
was perhaps also made by Blankenbecler and Sugar\cite{bs}.
By imposing  the unitarity condition, they show that 
the Bethe-Salpeter equation
can be reduced into a covariant
three-dimensional equation which is linear and
can be managed in practice. Compared with the dispersion relation approach,
the challenge here is account for the off-shell dynamics.
This approach was later further developed by Aaron, Amado, and 
Young (AAY)\cite{aay}.  With the assumption that all interactions
are due to the formation and decay of isobars,
they developed a set of covariant three-dimensional
equations for describing both the $\pi N$ elastic scattering and
$\pi N \rightarrow \pi\pi N$ reaction.
They however had only obtained\cite{aay,aay-1,aa-1,aa-2} a very qualitative
description of the $\pi N$ data and only investigated
very briefly the electromagnetic meson production reactions.
Their results suggested 
the limitation of the isobar model and the need of
additional mechanisms. For example, the $N^*$ excitation mechanisms are
not included in their formulation.
They then proposed\cite{aa-76} an approach
to include the additional mechanisms
phenomenologically in fitting  the data by
using  the "minimal" equations which are
rigorously constrained by the $\pi N$ and
$\pi\pi N$ unitarity conditions and have the correct analyticity of the
isobar model. 
The AAY approach was later applied mainly in the studies of $\pi NN$ 
systems, such as those by
Afnan and Thomas\cite{afnan-thomas} and
 by Matsuyama and Yazaki\cite{matsuyama}. Development in this
direction was well reviewed in Ref.\cite{garcilazo}.

The dynamical study of $\pi N$ scattering was pursued further in 1980's by
Pearce and Afnan\cite{pa-86,ap-87,afnan-88}. 
They derived the $\pi N$ scattering
equations by using a diagrammatic method, originally developed for
investigating the
$\pi NN$ problem\cite{garcilazo}, to sum the perturbation diagrams
which are selected by imposing 
the unitarity condition.
Furthermore, they relate the $\pi N$ scattering to the 
cloudy bag model
by extending the work of Thieberg, Thomas and Miller\cite{cbm-1,cbm-2,cbm-3}
to include the $\pi\pi N$ unitarity condition.

Since 1990 the $\pi N$ and $\gamma N$ reactions have  been 
investigated mainly by using either 
the three-dimensional reductions\cite{klein}
of the Bethe-Salpeter equation or 
the unitary transformation methods\cite{sl-1,fuda-1}.
These efforts were motivated mainly by the success of the meson-exchange
models of $NN$ scattering\cite{machleidt}, and have yielded
the meson-exchange models
developed by Pearce and Jennings\cite{pj-91},
National Taiwan University-Argonne National Laboratory (NTU-ANL) collaboration
\cite{ntuanl-1,ntuanl-2}, Gross and Surya\cite{gross},
Sato and Lee\cite{sl-1,sl-2},
Julich Group\cite{julich-1,julich-2,julich-3,julich-4},
Fuda and his collaborators\cite{fuda-1,fuda-2}, and 
Utretch-Ohio collaboration\cite{pasc-1,pasc-2}. 
All of these dynamical
models can describe well the data in the $\Delta$ region, but
have not been fully developed in the higher mass $N^*$ region. 
The main challenge is to include correctly the coupling with 
the $\pi\pi N$ channels.

We now return to discussing  
the two-pion production channel
which is an essential part of our formulation.
Most of the recent two-pion production calculations are the extensions of the
isobar model of L\"{u}ke and S\"{o}ding\cite{luke}. The production mechanisms
are calculated from tree-diagrams of appropriately chosen Lagrangians.
The calculations of Valencia Group\cite{oset} included the tree diagrams
calculated 
from Lagrangians with $\gamma$, $N$, $\pi$, $\rho$, $\Delta$(1232),
$N^*(1440)$, and $N^*(1520)$ fields. To
describe the total cross section data in
all charged $\pi\pi N$ channels, they also
included\cite{oset-1} the production of
$\Delta(1700)$ and $\rho$ effect arising from $N^*(1520)$.

The model developed by Ochi, Hirata, Katagiri, and 
Takaki\cite{hirata-1,hirata-2,hirata-3} contains the
 tree diagrams calculated from
Lagrangians with $\gamma$, $\pi$, $\rho$, $\omega$,
 $N$, $\Delta$ and $N^*(1520)$ fields.
An important feature of this model is to describe the excitation of 
$N^*(1520)$ within an isobar model with three channels $\pi N$,
$\rho N$, and $\pi\Delta$.
They found that the invariant mass distributions of all charged channels
of $\gamma p \rightarrow \pi\pi N$ can be better described if the pseudo-scalar
$\pi NN$ coupling is used.
They also found that
the $N^*(1520)\rightarrow \rho N$ decay
is the essential
mechanism to explain the differences between the invariant mass distributions 
of $\pi^+\pi^0$ and $\pi^0\pi^0$.
Similar tree-diagram calculations of
two pion photo-production  have also been
performed by  Murphy and Laget\cite{laget}.

The analyses\cite{mokeev-1,mokeev-2,mokeev-3} 
of two pion production 
by using the JLab-Moscow State University (JLAB-MSU) isobar model 
considered only the minimum set of the tree diagrams proposed in
 the original work of L\"{u}ke and S\"{o}ding\cite{luke}. However, they made two 
improvements. They included all 3-star and 4-star
resonances listed by the Particle Data Group
and used the absorptive model developed by
Gottfried and Jackson\cite{gott} to account for
 the initial and final state interactions.
They found that
the $\pi N\Delta$ form factor is needed to get agreement with the data of
$\gamma p \rightarrow \pi^-\Delta^{++}$, while the initial and final
state interactions are not so large.
In analyzing the two-pion electro-production data, they further included
a $\pi\pi N$ phase-space term 
with its magnitude adjusted to fit the data.
This term was later replaced by a phenomenological particle-exchange
amplitude which improves significantly the fits to the data.
With this model, they had identified\cite{mokeev-3}
 a new $N^*$($\frac{3}{2}^+, 1720$) and
the production of
the  isobar channel $\pi^+ D_{13}(1520)$ which has never been
considered before.

The common feature of all of the two-pion production calculations described 
above is that the coupled-channel effects due to the unitarity condition, such
as that given in Eqs.(\ref{eq:unitarity-0})-(\ref{eq:unitarity-1}), 
are not included. The problems arise from this simplification
were very well studied by Aaron and Amado\cite{aa-76}, and will be discussed
later in this paper.
While the results from these tree-diagram models
are very useful for identifying the reaction mechanisms, their findings
concerning $N^*$ properties must be
further examined. To make progress, it is necessary to
develop a coupled-channel
formulation within which the $\pi\pi N$ channel is explicitly included.
In this paper, we report our effort in this direction.

We have developed
a dynamical coupled-channel model by extending the
model developed in Refs.\cite{sl-1,sl-2} to include the
higher mass $N^*$ and all relevant reaction channels seen in 
Fig.\ref{fig:totcross}. Our presentations will only include
two-particle channels $MB=\gamma N$, $\pi N$, $\eta N$ and three-particle
channel $\pi\pi N$ which has resonant components
$\pi\Delta$, $\rho N$, and $\sigma N$.
But the formulation can be easily extended to include
other two-particle channels such as $\omega N$, $K\Lambda$ and $K\Sigma$
and three-particle channels such as
$\pi\eta N$ and $K\bar{K} N$. 

Our main purpose here is to  give a complete and detailed presentation of our
model and the numerical methods needed to solve the resulting coupled-channel
equations.
A complete coupled-channel analysis requires a simultaneous fit to
all of the  meson production data from $\pi N$ and $\gamma N$ reactions,
such as the total cross section data  illustrated in Fig.\ref{fig:totcross}
and the very extensive data from recent high precision experiments on
 photo-production and electro-production reactions.
Obviously, this is a rather complex
problem which can not be accomplished in this paper.
Instead, we will apply our approach only to address
the  theoretical questions concerning the effects due to
$\pi\pi N$ unitarity cuts. For this very limited purpose, 
we present results from our first calculations of
 $\gamma N \rightarrow \pi\pi N$ reactions.

In section II, we present the model Hamiltonian of our formulation.
It is derived from a set of Lagrangians, given explicitly in Appendix A,
by  applying the unitary transformation method which was explained
in detail in Refs.\cite{sl-1,sko}. The coupled-channel equations are 
then derived from the model Hamiltonian in section III with
details explained in Appendix B.
In section IV, we explain the procedures for performing numerical
 calculations within our formulation.
The numerical
methods for solving the coupled-channel equations with
$\pi\pi N$ cut are explained in section V.
Results of  $\gamma p \rightarrow \pi\pi N$ are presented 
and discussed in section VI. A summary and the plans for future developments
are given in section VII. For the completeness in explaining our 
numerical procedures, several appendices
are given to present
explicitly the formula for efficient calculations of a very large number
of partial-wave matrix elements which are the input to the coupled-channel
equations, and to explain how the constructed resonant amplitudes
are related to the information listed by the Particle Data Group (PDG)\cite{pdg}.

\section{Model Hamiltonian}

In this section we present a  model Hamiltonian for
constructing a coupled-channel reaction model with 
$\gamma N$, $\pi N$, $\eta N$ and $\pi\pi N$ channels.
Since significant parts of the $\pi\pi N$ production are known
experimentally to be through 
the unstable states $\pi\Delta$, $\rho N$, and
perhaps also $\sigma N$, we will also include  $bare$ $\Delta$,
$\rho$ and $\sigma$ degrees of freedom in our formulation. Furthermore,
we introduce $bare$ $N^*$ states to represent 
the quark-core components of the nucleon resonances.
The model is expected to be valid up to $W=2$ GeV below which three
pion production is very weak.

Similar to the model of Refs.\cite{sl-1,sl-2}(commonly called the SL model), 
our starting point is a set of Lagrangians 
describing the interactions between mesons 
($M =\gamma, \pi,\eta ,\rho, \omega, \sigma \cdot\cdot\cdot$) and 
baryons ($B = N, \Delta, N^* \cdot\cdot\cdot$). These Lagrangian
are constrained by various well-established symmetry properties,
such as the invariance under isospin, parity, and gauge
transformation. The chiral symmetry is also implemented as 
much as we can.
The considered Lagrangians are given in Appendix A. 
By applying the
standard canonical quantization, we obtain 
a  Hamiltonian of the following form 
\begin{eqnarray}
H &=&\int  {\it h(\vec{x},t=0)} d\vec{x} \nonumber \\
&=& H_0 + H_I\,,
\label{eq:sl-1}
\end{eqnarray}
where ${\it h(\vec{x},t)}$ is the Hamiltonian density constructed from
the starting Lagrangians and the conjugate momentum field operators.
In Eq.(\ref{eq:sl-1}), $H_0$ is the free Hamiltonian and
\begin{eqnarray}
H_I = \sum_{M,B,B^\prime} \Gamma_{MB\leftrightarrow B^\prime}
+\sum_{M,M',M''} h_{M'M''\leftrightarrow M} \,,
\label{eq:sl-2}
\end{eqnarray}
where 
 $ \Gamma_{MB \leftrightarrow B^\prime}$ describes
the absorption and emission of a meson($M$) by a baryon($B$) such
as $\pi N \leftrightarrow N$ and $\pi N \leftrightarrow \Delta$, and
$h_{M'M''\leftrightarrow M} $ describes the vertex interactions between
mesons such
as $\pi\pi \leftrightarrow \rho$ and $\gamma \pi \leftrightarrow \pi$.
Clearly, it is a non-trivial many body problem to
calculate meson-baryon scattering and meson production
reaction amplitudes from the Hamiltonian defined by
Eqs.(\ref{eq:sl-1})-(\ref{eq:sl-2}).
To obtain a manageable reaction model,
we apply a unitary transformation method\cite{sl-1,sko} 
to derive an effective Hamiltonian from Eqs.(\ref{eq:sl-1})-(\ref{eq:sl-2}).
The essential idea of the employed
unitary transformation method is to eliminate the unphysical vertex
interactions $MB \rightarrow B^\prime$ with masses $m_M +m_B < m_{B^\prime}$
from the Hamiltonian and absorb their effects into
 $MB\rightarrow M^\prime B^\prime$ two-body interactions. 
The resulting effective Hamiltonian is energy independent and hence
is easy to be used in developing reaction models and performing
many-particle calculations.
The details of this method 
have been explained in section II and the appendix of Ref.\cite{sl-1}. 

\begin{figure}
\centering
\includegraphics[width=10cm,angle=-0]{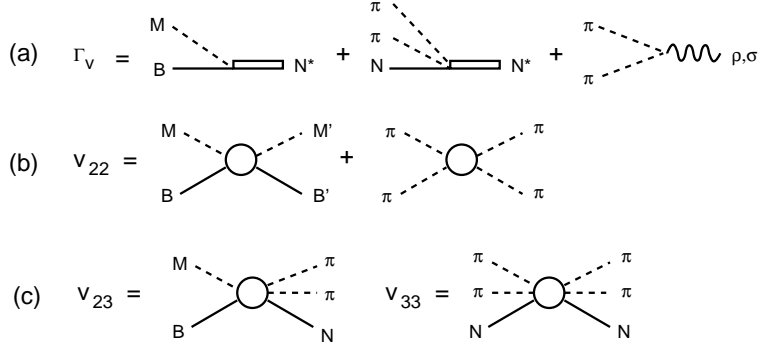}
\caption{Basic mechanisms of the Model Hamiltonian defined in Eqs.(8)-(10). }
\label{fig:h}
\end{figure}

Our main step is to derive from 
Eqs.(\ref{eq:sl-1})-(\ref{eq:sl-2}) an effective Hamiltonian
which contains interactions involving $\pi\pi N$ three-particle
states.  This is accomplished by
applying the unitary transformation 
method up to the
third order in interaction $H_I$ of Eq.(\ref{eq:sl-2}). The resulting
effective
Hamiltonian is of the following form
\begin{eqnarray}
H_{eff}= H_0 + V \,,
\label{eq:H-1}
\end{eqnarray}
with 
\begin{eqnarray}
H_0 = \sum_{\alpha} K_\alpha \,,
\end{eqnarray}
where $K_\alpha = \sqrt{m_\alpha^2+\vec{p_\alpha}^2}$
is the free energy operator of particle $\alpha$ with a mass $m_\alpha$,
and the interaction Hamiltonian is 
\begin{eqnarray}
V =\Gamma_V + v_{22}   + v' \,,
\label{eq:H-2}
\end{eqnarray} 
where 
\begin{eqnarray}
\Gamma_V&=& \{ \sum_{N^*}
(\sum_{MB} \Gamma_{N^*\rightarrow MB} +\Gamma_{N^* \rightarrow \pi\pi N})
+ \sum_{M^*} h_{M^*\rightarrow \pi\pi} \} + \{c.c.\} \,, 
\label{eq:gammav} \\ 
v_{22}&=& \sum_{MB,M^\prime B^\prime}v_{MB,M^\prime B^\prime}
+v_{\pi\pi} \,.
\label{eq:v22} 
\end{eqnarray}
Here ${c.c.}$ denotes the complex conjugate of the terms on its 
left-hand-side.
In the above equations, 
 $MB =  \gamma N, \pi N, \eta N, \pi\Delta, 
\rho N, \sigma N$ represent the considered meson-baryon states.
The resonance associated with
the $bare$ baryon state $N^*$ is
induced by the vertex interactions 
$\Gamma_{N^* \rightarrow MB }$ and $\Gamma_{N^* \rightarrow \pi\pi N}$. 
Similarly, the $bare$
meson states $M^* $ = $\rho$, $\sigma$ can develop into resonances through
the vertex interaction $h _{M^*\rightarrow \pi\pi}$. These vertex
interactions  are illustrated in
Fig.\ref{fig:h}(a).
Note that the masses $M^0_{N^*}$ and $m^0_{M^*}$ 
of the bare  states $N^*$ and $M^*$ 
are the parameters of the model which will be determined by fitting the
$\pi N$ and $\pi\pi$ scattering data. They differ
from the empirically determined
resonance positions by mass shifts which are due to the
coupling of the bare states with the meson-baryon
$scattering$ states. It is thus reasonable to speculate that
these bare masses can be
identified with the mass spectrum predicted by  the hadron structure 
calculations which do not account for the meson-baryon $continuum$ 
scattering states, such as the calculations
based on the constituent quark models which do not have
meson-exchange quark-quark interactions. It is however much more difficult,
but more interesting, to relate these bare masses to the $current$
Lattice QCD calculations which
can not account for the scattering states rigorously mainly
because of the limitation of the lattice spacing.

\begin{figure}
\includegraphics[width=12cm]{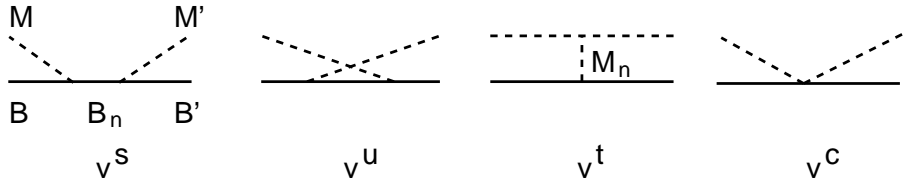}
\caption{Mechanisms for $v_{MB,M'B'}$ of Eq.(9):
(a) direct s-channel, (b) crossed u-channel,
 (c) one-particle-exchange t-channel, (d) contact interactions.}
\label{fig:mbmb}
\end{figure}

\begin{figure}
\centering
\includegraphics[width=12cm,angle=-0]{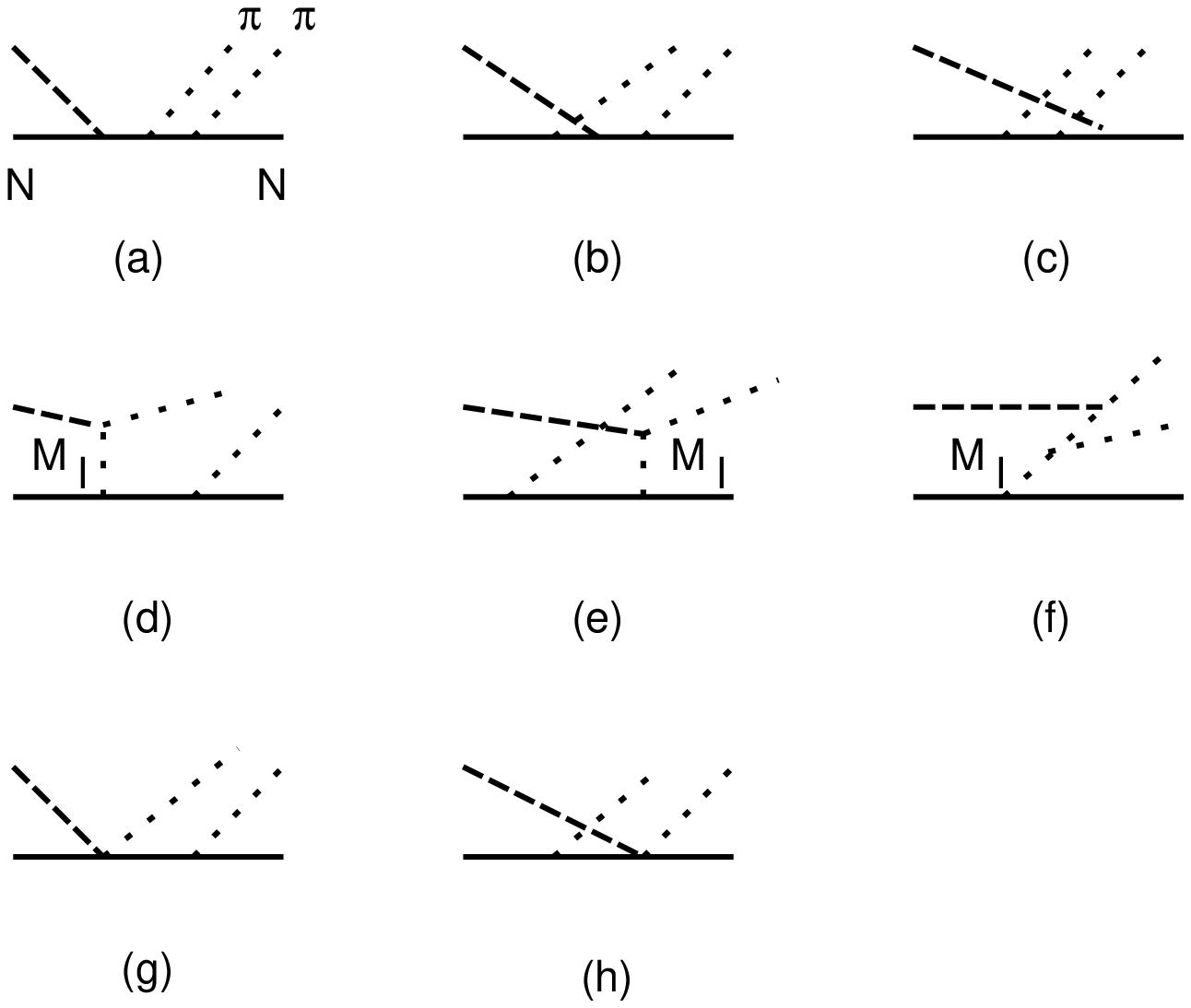}
\caption{ Examples of non-resonant mechanisms of $v_{MN,\pi\pi N}$
with $M = \pi$ or $ \gamma $ (denoted by long-dashed lines). $M_I$ denotes
the intermediate mesons ($\pi, \rho, \omega$).}
\label{fig:mbpipin}
\end{figure}

\begin{figure}
\centering
\includegraphics[width=12cm,angle=-0]{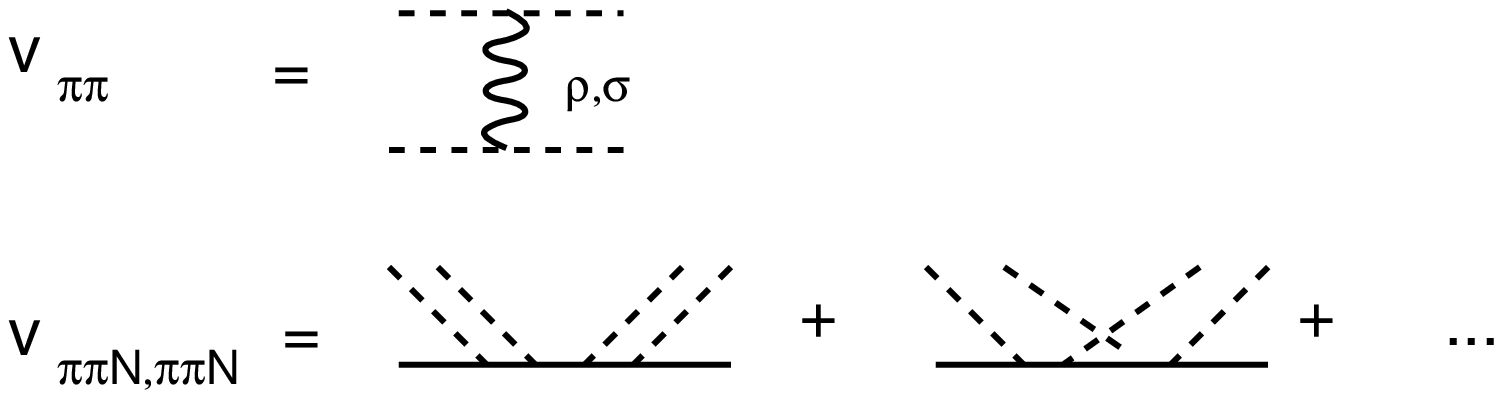}
\caption{ Examples of non-resonant mechanisms of $v_{\pi\pi} $
and $v_{\pi\pi N,\pi\pi N}$}
\label{fig:pipi-pipin}
\end{figure}

In Eq.(\ref{eq:v22}), $v_{MB,M^\prime B^\prime}$ is the
non-resonant meson-baryon interaction and $v_{\pi\pi}$ is the
non-resonant $\pi\pi$ interaction. They are illustrated in Fig.\ref{fig:h}(b).
The third term in Eq.(\ref{eq:H-2}) describes the non-resonant
interactions involving $\pi\pi N$ states
\begin{eqnarray}
v' =  v_{23} + v_{33} 
\label{eq:v23}
\end{eqnarray}
with 
\begin{eqnarray}
v_{23}&=&\sum_{MB}[( v_{MB,\pi\pi N}) + (c.c.)] \nonumber \\
v_{33}&=&  v_{\pi\pi N, \pi\pi N}\,. \nonumber
\end{eqnarray}
They are illustrated in Fig.\ref{fig:h}(c).
All of these interactions are defined by the tree-diagrams generated from
the considered Lagrangians. They are illustrated in Fig.\ref{fig:mbmb}
for two-body interactions $v_{MB,M'B'}$ and in Fig.\ref{fig:mbpipin}
for $v_{MB,\pi\pi N}$.
Some leading mechanisms of
$v_{\pi\pi}$ and $v_{\pi\pi N,\pi\pi N}$  are illustrated in
Fig.\ref{fig:pipi-pipin}.
The calculations of the
matrix elements of these interactions will be discussed later in
the section on our calculations and detailed in appendices.
Here we only mention that the matrix elements of
these interactions are calculated from the usual
Feynman amplitudes with their time components in the propagators
of intermediate states defined by the three momenta
of the initial and
final states, as specified by the unitary transformation methods.
Thus they are independent of the collision energy $E$.

\section{Dynamical Coupled-Channel Equations}

With the  Hamiltonian defined by Eqs.(\ref{eq:H-1})-(\ref{eq:v23})
, we follow the formulation of
Ref.\cite{gw} to define the scattering S-matrix
as
\begin{eqnarray}
S_{ab}(E) = \delta_{ab}-(2\pi) i T_{ab}(E) \,,
\label{eq:smatrix}
\end{eqnarray}
where the scattering T-matrix is defined by
\begin{eqnarray}
T_{ab}(E) =<a|T(E)|b> \nonumber 
\end{eqnarray}
with 
\begin{eqnarray}
T(E)=V + V\frac{1}{E-H_0 + i\epsilon} T(E)  \,.
\label{eq:loweq}
\end{eqnarray}
Since the interaction $V$, defined by Eqs.(\ref{eq:H-2})-(\ref{eq:v23}),
is energy independent, it is rather straightforward to follow
the formal scattering theory given in Ref.\cite{gw} 
to show that Eq.(\ref{eq:loweq}) leads to the following
unitarity condition 
\begin{eqnarray}
(T(E)-T^\dagger(E))_{ab} = -2\pi i\sum_{c}
T^\dagger_{ac}(E)\delta(E_c-E)T_{cb}(E) \,,
\label{eq:unitarity}
\end{eqnarray}
where $a,b,c$ are the
reaction channels in the considered energy region. 

Our task is to derive from Eq.(\ref{eq:loweq}) 
a set of dynamical coupled-channel equations for
practical calculations within the model space $N^*\oplus MB \oplus \pi\pi N$. 
In the derivations, the unitarity condition
Eq.(\ref{eq:unitarity}) must be maintained exactly.
We achieve this rather complex task by
applying the standard projection operator techniques\cite{feshbach},
 similar to that
employed in 
a study\cite{LM85} of $\pi NN$ scattering.
The details of our derivations 
are given in Appendix B.
To explain our coupled-channel equations, it is
sufficient to present the formula obtained from setting 
$\Gamma_{N^*\rightarrow \pi\pi N} =0 $ in our derivations.
The resulting model is defined
by Eqs.(B74)-(B96) of appendix B. Here we explain these equations and
discuss their dynamical content. 

\begin{figure}
\centering
\includegraphics[width=12cm,angle=-0]{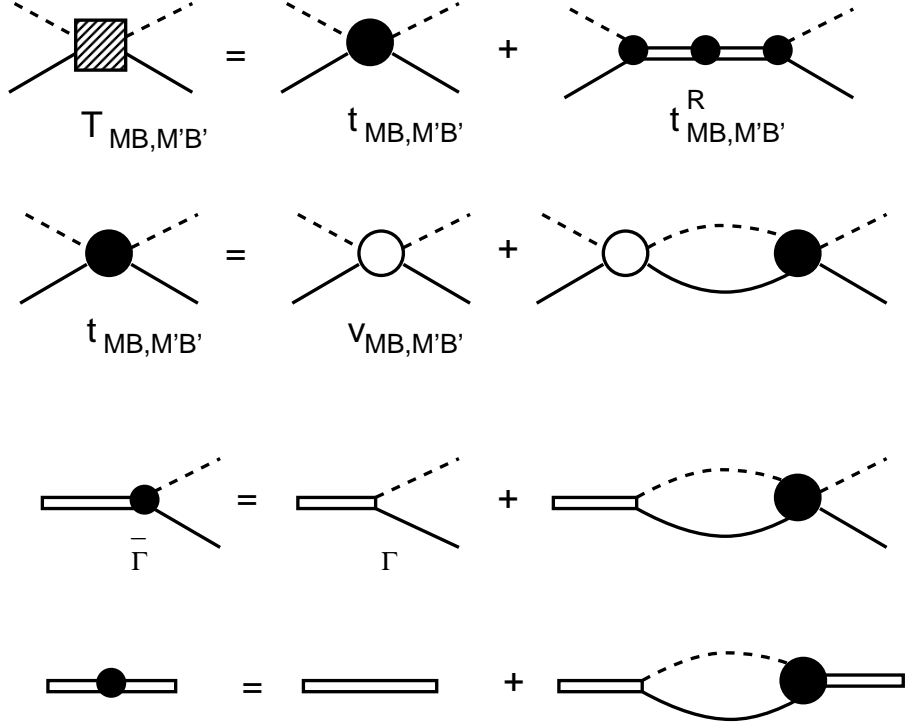}
\caption{ Graphical representations of Eqs.(14)-(19).}
\label{fig:tmatmbmb}
\end{figure}

The  resulting $MB \rightarrow M^\prime B^\prime$   amplitude 
$T_{MB \rightarrow M^\prime B^\prime}$  
in each partial wave is illustrated in Fig.\ref{fig:tmatmbmb}. It
can be written as
\begin{eqnarray}
T_{MB,M^\prime B^\prime}(E)  &=&  t_{MB,M^\prime B^\prime}(E)
+ t^R_{MB,M^\prime B^\prime}(E) \,,
\label{eq:tmbmb}
\end{eqnarray}

The second term in the right-hand-side of
Eq.(\ref{eq:tmbmb}) is the resonant term defined by
\begin{eqnarray} 
t^R_{MB,M^\prime B^\prime}(E)= \sum_{N^*_i, N^*_j}
\bar{\Gamma}_{MB \rightarrow N^*_i}(E) [D(E)]_{i,j}
\bar{\Gamma}_{N^*_j \rightarrow M^\prime B^\prime}(E) \,,
\label{eq:tmbmb-r} 
\end{eqnarray}
with
\begin{eqnarray}
[D(E)^{-1}]_{i,j}(E) = (E - M^0_{N^*_i})\delta_{i,j} - \bar{\Sigma}_{i,j}(E)\,,
\label{eq:nstar-g}
\end{eqnarray}
where $M_{N^*}^0$ is the bare mass of the resonant state $N^*$, and
the self-energies are 
\begin{eqnarray}
\bar{\Sigma}_{i,j}(E)= \sum_{MB}\Gamma_{N^*_i\rightarrow MB} G_{MB}(E)
\bar{\Gamma}_{MB \rightarrow N^*_j}(E) \,.
\label{eq:nstar-sigma}
\end{eqnarray}
The dressed vertex interactions in Eq.(\ref{eq:tmbmb-r}) and
Eq.(\ref{eq:nstar-sigma}) are 
 (defining
 $\Gamma_{MB\rightarrow N^*}=\Gamma^\dagger_{N^* \rightarrow MB}$)
\begin{eqnarray}
\bar{\Gamma}_{MB \rightarrow N^*}(E)  &=&  
  { \Gamma_{MB \rightarrow N^*}} + \sum_{M^\prime B^\prime}
t_{MB,M^\prime B^\prime}(E)
G_{M^\prime B^\prime}(E)
\Gamma_{M^\prime B^\prime \rightarrow N^*}\,, 
\label{eq:mb-nstar}
\\
\bar{\Gamma}_{N^* \rightarrow MB}(E)
 &=&  \Gamma_{N^* \rightarrow MB} +
\sum_{M^\prime B^\prime} \Gamma_{N^*\rightarrow M^\prime B^\prime}
G_{M^\prime B^\prime }(E)t_{M^\prime B^\prime,M B}(E) \,. 
\label{eq:nstar-mb}
\end{eqnarray}
The meson-baryon propagator $G_{MB}$ in the above equations 
takes the following form
\begin{eqnarray}
G_{MB}(E)=\frac{1}{E - K_B -K_M  -\Sigma_{MB}(E) + i\epsilon} \,,
\label{eq:g-mb-1}
\end{eqnarray}
where the  mass shift $\Sigma_{MB}(E)$ depends on the considered $MB$
channel.
It is $\Sigma_{MB}(E)=0$ for the
stable particle channels  $MB=\pi N, \eta N$.
For channels containing an
unstable particle,  such as $MB=\pi\Delta, \rho N, \sigma N$, we have
\begin{eqnarray}
\Sigma_{MB}(E) =[ <MB| g_V
\frac{P_{\pi\pi N}}{E-K_\pi-K_\pi -K_N+i \epsilon}
g^\dagger_V |MB> ]_{un-connected} 
\label{eq:sigma-m}
\end{eqnarray}
with
\begin{eqnarray}
g_V =\Gamma_{ \Delta\rightarrow \pi N}+
h_{\rho\rightarrow \pi\pi} + h_{\sigma \rightarrow \pi\pi} \,.
\label{eq:gv}
\end{eqnarray}
In Eq.(\ref{eq:sigma-m}) "$un-connected$" means that the
stable particle, $\pi$ or $N$, of the $MB$ state is a spectator in
 the $\pi\pi N$ propagation. Thus $\Sigma_{MB}(E)$ is just the mass
renormalization of the unstable particle in the $MB$ state.

\begin{figure}
\centering
\includegraphics[width=12cm,angle=-0]{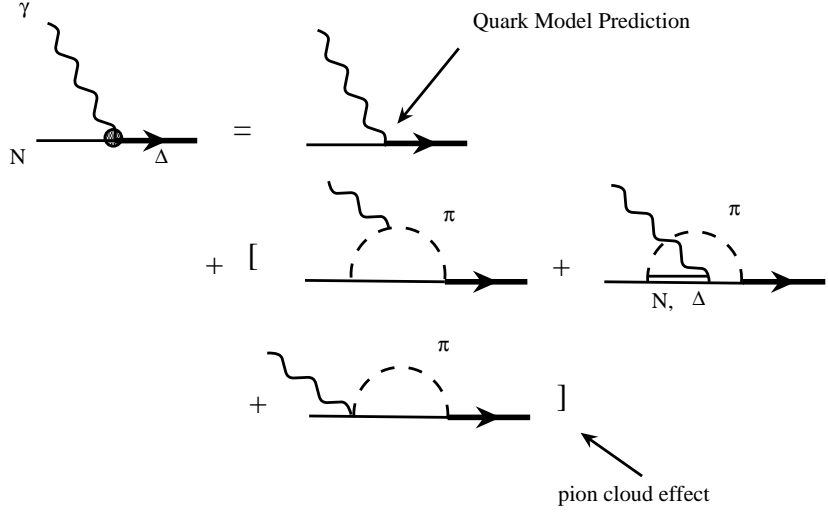}
\caption{Graphical representation of the dressed
$\bar{\Gamma}_{\Delta,\gamma N}$ interpreted in Refs.\cite{sl-1,sl-2}.}
\label{fig:mb-cloud}
\end{figure}

It is important to note that the
resonant amplitude $t^R_{M'B',MB}(E)$ is influenced by the non-resonant
amplitude $t_{M'B',MB}(E)$, as seen
in Eq.(\ref{eq:tmbmb-r})-(\ref{eq:nstar-mb}).
In particular, Eqs.(\ref{eq:mb-nstar})-(\ref{eq:nstar-mb})
describe the meson cloud effects on $N^*$ decays, as illustrated in
Fig.\ref{fig:mb-cloud} for the $\Delta \rightarrow \gamma N$ decay
interpreted in Refs.\cite{sl-1,sl-2}.
This feature of our formulation
is essential in interpreting the extracted resonance parameters.

Here we note that
the $N^*$ propagator $D(E)$ defined by 
Eq.(\ref{eq:nstar-g}) can be diagonalized
to write the resonant term Eq.(\ref{eq:tmbmb-r}) as
\begin{eqnarray}
t^R_{MB,M'B'}(k, k') = \sum_{\bar{N}^{*}}
\frac{\tilde{\Gamma}_{MB \rightarrow \bar{N}^{*}}(k)
\tilde{\Gamma}_{\bar{N}^{*}\rightarrow M'B'}(k')}
{E - {M}_{\bar{N}^*}(E) +\frac{i}{2} \Gamma^{tot}_{\bar{N}^{*}}(E)} \,,
\label{eq:digres}
\end{eqnarray}
where $\tilde{\Gamma}_{MB \rightarrow \bar{N^*}}$ and mass parameters
${M}_{\bar{N}^*}(E)$ and $\Gamma^{tot}_{\bar{N^*}}(E)$
are of
course related the dressed vertexes $\bar{\Gamma}_{N^*\rightarrow MB}$
and self energies $\Sigma_{i,j}$ defined in
Eqs.(\ref{eq:nstar-sigma})-(\ref{eq:nstar-mb}). Eq.(\ref{eq:digres}) is similar to the usual
Breit-Wigner form and hence can be used to relate our model
to the empirical resonant parameters listed by Particle Data Group.
This non-trivial subject is being investigated in Ref. \cite{suzuki}.
%Eq.(\ref{eq:digres}) will be used in our later calculations.

The non-resonant amplitudes $ t_{MB,M^\prime B^\prime}$ 
in Eq.(\ref{eq:tmbmb}) and Eqs.(\ref{eq:mb-nstar})-(\ref{eq:nstar-mb})
are defined by the following coupled-channel equations
\begin{eqnarray}
t_{MB,M^\prime,B^\prime}(E)&= &V_{MB,M^\prime B^\prime}(E) 
+ \sum_{M^{\prime\prime}B^{\prime\prime}}
V_{MB,M^{\prime\prime}B^{\prime\prime}}(E)
G_{M^{\prime\prime}B^{\prime\prime}}(E) 
t_{M^{\prime\prime}B^{\prime\prime},M^\prime B^\prime}(E)
\label{eq:cc-mbmb}
\end{eqnarray}
with
\begin{eqnarray}
V_{MB,M^\prime B^\prime}(E)= v_{MB,M^\prime B^\prime}
+Z_{{M}{B},{M}^\prime {B}^\prime}(E) \,.
\label{eq:veff-mbmb}
\end{eqnarray}
Here $Z_{{M}{B},{M}^\prime {B}^\prime}(E)$
contains the effects due to the coupling with $\pi\pi N$ states.
It has the following form
\begin{eqnarray}
Z_{{M}{B},{M}^\prime {B}^\prime}(E)
&=& < {M}{B} \mid F
\frac{ P_{\pi \pi N}}
{E- H_0 - \hat{v}_{\pi \pi N}+ i\epsilon}
F^\dagger  \mid {M}^\prime {B}^\prime > \nonumber \\
& & \nonumber \\
& & -[ \delta_{MB,M'B'} \Sigma_{MB}(E) ]
\label{eq:z-mbmb0}
\end{eqnarray}
with
\begin{eqnarray}
\hat{v}_{\pi\pi N} &=& v_{\pi N,\pi N}+ v_{\pi\pi} +v_{\pi\pi N,\pi\pi N}
\label{eq:hatv-pipin}\,, \\
F &=& g_V + v_{MB,\pi\pi N}\,,
\label{eq:vertex-f}
\end{eqnarray}
where $g_V$ has been defined in Eq.(\ref{eq:gv}).
Note that the second term in Eq.(\ref{eq:z-mbmb0})
is the effect which is already included in the mass shifts
$\Sigma_{MB}$ of the propagator Eq.(\ref{eq:g-mb-1})
and must be
removed to avoid double counting. 

\begin{figure}
\centering
\includegraphics[width=12cm,angle=-0]{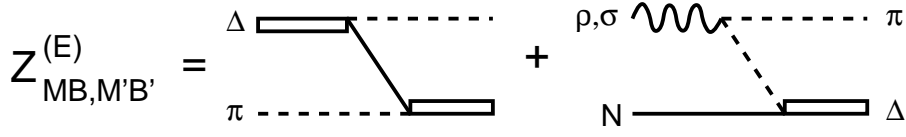}
\caption{One-particle-exchange interactions
 $Z^{(E)}_{\pi\Delta,\pi\Delta}(E)$, $Z^{(E)}_{\rho N,\pi\Delta}$
and $Z^{(E)}_{\sigma N,\pi\Delta}$ of
Eq.(30).}
\label{fig:z}
\end{figure}

\begin{figure}
\centering
\includegraphics[width=12cm,angle=-0]{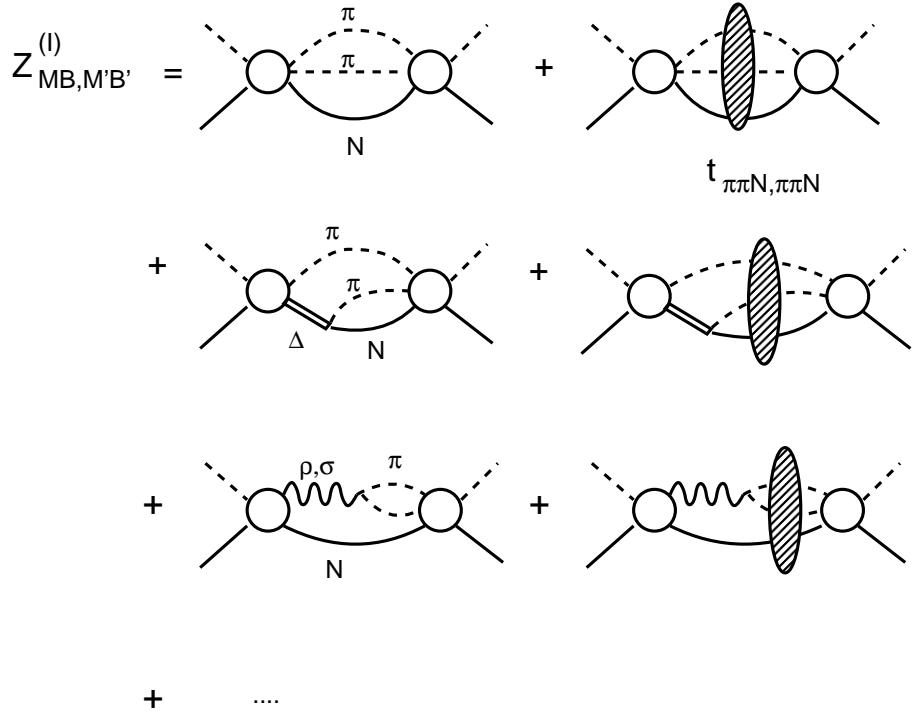}
\caption{Examples of mechanisms included in $Z^{(I)}_{MB,M'B'}(E)$
of Eq.(31).}
\label{fig:zi}
\end{figure}

\begin{figure}
\centering
\includegraphics[width=12cm,angle=-0]{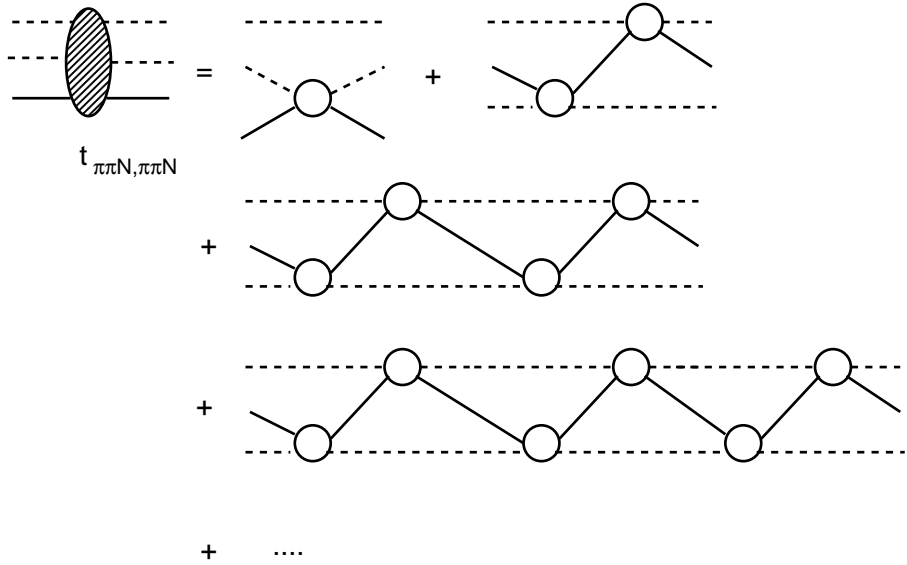}
\caption{ Some of the leading order terms of $t_{\pi\pi N,\pi\pi N}$
of Eq.(32). The open circle represents the direct s-channel
 interaction $v^s$ illustrated in
Fig.3 for the $MB=M'B'=\pi N$ case. }
\label{fig:tpinpin}
\end{figure}

The appearance of the projection operator $P_{\pi\pi N}$ in 
Eqs.(\ref{eq:sigma-m}) and (\ref{eq:z-mbmb0}) is the
consequence of the  unitarity condition Eq.(\ref{eq:unitarity}).  
To isolate the effects entirely due to the vertex interaction 
$g_V=\Gamma_{ \Delta\rightarrow \pi N}+
h_{\rho\rightarrow \pi\pi} +h_{\sigma\rightarrow \pi\pi}$, we use
the operator relation Eq.(B33) of Appendix B
to decompose the $\pi\pi N$ propagator
of Eq.(\ref{eq:z-mbmb0}) to write
\begin{eqnarray}
Z_{MB,M^\prime B^\prime}(E)
= Z^{(E)}_{MB,M^\prime B^\prime}(E)
+ Z^{(I)}_{MB,M^\prime B^\prime}(E) \,.
\label{eq:z-mbmb-s}
\end{eqnarray}
The first term is
\begin{eqnarray}
 Z^{(E)}_{MB,M^\prime B^\prime}(E)
&=&  < MB \mid g_V
\frac{ P_{\pi \pi N}}
{E- H_0  + i\epsilon}  g^\dagger_V
\mid  M^\prime B^\prime > - [ \delta_{MB,M'B'} \Sigma_{MB}(E) ]\,.
\label{eq:z-mbmb-e}
\end{eqnarray}
Obviously, $Z^{(E)}_{MB,M^\prime B^\prime}(E)$ is the one-particle-exchange
interaction between unstable particle channels $\pi\Delta$, $\rho N$, and
$\sigma N$, as illustrated in Fig.\ref{fig:z}.
The second term of Eq.(\ref{eq:z-mbmb-s}) is
\begin{eqnarray}
 Z^{(I)}_{MB,M^\prime B^\prime}(E)
&=& < MB \mid F
\frac{ P_{\pi \pi N}} {E- H_0 +  i\epsilon}
t_{\pi \pi N,\pi\pi N}(E)
\frac{ P_{\pi \pi N}} {E- H_0 +  i\epsilon}
F^\dagger
\mid M^\prime B^\prime > \nonumber \\
& & + <MB \mid g_V
\frac{ P_{\pi \pi N}}
{E- H_0  + i\epsilon} v^\dagger_{MB,\pi\pi N} 
\mid B^\prime M^\prime > \nonumber \\
& &+ < MB \mid v_{MB,\pi\pi N}
\frac{ P_{\pi \pi N}}
{E- H_0  + i\epsilon}  g^\dagger_V
\mid  M^\prime B^\prime> \nonumber \\
& & + < MB \mid v_{MB,\pi\pi N}
\frac{ P_{\pi \pi N}}
{E- H_0  + i\epsilon} v^\dagger_{MB,\pi\pi N} 
\mid M^\prime B^\prime > \,.
\label{eq:z-mbmb-i}
\end{eqnarray}
Some of the leading terms of $Z^{(I)}_{MB,M'B'}(E)$ are illustrated in
Fig.\ref{fig:zi}.
Here $t_{\pi\pi N,\pi\pi N}(E)$ is a three-body scattering amplitude
defined by
\begin{eqnarray}
t_{\pi\pi N, \pi\pi N}(E) = \hat{v}_{\pi\pi N} +
\hat{v}_{\pi\pi N}
\frac{1}{E-K_\pi-K_\pi-K_N-\hat{v}_{\pi\pi N} + i\epsilon} \hat{v}_{\pi\pi N}
\label{eq:tpinpin}
\end{eqnarray}
where $\hat{v}_{\pi\pi N}$ has been defined in Eq.(\ref{eq:hatv-pipin}).
Few leading terms of Eq.(\ref{eq:tpinpin}) due to 
the  direct s-channel interaction $v^s$ (illustrated in Fig.3) of
$v_{\pi N,\pi N}$
are shown in 
Fig.\ref{fig:tpinpin}.
These terms involve
the $\pi\pi N$ propagator ${1}/({E-K_\pi-K_\pi-K_N + i\epsilon}) $
and obviously  can generate $\pi\pi N$ cut effects which are due
to the  $\pi\pi N$ vertex. 
This observation indicates that the $\pi N$ scattering equation of
Aaron, Amado, and Young\cite{aay}
 can be related to our formulation if  the
interactions which are $only$ determined by the  $\pi \pi N$ vertex
 are kept in the equations 
presented above. We however will not discuss this  
issue in this paper.
The relations between our formulation and the $AAY$ model 
can be better understood
 in our next 
publication\cite{jlms}
where we will determine the strong interaction parts of our
Hamiltonian by fitting $\pi N$ reaction data up to invariant mass
$W = 2 $ GeV.

The amplitudes $T_{MB,M'B'}=t_{MB,M'B'}+t^R_{MB,M'B'}$ defined by
Eq.(\ref{eq:tmbmb}) can be used directly to calculate the cross sections
of $\pi N \rightarrow \pi N, \eta N$ and 
$\gamma N \rightarrow \pi N, \eta N$ reactions. 
They are also the input to the calculations of the
two-pion production amplitudes.
The two-pion production amplitudes resulted from our derivations given 
in Appendix B are illustrated in 
Fig.\ref{fig:mb-pipin}. They can be cast exactly into the following
form
\begin{eqnarray}
T_{\pi\pi N,MB}(E) = T^{dir}_{\pi\pi N,MB}(E)+
T^{\pi\Delta}_{\pi\pi N,MB}(E)+T^{\rho N}_{\pi\pi N,MB}(E)
+T^{\sigma N}_{\pi\pi N,MB}(E)
\label{eq:tpipin-1}
\end{eqnarray}
with
\begin{eqnarray}
T^{dir}_{\pi\pi N,MB}(E)
&=&
<\psi^{(-)}_{\pi\pi N}(E)|
 \sum_{M'B'}v_{\pi\pi N,M'B'}[\delta_{M'B',MB}  \nonumber \\
& & +G_{M'B'}(E)(t_{M'B',MB}(E) 
+t^R_{M'B',MB})] | MB>
\label{eq:tpipin-dir} \,, \\
T^{\pi \Delta}_{\pi\pi N,MB}(E)
&=&
 <\psi^{(-)}_{\pi\pi N}(E)|
\Gamma^\dagger_{\Delta\rightarrow \pi N}
G_{\pi\Delta}(E)[t_{\pi\Delta, MB}(E)+t^R_{\pi\Delta, MB}(E)] | MB>
\label{eq:tpipin-pid} \,, \\
& & \nonumber \\
T^{\rho N}_{\pi\pi N,MB}(E)
&=&
 <\psi^{(-)}_{\pi\pi N}(E)|
h^\dagger_{\rho\rightarrow \pi\pi}
G_{\rho N}(E)[t_{\rho N, MB}(E)+t^R_{\rho N, MB}(E)] | MB>
\label{eq:tpipin-rhon} \,, \\
& & \nonumber \\
T^{\sigma N}_{\pi\pi N,MB}(E)
&=&
 <\psi^{(-)}_{\pi\pi N}(E)|
h^\dagger_{\sigma \rightarrow \pi\pi}
G_{\sigma N}(E)[t_{\sigma N, MB}(E)+t^R_{\sigma N, MB}(E)] | MB> \,.
\label{eq:tpipin-sigman-ext}
\end{eqnarray}
In the above equations, the $\pi\pi N$ scattering wave function is defined by
\begin{eqnarray}
<\psi^{(-)}_{\pi\pi N}(E)|&=&
<\pi\pi N|\Omega^{(-)\dagger}_{\pi\pi N}(E) \,,
\end{eqnarray}
where the scattering operator is defined by
\begin{eqnarray}
\Omega^{(-)\dagger}_{\pi\pi N}(E)
 =
<\pi\pi N|[1+ t_{\pi\pi N, \pi\pi N}(E)\frac{1}
{E-K_\pi-K_\pi-K_N + i\epsilon}]\,.
\end{eqnarray}
Here the three-body scattering amplitude
 $t_{\pi\pi N, \pi\pi N}(E)$ is 
 determined by the
non-resonant interactions $v_{\pi\pi}$, $ v_{\pi N,\pi N}$ and 
$v_{\pi\pi N, \pi\pi N}$, as defined by Eq.(\ref{eq:tpinpin}).

\begin{figure}
\centering
\includegraphics[width=12cm,angle=-0]{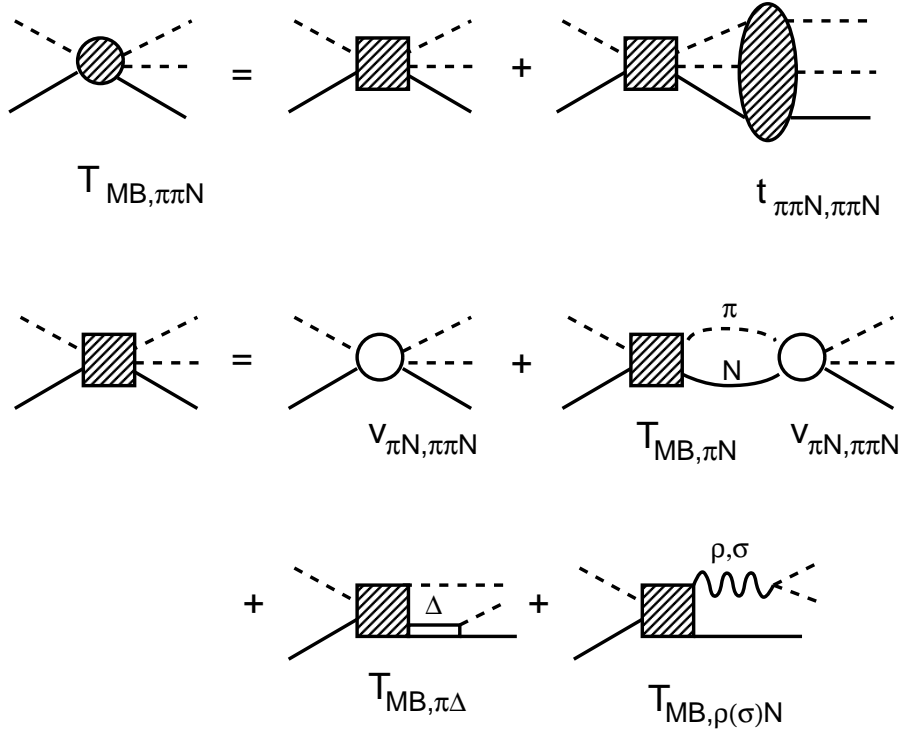}
\caption{Graphical representations of  $T_{\pi\pi N,MB}$
defined by Eqs.(33)-(37).}
\label{fig:mb-pipin}
\end{figure}

We note here that the direct production amplitude
$T^{dir}_{\pi\pi N,MB}(E)$ of Eq.(\ref{eq:tpipin-dir}) is due to 
$v_{\pi\pi N, MB}$ interaction illustrated in Fig.\ref{fig:mbpipin}, while
the other three terms are through the unstable $\pi\Delta$, $\rho N$,
and $\sigma N$ states.
Each term has the contributions from the
non-resonant amplitude $t_{M'B',MB}(E)$
and resonant term $t^R_{M'B',MB}(E)$. 
As seen in Eq.(\ref{eq:tmbmb-r})-(\ref{eq:nstar-mb}), the
resonant amplitude $t^R_{M'B',MB}(E)$ is influenced by the non-resonant
amplitude $t_{M'B',MB}(E)$.
This an important consequence of unitarity condition Eq.(\ref{eq:unitarity}).

\section{Calculations}

The $N^*$ information can be accurately extracted only when the
extensive meson production data of $\pi N$ and $\gamma N$ reactions
are analyzed simultaneously. Obviously, 
this is a rather complex task by using
the dynamical coupled-channel formulation described in section III.
In addition, it is a highly non-trivial numerical task to solve the
coupled-channel equation Eq.(\ref{eq:cc-mbmb}) which contains a
logarithmically divergent driving term
$Z_{MB,M'B'}{(E)}$ defined by Eqs.(\ref{eq:z-mbmb-s})-(\ref{eq:z-mbmb-i}).
As a first step, we 
focus in this work on the development of numerical methods for solving
this coupled-channel equation. This then allows us to perform
two-pion photo-production calculations to investigate 
the effects due to the 
$\pi\pi N$  cut effects which are not included in the
recent two-pion production
calculations, as briefly reviewed in section I. 

To proceed, we first note that the matrix elements of
$Z^{(I)}_{MB,M'B'}$, as defined by Eq.(\ref{eq:z-mbmb-i}),
is  expected to be 
weaker than the other driving terms $v_{MB,M'B'}$ and $Z^{(E)}_{MB,M'B'}$
because it involves more intermediate states.
For our present purpose of developing
numerical methods, this rather complex term can be neglected
in solving the coupled-channel Eq.(\ref{eq:cc-mbmb}).
For simplicity, we also neglect the
non-resonant interactions on the final $\pi\pi N$ state
by  setting $<\psi^{(-)}_{\pi\pi N}(E) | \rightarrow < \pi\pi N|$ in
the calculation of two-pion production amplitudes defined by
Eqs.(\ref{eq:tpipin-dir})-(\ref{eq:tpipin-sigman-ext}).

To make contact with recent experimental developments, 
we focus on the $\gamma N \rightarrow \pi \pi N$
process. Our task is therefore to develop numerical
methods for solving the following equations
\begin{eqnarray}
T_{\pi\pi N,\gamma N}(E) &=& \hat{T}^{dir}_{\pi\pi N,\gamma N}(E)
+\hat{T}^{\pi \Delta}_{\pi\pi N,\gamma N}(E)
+\hat{T}^{\rho N}_{\pi\pi N,\gamma N}(E)
+\hat{T}^{\sigma N}_{\pi\pi N,\gamma N}(E) 
\label{eq:hat-t1}
\end{eqnarray}
with
\begin{eqnarray}
\hat{T}^{dir}_{\pi\pi N,\gamma N}(E) &=&  
 <\pi\pi N| v_{\pi\pi N, \gamma N} + v_{\pi\pi N, \pi N}G_{\pi N}(E)
[\hat{t}_{\pi N,\gamma N}+t^{R}_{\pi N,\gamma N}]  | \gamma N> \,,
\label{eq:hat-t-dir}\\
\hat{T}^{\pi \Delta}_{\pi\pi N,\gamma N}(E) &=&
<\pi\pi N|
\Gamma^\dagger_{\Delta\rightarrow \pi N}
G_{\pi\Delta}(E)[\hat{t}_{\pi\Delta,\gamma N}(E) + 
t^{R}_{\pi \Delta, \gamma N}]| \gamma N> \,,
\label{eq:hat-tpd} \\
\hat{T}^{\rho N}_{\pi\pi N,\gamma N}(E) &=&
 <\pi\pi N|
h^\dagger_{\rho\rightarrow \pi\pi}
G_{\rho N}(E)[\hat{t}_{\rho N, \gamma N}(E) 
+t^{R}_{\rho N, \gamma N}]| \gamma N> \,,
\label{eq:hat-trn} \\
\hat{T}^{\sigma N}_{\pi\pi N,\gamma N}(E) &=&
 <\pi\pi N|
h^\dagger_{\sigma \rightarrow \pi\pi}
G_{\sigma N}(E)[\hat{t}_{\sigma N, \gamma N}(E) 
+ t^{R}_{\sigma N, \gamma N} ]| \gamma N> \,.
\label{eq:hat-tsn}
\end{eqnarray}
Here the non-resonant scattering amplitudes $\hat{t}_{MB,M'B'}$
is obtained from solving Eq.(\ref{eq:cc-mbmb}) with one of its
driving term $Z^{(I)}_{MB,M'B'}$ set to zero.
To  the first order in
electromagnetic  coupling, the matrix elements of these non-resonant
amplitudes are calculated from the following coupled-channel equations
\begin{eqnarray}
\hat{t}_{MB,M'B'}(\vec{k},\vec{k'},E) &=&
 \hat{V}_{MB,M'B'}(\vec{k},\vec{k'},E) \nonumber \\
& &+\sum_{M''B''} \int d\vec{k'}\hat{V}_{MB,M''B''}(\vec{k},\vec{k''},E)
G_{M''B''}(\vec{k''},E) \hat{t}_{M''B'',M'B'}(\vec{k''},\vec{k'},E)\,, \nonumber
\\
& & \label{eq:hat-tmbmb} \\
\hat{t}_{MB, \gamma N}(\vec{k},\vec{q},E) &=& 
v_{MB,\gamma N}(\vec{k},\vec{q}) \nonumber \\
& & + \sum_{M'B'}\int d\vec{k'}
\hat{t}_{MB,M'B'}(\vec{k},\vec{k'},E)G_{M'B'}(\vec{k'},E) 
v_{M'B',\gamma N}(\vec{k},\vec{q})
\label{eq:hat-tmbgn} 
\end{eqnarray}
with
\begin{eqnarray}
\hat{V}_{MB,M'B'}(\vec{k},\vec{k'},E) = v_{MB,M'B'}(\vec{k},\vec{k'})
 + Z^{(E)}_{MB,M'B'}(\vec{k},\vec{k'},E)
\label{eq:hat-v}
\end{eqnarray}
where $MB = \pi N, \eta N, \pi \Delta, \rho N, \sigma N$.
Despite the neglect of some of the terms of the formulation 
presented in section III, the calculations based on the above equations
are already far more complex than all of existing
calculations of two-pion production based on the
tree-diagram models or
K-matrix coupled-channel models. This is however a necessary step to
correctly 
account for the meson-baryon scattering wavefunctions in the
short range region where we want to extract and interpret the $N^*$
parameters using the data of meson production reactions,
as discussed in section I.

In the following subsections, we describe our numerical procedures for
solving Eqs.(\ref{eq:hat-tmbmb})-(\ref{eq:hat-v}) to get the
non-resonant amplitudes $\hat{t}_{MB,M'B'}$,  calculating the
resonance amplitudes $t^R_{MB,M'B'}$, and evaluating
the two-pion production amplitudes
Eqs.(\ref{eq:hat-t1})-(\ref{eq:hat-tsn}).

\subsection{Non-resonant amplitudes}

We solve Eq.(\ref{eq:hat-tmbmb}) in the partial-wave representation. 
To proceed, we follow the convention of Goldberger and Watson\cite{gw}
to normalize the plane-wave state $|\vec{k}>$ by setting
$ <\vec{k}|\vec{k'}> = \delta(\vec{k}-\vec{k'})$. In the
center of mass frame, Eq.(\ref{eq:smatrix})
 then leads to
the following formula of the cross section of 
$M(\vec{k}) + B (-\vec{k}) \rightarrow
M(\vec{k'}) + B (-\vec{k'})$ for stable particle channels $MB, M'B' =
\gamma N, \pi N, \eta N$
\begin{eqnarray}
& &\frac{d\sigma}{d\Omega}
= \frac{(4\pi)^2}{k^2}\rho_{M'B'}(k')\rho_{MB}(k)\frac{1}{(2j_M+1)(2j_B+1)}
\sum_{m_{j_M},m_{j_B}}\sum_{m'_{j_M},m'_{j_B}} |<M'B'|T(E)|MB>|^2 \nonumber \\
\label{eq:crst}
\end{eqnarray}
with
\begin{eqnarray}
& &<M'B'|T(E)|MB> = \nonumber \\
& &  < j'_M m'_{j_M}, i'_M m'_{i_M}; j'_B m'_{j_B},\tau'_B m'_{\tau_B}
|T_{M'B',MB}(\vec{k'},\vec{k},E)
| j_M m_{j_M}, i_M m_{i_M}; j_B m_{j_B},\tau_B m_{\tau_B}>\,,
 \nonumber \\
\label{eq:crst-1}
\end{eqnarray}
where $[(j_M,m_{j_M}), (i_M, m_{i_M})]$ 
and $[( j_B m_{j_B}), (\tau_B m_{\tau_B})]$  are
the spin-isospin quantum numbers of mesons and baryons, respectively.
The incoming and outgoing momenta
$k$ and $k'$ are defined by the collision energy $E$ 
\begin{eqnarray}
E=E_M(k)+E_B(k) 
= E_{M'}(k') + E_{B'}(k') \, ,
\end{eqnarray}
and the phase-space factor is
\begin{eqnarray}
\rho_{MB}(k)&=&\pi\frac{k E_M(k)E_B(k)}{E} \,.
\label{eq:ps}
\end{eqnarray}
The partial-wave expansion of the scattering amplitude is defined as
\begin{eqnarray}
& &
T_{M'B',MB}(\vec{k'},\vec{k},E)
=\sum_{JM,TM_T}\sum_{LS,L'S'} 
|{\it Y}^{JM,TM_T}_{L'(j'_M j'_B)S'}(\hat{k'})>
T^{JT}_{L'S'M'B',LSMB}(k',k,E) 
<{\it Y}^{JM,TM_T}_{L(j_M j_B)S}(\hat{k})| \nonumber \\
\label{eq:pwa-mbmb}
\end{eqnarray}
where the total angular vector in the spin-isospin space is defined by
\begin{eqnarray}
|{\it Y}^{JM,TM_T}_{L(j_M j_B)S}(\hat{k})>
&=& \sum_{all\,\,\, m} |j_M m_{j_M}, i_M, m_{i_M}; j_B m_{j_B},\tau_B m_{\tau_B}>
<TM_T|i_M\tau_B m_{i_M}m_{\tau_B}> \nonumber \\
& &\times< JM|LSm_Lm_S><Sm_S|j_Mj_Bm_{j_M}m_{j_B}> Y_{L m_L}(\hat{k})\,.
 \nonumber \\
\label{eq:ylm}
\end{eqnarray}
Clearly, Eqs.(\ref{eq:pwa-mbmb})-(\ref{eq:ylm}) lead to
\begin{eqnarray}
& &T^{JT}_{L'S'M'B',LSMB}(k',k,E) \nonumber \\
& & =\int d\hat{k'}\int d\hat{k}
<{\it Y}^{JM,TM_T}_{L'(j'_M j'_B)S'}(\hat{k'})|
T_{M'B',MB}(\vec{k'},\vec{k};E)|{\it Y}^{JM,TM_T}_{L(j_M j_B)S}(\hat{k})>
\,. \nonumber \\
\end{eqnarray}

By also expanding the driving term 
$\hat{V}_{MB,M'B'}(\vec{k},\vec{k'},E)$ of Eq.(\ref{eq:hat-tmbmb}) into the
partial-wave form similar to
Eq.(\ref{eq:pwa-mbmb}), we then obtain
a set of coupled 
one-dimensional integral equations 
\begin{eqnarray}
\hat{t}^{JT}_{L' S' M'B', L S MB}(k',k,E)
&=&
\hat{V}^{JT}_{L' S' M'B', L S MB}(k',k,E)
\nonumber \\
&+&\sum_{M^{\prime\prime}B^{\prime\prime}}
\sum_{ L^{\prime\prime}S^{\prime\prime}}
\int k^{\prime\prime 2}dk^{\prime\prime }
\hat{V}^{JT}_{L' S' M'B', L'' S''M^{\prime\prime}B^{\prime\prime}}(k',k^{\prime\prime},E) 
\nonumber \\
& &\times G_{M''B''}(k'',E)
\hat{t}^{JT}_{L'' S''M^{\prime\prime}B^{\prime\prime}, L S MB}(k'',k,E) \,,
\label{eq:pw-tmbmb}
\end{eqnarray}
where 
the driving term 
is  
\begin{eqnarray}
V^{JT}_{L' S' M'B', LSMB}(k',k)
=v^{JT}_{L' S' M'B', LSMB}(k',k)
+Z^{(E)JT}_{L' S' M'B', LSMB}(k',k,E)\,.
\end{eqnarray}
The above partial-wave matrix elements of the 
non-resonant interaction
$v_{M'B',MB}$ and one-particle-exchange interaction $Z^{(E)}_{M'B',MB}(E)$ 
are given in Appendices C and E, respectively. There the numerical methods
for evaluating them are also discussed in some details; 
in particular on the use of the transformation from the
helicity representation to the partial-wave representation.

The propagators in Eq.(\ref{eq:pw-tmbmb}) are given in Appendix B.
Taking the matrix elements of Eqs.(B84)-(B90),
we have
\begin{eqnarray}
G_{MB}(k,E) = \frac{1}{E-E_M(k)-E_B(k)+ i\epsilon}
\end{eqnarray}
for stable particle channels $MB=\pi N, \eta N$, and
\begin{eqnarray}
G_{MB}(k,E) = \frac{1}{E-E_M(k)-E_B(k)-\Sigma_{MB}(k,E)} 
\end{eqnarray}
for unstable particle channels $MB=\pi\Delta, \rho N, \sigma N$ with
\begin{eqnarray}
\Sigma_{\pi\Delta}(k,E)&=& \int q^2 dq\frac{|f_{\Delta,\pi N}(q)|^2}{
E-E_\pi(k) -[(E_N(q)+E_\pi(q))^2+k^2]^{1/2} + i\epsilon} \,, 
\label{eq:sig-pid} \\
\Sigma_{\rho N}(k,E)&=&
\int q^2 dq\frac{|f_{\rho,\pi\pi}(q)|^2}{
E-E_N(k) - [(2E_\pi(q))^2+k^2]^{1/2} + i\epsilon} \,, 
\label{eq:sig-rn} \\
\Sigma_{\sigma N}(k,E)&=&
\int q^2 dq\frac{|f_{\sigma,\pi\pi}(q)|^2}{
E-E_N(k) -[(2 E_\pi (q))^2+k^2]^{1/2} + i\epsilon} \,,
\label{eq:sig-sn}
\end{eqnarray}
where the vertex function $f_{\Delta,\pi N}(q)$ is from Ref.\cite{sl-1},
$f_{\rho,\pi \pi}(q)$ and $f_{\sigma,\pi \pi}(q)$ are from the isobar
fits\cite{johnstone} to 
the $\pi\pi $ phase shifts. They are given in Eqs.(D7)-(D9) of Appendix D.

\begin{figure}
\centering
\includegraphics[width=8cm,angle=-0]{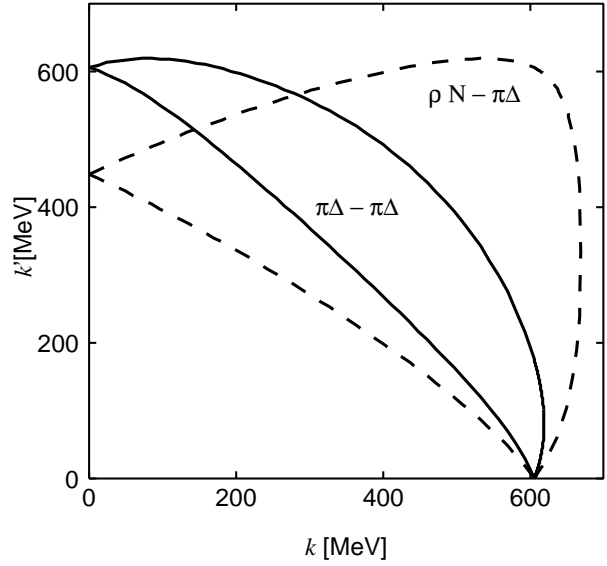}
\caption{Logarithmically divergent moon-shape regions
of the matrix elements of
$Z^{(E)}_{\pi \Delta,\pi \Delta }(k',k,E)$ (solid curves) and
$Z^{(E)}_{\rho N,\pi \Delta }(k',k,E)$ (dashed curves).  }
\label{fig:moon-aos}
\end{figure}

\begin{figure}
\centering
\includegraphics[width=8cm,angle=-0]{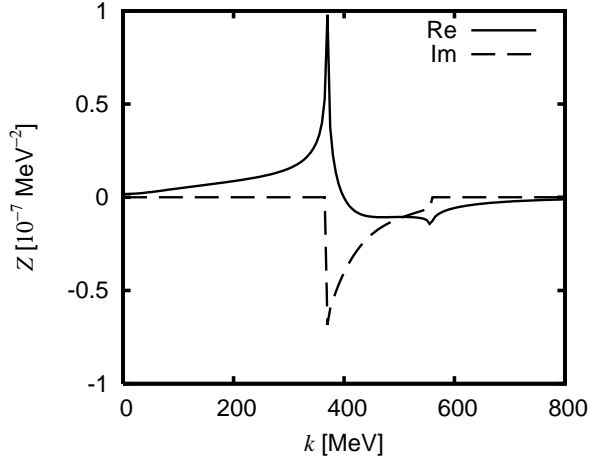}
\caption{ Matrix elements of the one-particle-exchange term
$Z^{(E)}_{\pi \Delta,\pi \Delta }(k,k',E)$ 
for $L=L'=1, J=5/2, T=1/2$ at $k'=300$ MeV/c and
$E=1.88$ GeV. }
\label{fig:zpdpd}
\end{figure}

To solve the coupled-channel integral equation Eq.(\ref{eq:pw-tmbmb}),
 we note that the matrix elements of
their particle-exchange 
driving terms $Z^{(E)}_{\pi\Delta,\pi\Delta}(k,k',E)$ and
$Z^{(E)}_{\rho N,\pi\Delta}(k,k',E)$ (Fig.\ref{fig:z}) contain 
singularities due to the $\pi\pi N$ cuts. This can be seen
in Eq.(E5) of Appendix E which is the essential component of 
their partial-wave matrix elements Eq.(E2). Qualitatively, they are
of the following form 
\begin{eqnarray}
Z^{(E)JT}_{L'S'\pi \Delta,LS \pi \Delta} (k,k',E) &\sim& 
\sum_{{\it l}} \int_{-1}^{+1} dx
\frac{A^{JT}_{\pi\Delta,\pi\Delta}(L'S',LS,{\it l},\vec{k},\vec{k'})
P_{\it l}(x)}
{E - E_\pi(k)-E_\pi(k')-E_N(\vec{k}+\vec{k'})+ i \epsilon}
\\
Z^{(E)JT}_{L'S'\rho N ,LS \pi \Delta}(k,k',E) &\sim&
\sum_{{\it l}} \int_{-1}^{+1} dx
\frac{A^{JT}_{\rho N,\pi\Delta}(L'S',LS,{\it l},\vec{k},\vec{k'}) 
P_{\it l}(x)}
{E - E_\pi(k)-E_N(k')-E_\pi(\vec{k}+\vec{k'})+ i \epsilon}
\end{eqnarray}
where $A^{JT}$ is a non-singular function, $P_{\it l}(x)$ is the Legendre
polynomial, 
and $x=\hat{k}\cdot\hat{k'}$.
One can easily see
 that these two driving terms diverge logarithmically 
in some momentum regions. 
For $E=1.88$ GeV, they are within
the moon-shape regions of Fig.\ref{fig:moon-aos}. Their boundary curves
are defined by $E - E_\pi(k)-E_\pi(k')-E_N({k}\pm{k'})=0$
for $Z^{(E)}_{\pi\Delta,\pi\Delta}$
and  by
$E - E_\pi(k)-E_N(k')-E_\pi(k\pm {k'})=0$ for $Z^{(E)}_{\rho N,\pi\Delta}$.
In Fig.\ref{fig:zpdpd}, we show the rapid change of the
matrix element
 $Z^{(E)}_{\pi\Delta,\pi\Delta}(k,k';E)$ at $E=1.88$ GeV 
and $k'=300$ MeV/c when the momentum $k$ is varied to
cross the moon-shape region.
In particular, the imaginary part (dashed line) is non-zero only in
a narrow region. The matrix elements
of $Z^{(E)}_{\rho N,\pi\Delta}(k,k';E)$ have the similar singular structure.

With the singular structure illustrated in Fig.\ref{fig:zpdpd}, 
Eq.(\ref{eq:pw-tmbmb}) can not be solved by the standard
subtraction method.
To get $\pi N \rightarrow \pi N, \eta N$ and
$\gamma N \rightarrow \pi N, \eta N$ on-shell scattering amplitudes,
it is sufficient to apply the well-developed method of contour rotation
to solve Eq.(\ref{eq:pw-tmbmb}) on the complex momentum axis defined
by $k_\theta=ke^{-i\theta}$ with $\theta> 0$. 
However, the resulting half-off-shell 
transition amplitudes $\hat{t}_{MB,\gamma N}(k_\theta,q;E)$ with 
$MB=\pi\Delta, \rho N, \sigma N$, defined on the complex
momentum $k_\theta$, can not be used
directly 
to evaluate the matrix elements 
Eqs.(\ref{eq:hat-t-dir})-(\ref{eq:hat-tsn}) for
calculating the two-pion production amplitudes.
Considerable effort is needed to find  an appropriate contour integration
for getting the desired matrix elements on the real momentum axis.
The situation is similar to the calculations of
deuteron breakup in $\pi d$ or
$p d$ reactions, as well discussed in the literatures\cite{thomas-review}.
We overcome this difficulty by applying
the spline-function
method developed in the study of 
$\pi NN$ reactions\cite{AM-1,AM-2}.
This method is explained in
details in the next section.

The solutions of
 Eq.(\ref{eq:pw-tmbmb}) are then used to calculate the non-resonant
photo-production amplitudes Eq.(\ref{eq:hat-tmbgn}). 
Here we use the helicity-LSJ mixed-representation
that the initial $\gamma N$ state is specified by their
 helicities, $\lambda_\gamma,\lambda_N$, but the
final $MB$ is defined by the $(LS)J$ angular momentum variables
\begin{eqnarray}
v_{MB,\gamma N}(\vec{k},\vec{q})
&=&\sum_{JM,TM_T}\sum_{LS}\sum_{\lambda_\gamma\lambda_N}
|{\it Y}^{JM,TM_T}_{L(j_M j_B)S}(\hat{k})>
v^{JT}_{LSMB,\lambda_\gamma\lambda_Nm_{\tau_N}}(k,q,E) \nonumber \\
& &\times \sqrt{\frac{2J+1}{4\pi}}
D^{J}_{M,(\lambda_\gamma-\lambda_N)}(\phi_q,\theta_q,-\phi_q)
<\lambda_\gamma,\lambda_N m_{\tau_N} | \,,
\label{eq:mixed-rep}
\end{eqnarray}
where $D^J_{m,m'}(\phi,\theta,-\phi) = e^{i(m+m')\phi}d^j_{m,m'}(\theta)$
with $d^j_{m,m'}(\theta)$ being the Wigner rotation function.
Eq.(\ref{eq:hat-tmbgn}) then leads to
\begin{eqnarray}
\hat{t}^{JT}_{L S MB,\lambda_\gamma\lambda_Nm_{\tau_N}}(k,q,E)&=&
{\it v}^{JT}_{L S MB,\lambda_\gamma\lambda_Nm_{\tau_N}}(k,q,E)
+\sum_{M'B'}
\sum_{L^{\prime}S^{\prime}}
\int k^{\prime 2}dk^{\prime }
 \hat{t}^{JT}_{L S MB, L' S'M'B'}(k,k^{\prime},E) \nonumber \\
& & \times G_{M'B'}(k',E)
{\it v}^{JT}_{L' S' M'B' , \lambda_\gamma\lambda_Nm_{\tau_N}}(k',q,E)\,.
 \nonumber \\
\label{eq:pw-tmbgn}
\end{eqnarray}
The matrix elements ${\it v}^{JT}_{L S MB,\lambda_\gamma\lambda_N m_{\tau_N}}(k,q,E)$
considered in our calculations are given in Appendix F. This unconventional
representation, which  is convenient for calculations, can
be related to the usual multipole expansion,
as also given in appendix G.

\subsection{Resonant amplitudes}
Our next step is to calculate the resonant term defined by
Eq.(\ref{eq:tmbmb-r}). Here we need to perform calculations
using bare $N^*\rightarrow MB$ vertex functions generated from
some hadron models.
Obviously, this is a non-trivial task and beyond the scope of this work.
In particular, one needs to analyze the consistency between the employed
hadron model and our reaction model.   
Instead, we use the diagonalized form  Eq.(\ref{eq:digres}) and
 simply make some plausible assumptions to calculate the resonant
amplitude $t^R_{MB,M'B'}$ by using
the information listed by  Particle Data Group (PDG)\cite{pdg}.
In the center of mass frame we write
Eq.(\ref{eq:digres}) for $\gamma N \rightarrow MB$ transition
in the helicity-LSJ mixed-representation as
\begin{eqnarray}
t^{R,JT}_{LSMB,\lambda_\gamma\lambda_Nm_{\tau_N}}(k, q, E) = \sum_{N^*}
[\tilde{\Gamma}^{JT}_{N^*,LSMB}(k)]^*
\frac{1}{E - M_{N^*}  +\frac{i}{2} \Gamma_{N^*}(E)}
\tilde{\Gamma}^{JT}_{N^*,\lambda_\gamma\lambda_Nm_{\tau_N}}(q) \,,
\label{eq:pdgres}
\end{eqnarray}
where $M_{N^*}$ is the resonance position.  
The calculations of the decay functions $\tilde{\Gamma}^{JT}_{N^*,LSM'B'}(k')$  
and $\tilde{\Gamma}^{JT}_{N^*,\lambda_\gamma\lambda_Nm_{\tau_N}}(q)$
are explained in appendix I. 
They are
\begin{eqnarray}
\tilde{\Gamma}^{JT}_{N^*,LSMB}(k) 
&=& \frac{1}{(2\pi)^{3/2}}\frac{1}{\sqrt{2E_M(k)}}
\sqrt{\frac{m_B}{E_B(k)}}
\sqrt{\frac{8\pi^2M_{N^*}}{m_Bk_R}}\,[G^{JT}_{LS,MB}]\, f^{JT}_{LS}(k,k_R)
(\frac{k}{k_R})^L
\label{eq:gmb}\\
\tilde{\Gamma}^{JT}_{N^*,\lambda_\gamma\lambda_N m_{\tau_N}}(q)
& =&\frac{1}{(2\pi)^{3/2}}\sqrt{\frac{m_N}{E_N(q)}}\frac{1}{\sqrt{2q}}
[\sqrt{2q_R} A^{JT}_{\lambda, m_{\tau_N}}] g^{JT}_\lambda(q,q_R) 
\delta_{\lambda, (\lambda_\gamma-\lambda_N)} \,,
\label{eq:ggn}
\end{eqnarray}
where $k_R$ and $q_R$ are  defined by 
$M_{N^*}=E_B(k_R)+E_M(k_R) = q_R+E_N(q_R)$. The form factors are normalized
such that $f^{JT}_{LS}(k_R,k_R)=1$ and $g^{JT}_\lambda(q_R,q_R)=1$.
For simplicity, we 
choose $f^{JT}_{LS}(k,k_R)=(\Lambda^2/((k-k_R)^2+\Lambda^2))^2$ and
$g^{JT}_\lambda(q,q_R)=(\Lambda^2/((q-q_R)^2+\Lambda^2))^2$
with $\Lambda=650$ MeV/c. 
As explained in Appendix I, the forms Eqs.(\ref{eq:gmb})-(\ref{eq:ggn}) are chosen such that
the coupling strength $G^{JT}_{LS,MB}$ is related to the
partial decay width $\Gamma_{MB}(N^*_{JT})$ of the considered $N^*\rightarrow MB$
\begin{eqnarray} 
\Gamma_{MB}(N^*_{JT})=\sum_{LS}|G^{JT}_{LS,MB}|^2 \,,
\end{eqnarray}
and the $\gamma N \rightarrow N^*$
helicity amplitude $A^{JT}_{\lambda m_{\tau_N}}$ is
related to the partial decay width by
\begin{eqnarray}
\Gamma_{\gamma, m_{\tau_N}}(N^*_{JT})
=\frac{q^2_R}{4\pi}\frac{m_N}{M_{N^*}}\frac{8}{2J+1}
[|A^{JT}_{3/2,m_{\tau_N}}|^2+|A^{JT}_{1/2,m_{\tau_N}}|^2] \,.
\label{eq:gn-heli}
\end{eqnarray}
Eq.(\ref{eq:gn-heli}) is defined in the $N^*$ rest frame and
the photon momentum  $\vec{q}$ is in the quantization z-direction.
 
The total width $\Gamma_{N^*}(E)$ in Eq.(\ref{eq:pdgres})
 is parameterized by using the variables of
$N^* \rightarrow \pi N$ decay as
\begin{equation}
\Gamma_{N^*} (E) = \Gamma^{tot}_{N^*}
\frac{\rho(k_\pi)}{\rho(k_{0\pi})}
\left( \frac{k_\pi}{k_{0\pi}} \right)^{2L_\pi}
\left[ \frac{\Lambda^2}{( k_\pi-k_{0\pi})^2+\Lambda^2 }
\right]^{L_\pi+4},
\label{eq:nsdecay}
\end{equation}
where $\Gamma^{tot}_{N^*}$ is the value given by the
Particle Data Group,
$L_\pi$ is the orbital angular momentum of the
considered $\pi N$ state and
\begin{equation}
\rho(k) = \pi\frac{k E_N(k) E_\pi(k)}{E_N(k) + E_\pi(k)}.
\label{eq:pdgres-1}
\end{equation}
In the above equations, $k_\pi$ is the pion momentum
at energy E while $k_{0\pi}$ is evaluated at $E=M_{N^*}$.
We set the form factor parameter $\Lambda = 650$ MeV/c.
Our main results on the effects due to the
$\pi\pi N$ cut are not changed much if we vary the cutoff
$\Lambda$ in Eqs.(\ref{eq:gmb})-(\ref{eq:nsdecay}).

\subsection{$\gamma N \rightarrow \pi\pi N$ cross sections}

Our last step is to calculate the  cross sections
of $\gamma(q) + N (p) \rightarrow \pi(k_1) +\pi(k_2)+ N (p')$. 
With the S-matrix defined by Eq.(\ref{eq:smatrix}) and the 
normalization $<\vec{k}|\vec{k'}>=\delta(\vec{k}-\vec{k'})$, 
we have 
\begin{eqnarray}
d\sigma &=&
\frac{(2\pi)^4}{v_{rel}} \delta^{(4)}(p+q-k_1-k_2-p') 
d\vec{k}_1d\vec{k}_2 d\vec{p'} \nonumber \\
& &\times \frac{1}{4}
\sum_{\lambda_\gamma\lambda_N}\sum_{m'_{j_N}}
|<\vec{k}_1,m_{i_1},\vec{k}_2,m_{i_2},\vec{p'}m'_{j_N} m'_{\tau_N} |
 T_{\pi\pi N,\gamma N}(E) 
| \vec{q}\lambda_\gamma,\vec{p}\lambda_N m_{\tau_N}> |^2 \,,
\label{eq:crst-gnpipin}
\end{eqnarray}
where $m_{i_1}$ and $m_{i_2}$ are the isospin quantum number of the outgoing
two pions, $m'_{j_N}$ and $m'_{\tau_N}$ are the spin-isospin quantum numbers of
the outgoing nucleon. The initial $\gamma N$ state is
specified by their
helicities $\lambda_\gamma, \lambda_N$ and the nucleon isospin $\tau_N$.
With some straightforward derivations, the differential cross section with 
respect to the $\pi\pi$ invariant mass $M_{\pi\pi}$ can be written
in the center of mass ( $\vec{p}=-\vec{q}$ and 
$\vec{k}=(\vec{k}_1+\vec{k}_2)=-\vec{p'}$)as
\begin{eqnarray}
\frac{d\sigma}{dM_{\pi\pi}}&=& 
\int d\Omega_k\int d\Omega_{k_{12}}
\frac{d\sigma}{d\Omega_k d\Omega_{k_{12}}dM_{\pi\pi}}
\label{eq:crst:dcrstm12}
\end{eqnarray}
with
\begin{eqnarray}
\frac{d\sigma}{d\Omega_kd\Omega_{k_{12}}dM_{\pi\pi}} &=&
(2\pi)^4[\frac{E_N(p)}{E}][\frac{E_N(p')E_\pi(k_1)E_\pi(k_2)}{E}]
[\vec{k}\cdot\vec{k_{12}}]
\nonumber \\
& &\times\frac{1}{4}\sum_{\lambda_\gamma\lambda_N}\sum_{m_{j'_N}}
|<\vec{k}_1,m_{i_1},\vec{k}_2,m_{i_2},\vec{p'}m'_{j_N} m'_{\tau_N} | 
T_{\pi\pi N,\gamma N}(E)
| \vec{q}\lambda_\gamma,\vec{p}\lambda_N m_{\tau_N}> |^2
\label{eq:dcrstpipin}\,, \nonumber \\
\end{eqnarray}
where $\vec{k}_1$ and $\vec{k}_2$ are related to
the relative momentum $\vec{k}_{12}$ 
and center of mass momentum $\vec{k}$ of the
$\pi\pi$ subsystem by a Lorentz boost
\begin{eqnarray}
\vec{k}_1&=& \vec{k}_{12}+\frac{\vec{k}}{M_{\pi\pi}}[E_\pi(k_{12})
+\frac{\vec{k}\cdot\vec{k}_{12}}{E_{\pi\pi}(k) +M_{\pi\pi}}]\,, \\
\vec{k}_2&=& -\vec{k}_{12}+\frac{\vec{k}}{M_{\pi\pi}}[E_\pi(k_{12})
-\frac{\vec{k}\cdot\vec{k}_{12}}{E_{\pi\pi}(k) +M_{\pi\pi}}] 
\end{eqnarray}
with
\begin{eqnarray}
M_{\pi\pi} &=& 2 E_\pi(\vec{k_{12}})\,, \\
E_{\pi\pi}(k) &=&E_\pi(\vec{k}_1)+E_\pi(\vec{k}_2) \nonumber \\
&=& \sqrt{M^2_{\pi\pi}+\vec{k}^2}\,, \\
E&=&E_N(k)+E_{\pi\pi}(k) \,.
\end{eqnarray}
The above equations lead to
\begin{eqnarray}
k&=&\sqrt{(\frac{E^2-m_N^2+M_{\pi\pi}^2}{2E})^2-M_{\pi\pi}^2} \,, \\
k_{12}&=&\sqrt{\frac{M^2_{\pi\pi}}{4} - m_\pi^2} \,.\\
\label{eq:k12}
\end{eqnarray}

The matrix element 
$<\vec{k}_1,m_{i_1},\vec{k}_2,m_{i_2},\vec{p'}m'_{j_N} m'_{\tau_N} | 
T_{\pi\pi N,\gamma N}(E)
| \vec{q}\lambda_\gamma,\vec{p}\lambda_N m_{\tau_N}> $ can be calculated from
the partial-wave matrix elements of $\hat{t}_{MB,\gamma N}(E)$,
$\hat{t^R}_{MB,\gamma N}(E)$, and the
vertex interactions $\Delta\rightarrow \pi N$
and $\rho, \sigma \rightarrow \pi\pi$. As an example, the matrix element
of the term  $T^{\rho N}_{\pi\pi N,\gamma N}(E)$ defined by 
Eq.(\ref{eq:hat-trn})
can be calculated from 
\begin{eqnarray}
& &<\vec{k}_1,m_{i_1},\vec{k}_2,m_{i_2},\vec{p'}m'_{j_N} m'_{\tau_N} | 
\hat{T}^{\rho N}_{\pi\pi N,\gamma N}(E)
| \vec{q}\lambda_\gamma,\vec{p}\lambda_N m_{\tau_N}> \nonumber \\
& & = \sum_{ m_{j_\rho},m_{i_\rho}}\sum_{{\it l} s}\sum_{JM_J,T M_T} \sum_{LS}
<j_\pi m_{j_{\pi 1}}, i_\pi m_{i_1};
j_\pi m_{j_{\pi 2}}, i_\pi m_{i_2}|Y^{j_\rho m_{j_\rho}, i_\rho m_{i_\rho}}
_{{\it l},(j_\pi j_\pi)s}(\hat{k_{12}})> \nonumber \\
& & \times
 <j_\rho m_{j_\rho}, i_\rho m_{i_\rho};
j_N m'_{j_N},\tau_N m'_{\tau_N} |Y^{JM_J,TM_T}_{L(j_\rho j_N)S}(\vec{k})>
\frac{f_{\rho,\pi\pi}(k_{12})}
{E-E_N(k)-E_\rho(k) -\Sigma_{\rho N}(k,E) }
 \nonumber \\
& &\times [\hat{t}^{JT}_{LS\rho N, \lambda_\gamma \lambda_Nm_{\tau_N}}(k,q,E)
+\hat{t}^{R,JT}_{LS\rho N, \lambda_\gamma \lambda_Nm_{\tau_N}}(k,q,E) ]
\label{eq:endpipin}
\end{eqnarray}
where $|Y^{J M,TM_T}_{L,(j_1 j_2)S}(\hat{p})> $ 
has been defined in Eq.(\ref{eq:ylm}), $j_\pi=m_{j_\pi}=0$
and hence only $s=0$ and ${\it l}= j_\rho $ are allowed in the sum.

Expressions similar to Eqs.(\ref{eq:dcrstpipin})-(\ref{eq:endpipin})
 can be easily
obtained for the differential cross sections with respect to
the $\pi N$ invariant mass $M_{\pi N}$ by changing the labels of variables.

\section{Numerical Methods }

To illustrate the numerical method we have developed for 
 solving the coupled-channel equation Eq(\ref{eq:pw-tmbmb}) with
a singular particle-exchange driving term $Z^{E}_{MB,M'B'}$,
it is sufficient to consider 
the Alt-Grassberger-Sandhas (AGS) integral equation\cite{ags}
within  a simple 
 three identical bosons model of Amado\cite{amado-63}.
This model
describes the scattering of
a boson $b$ from a two-boson bound state $d$ via a $d\rightarrow bb$
form factor
$ g(q)=g_0/(q^2+\beta^2)$ with $q$ denoting the 
relative momentum between the two outgoing bosons.
The form factor 
is normalized as
$\int k^2dk g^2(k)/(B+\frac{k^2}{m})^2=1$ with $B$ being the binding energy
of the two-boson subsystem.

After partial wave projection, the AGS equation in each partial-wave
is  
\beq
 X(p',p_0,E)
 =Z(p',p_0,E) + \int p^2 dp Z(p',p,E)\tau(p,E) X(p,p_0,E) \ ,
\label{eq:ags}
\eeq
where $ X(p,p_0,E)$ is the half-off-shell
$bd\rightarrow bd$ scattering amplitude.
The one-particle exchange driving term $Z(p',p,E)$ and the
propagator $\tau(p,E)$ 
are calculated by using the familiar 
non-relativistic kinematics.
In the center of mass system, they are
\begin{eqnarray}
Z(p',p,E) &=& \frac12 \int_{-1}^1 dx P_L(x)
\frac{g_0}{(|\vec{p'}+\frac12\vec{p}|^2+\beta^2)}
\frac{g_0}{(|\vec{p}+\frac12\vec{p'}|^2+\beta^2)}
\label{eq:amado-z} \nonumber \\
&\times& \frac1{E -\frac{p^2}{2m}-\frac{p^{'2}}{2m}
-\frac{(\vec{p}+\vec{p'})^2}{2m}+i\epsilon}\,, \\
\tau^{-1}(p,E) &=& (E_2(p,E)+B) \nonumber \\
& & \times \Bigl[1-(E_2(p,E)+B)\int k^2dk
\frac{g^2(k)}{(B+\frac{k^2}{m})^2(E_2(p,E) -\frac{k^2}{m}+i\epsilon)}
\Bigr]\,,
\label{eq:amado-tau}
\end{eqnarray}
where $L$ is the orbital angular momentum, $E_2(p,E)=E-3p^2/4m$,  
and $P_L(x)$ is the Legendre polynomial with 
$x=\hat{p'}\cdot\hat{p}$.

\begin{figure}
\centering
\includegraphics[width=8cm,angle=-0]{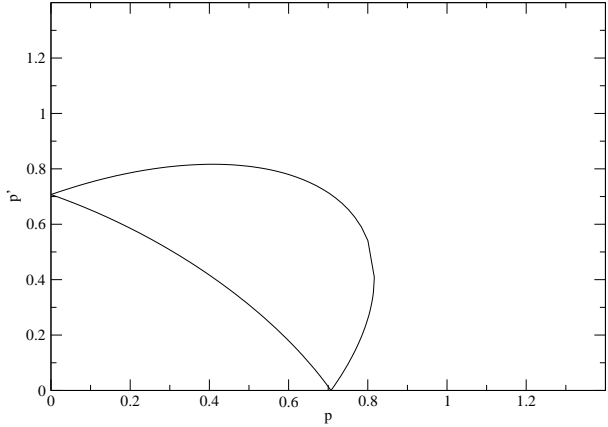}
\caption{ Logarithmically divergent moon-shape region
of the matrix elements 
$Z(p',p,E)$ of Eq.(85) of the Amado Model.
$p$, $p'$ and $E $ are in unit of $\hbar=2m=1$ with $E=1$. }
\label{fig:moon}
\end{figure}

Besides the 2-body bound state pole at $E_2(p,E)+B=E-3p^2/(4m)+B=0$, 
the interaction $Z(p',p,E)$ in the kernel of Eq.(\ref{eq:ags}) has logarithmic
singularity for energies above the three-particle breakup
threshold. 
With the parameters 
$\hbar=2m=1,\ B=1.5,\ \beta=5,$
and the total energy  $E=1$,
one can see from the energy denominator of Eq.(\ref{eq:amado-z})
that
the interaction $Z(p',p,E)$ is singular in the moon-shape region of
Fig.\ref{fig:moon}.
Since the singularity depends on both $p$ and $p'$,
it is  difficult to solve the integral equation Eq.(\ref{eq:ags})
by using the standard subtraction methods.
Although there are well-known methods of contour-deformation
to avoid the singularity, we will solve the equation without
contour-deformation by employing the interpolating function. 
Because mathematical problems of the singular integral equation (\ref{eq:ags})
are well discussed in Ref.\cite{r1} for example, we will concentrate
on the practical  numerical procedures.

Let us choose appropriate grid points $\{p_i\}$ and write
the unknown function $X(p,p_0,E)$ in terms of an interpolation
function $S_i(p)$ 
\beq
 X(p,p_0,E)=\sum_i S_i(p)X(p_i,p_0,E) \ .
\label{eq:ags-z}
\eeq
By inserting Eq.(\ref{eq:ags-z})) into eq.(\ref{eq:ags}), 
one obtains the matrix equation
\beq
 X(p_j,p_0,E) = Z(p_j,p_0,E)
 + \sum_i K_{ji}X(p_i,p_0,E) \ ,
\label{eq:ags-sp}
\eeq
where
\beq
 K_{ji} = \int p^2dp Z(p_j,p,E)\tau(p,E)S_i(p) 
        = \sum_n \int_{p_n}^{p_{n+1}}
            p^2dp Z(p_j,p,E)\tau(p,E)S_i(p) \ .
\label{eq:ags-sp-k}
\eeq
The integration in Eq.(\ref{eq:ags-sp-k}) can be carried out as precisely
as necessary since the interpolation functions $S_i(p)$
are known and the logarithmic singularity can be integrated
as $\int dx \ln(x) = x\ln(x)-x$. The integration over
the 2-body bound state pole of $\tau(p,E)$ can be worked out by using
the standard technique of pole subtraction.

\begin{figure}
\centering
\includegraphics[width=8cm,angle=-0]{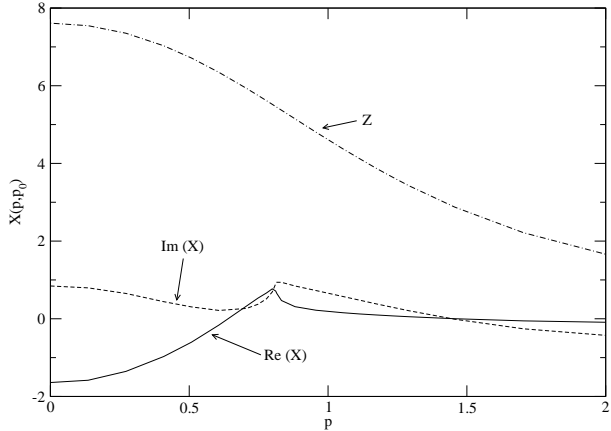}
\caption{Half-off-shell amplitude $X(p,p_0,E)$ of Eq.(85) of the
Amado Model. $p$ and $p_0$ are
 in unit of $\hbar=2m=1$ with $E=1$. The dot-dashed curves are from
deriveing term $Z(p,p_0,E)$ of Eq.(85).}
\label{fig:3boson-x}
\end{figure}

The choice of interpolation functions $S_i(p)$ depends on
the property of the function to be interpolated.
For example, the Lagrange interpolation polynomials are employed
in Ref.\cite{r1} with some care near the breakup threshold.
In the case of polynomial interpolation, however,
some changes in a small region may give rise to global
effects. Therefore it is better to use the spline interpolation
which depends locally on the grids points, i.e., the function
$S_i(p)$  dominates around the grid point $p_i$.
Moreover, the spline interpolation is known to be less oscillating
compared to the polynomial interpolation.

Spline functions are defined in terms of  piecewise polynomials
which are connected smoothly over the whole region.
Since cubic splines are mostly employed, we will explain it
in some detail. There are several kinds of spline functions
depending on the condition of continuity. Among them,
natural splines and Hermitean splines are very useful.
Their characteristic properties are:

(1) natural splines: first and second derivatives are
continuous at the grid points. It is a global spline
in the sense that the function $S_i(p)$ depends on the
whole grid points. It is known that the natural spline
interpolation has a minimum curvature property.

(2) Hermitean splines: Only first derivatives are continuous
at the grid points. It is a local spline in the sense that
the function $S_i(p)\ (p_i \leq p \leq p_{i+1})$ depends on 
4 grid points $\{p_{i-1},\ p_{i},\ p_{i+1},\ p_{i+2}\}$.

Since the practical ways of calculating the spline functions
$S_i(p)$ are well described in Ref.\cite{r2} for natural splines
and in Ref.\cite{r3} for Hermitean splines,
we will not repeat them here.

The choice of spline functions certainly depends on the behavior
of the solution $X(p,p_0)$. As is well-known, there appears
a square-root singularity at the breakup threshold \cite{r1}.
More precisely, the amplitude $X(p,p_0,E)$ goes like
$(p_B-p)^{\ell+1/2}$\ ( $\ell$ is an angular momentum of 
the 2-body bound state) below the breakup threshold $p_B$.
Therefore, in the case of $\ell=0$, the derivative is not
continuous at $p_B$ and there appears a sharp change
of the amplitude. The straightforward application of
the spline interpolation is not suitable since it requires
the smooth continuation. 
One of the ways to take into account this singular
threshold behavior is to divide the whole region $[0,\infty]$
into two regions $[0,p_B]\ [p_B,\infty]$, and employ Hermitean
spline interpolation in each region. It is also recommended that
the grid points are suitably modified to account for
the singularity near the breakup threshold, i.e.,
$p'=\sqrt{p_B^2-p^2} \ (p \leq p_B)$ and
$p'=\sqrt{p^2-p_B^2} \ (p \geq p_B)$.
In order to check the spline interpolation for the square-root
singularity, it is a good exercise to fit the simple model
function
\beq
f(x)=\left\{
 \begin{array}{ll}
 (1-x^2)^{\ell+1/2} & \qquad 0 \leq x \leq 1 \\
 (x^2-1)^{\ell +1/2}e^{-x} & \qquad 1 \leq x < \infty \ 
 \end{array} \right. 
\eeq
and to examine the accuracy of the interpolation.
This exercise also will give some idea about the distribution
of the grid points.

Now we will explain how the spline function method works in
a calculation of Eq.(\ref{eq:ags}) for the Amado model
with the parameters $\hbar=2m=1,\ B=1.5,\ \beta=5,$
and the total energy  $E=1$. As discussed above, the interaction
$Z(p',p,E)$ given in Eq.(\ref{eq:amado-z}) is singular in the 
moon-shape region of Fig.\ref{fig:moon}.
To choose the grid points for solving Eq.(\ref{eq:ags}) with
the input Eqs.(\ref{eq:amado-z})-(\ref{eq:amado-tau}), we  first
identify some typical momenta: the on-shell momentum 
$p_0=\sqrt{4m(E+B)/3}$ of $bd$ elastic scattering, 
the tips of the moon-shape region on the coordinate 
axes $p_{end}=\sqrt{mE}$, the breakup threshold $p_B=\sqrt{4mE/3}$, and
$p_b=\sqrt{mE/3}$ at which the moon-shape boundary has its maximum value
from each coordinate axis.
We then choose
$p_a=0,\ p_b=\sqrt{mE/3}=0.408,
\ p_c=p_{end}=\sqrt{mE}=0.707,
\ p_d=p_B=\sqrt{4mE/3}=0.816,
\ p_f=p_0=\sqrt{4m(E+B)/3}=1.291,
\ p_g=p_{\rm max}=20$
and $p_e=(p_d+p_f)/2=1.053$.
These momenta are chosen to make 6 regions as
$R_a=[p_a,\ p_b],\ R_b=[p_b,\ p_c], R_c=[p_c,\ p_d],
R_d=[p_d,\ p_e], R_e=[p_e,\ p_f], R_f=[p_f, p_g]$.
In addition to the grid points of those typical momenta,
we prepare $\{2,2,4,3,3,9\}$ grid points in each region
respectively, and thus 30 mesh points are used in solving
the matrix equation Eq.(\ref{eq:ags-sp}) .
They are distributed in equal space
for $R_a,\ R_b$ and $R_e$, while modified grid points
$p_i'$ are equally spaced near the breakup threshold for
$R_c$ and $R_d$. 
In the region $R_f$, grid points are distributed
as geometrical series with the ratio $r=1.5$ 
;i.e., $ p= 1.291,\ 1.456,\ 1.704,\ 2.075,\ 2.632,\
3.468,\ 4.722,\ 6.602,\ 9.423,\ 13.65,\ 20$
.

In order to evaluate the integral Eq.(\ref{eq:ags-sp-k}) accurately,
we have employed 4-point Gauss-Legendre integration formula
for each interval $[p_n,\ p_{n+1}]$ which has no singularity.
For the interval including the logarithmic singularity,
we have changed the integration variable by explicitly
taking account the location of the singularity as
\beq
 \int_{p_n}^{p_{n+1}} dp F(p)
 = \int_{-t_1}^{t_2} dt 3t^2F(p_s+t^3) \ ,
\eeq
where $p_s \ (p_n < p_s <p_{n+1})$ is the singular point.
The variable is changed as $p=p_s+t^3$ and $t_1=(p_s-p_n)^{1/3},
\ t_2=(p_{n+1}-p_s)^{1/3}$. This manipulation explicitly removes the
logarithmic divergence from the integrand.

Thus, we have prepared two kinds of mesh points, i.e., one is the
grid points $\{p_i\}$ at which the solution $X(p_i,p_0,E)$ is to be found
by solving the matrix equation Eq.(\ref{eq:ags-sp}),
and the other is to carry out the integration of Eq.(\ref{eq:ags-sp-k}) 
as precisely
as required.

The calculated amplitude $X(p,p_0,E)$ for zero total angular momentum
are the solid curve (real part)  and dashed curve (imaginary part)
shown in Fig.\ref{fig:3boson-x},
which can be compared with the similar calculation of
Ref. \cite{r5}. The amplitude $X(p,p_0,E)$
is dimensionless and
normalized as $X(p_0,p_0)=(\eta e^{2i\delta}-1)/(2i)$ at
the on-shell point.
One can see clearly the square-root singularity at the breakup
threshold. We have also carried out the calculation with
natural splines. Although natural splines are not suitable for
the square-root singularity, it is practically possible to
imitate the singularity by distributing many grid points
around the breakup threshold. For example, the elastic amplitudes
calculated by two different splines agree within
the accuracy of 1\%,
since the on-shell point is away from the breakup threshold.
In practice, both amplitudes coincides fairly well except for
the small region around the breakup threshold.
In Fig.\ref{fig:3boson-x}, we also show the contribution (dot-dashed curve)
from the
driving term $Z(p,p_0,E)$ defined by Eq.(\ref{eq:amado-z}). Its differences
with the solid and dashed curves clearly show that the multiple
scattering effects are very important.

The method described above can be readily extended to solve 
the coupled-channel equation Eq.(\ref{eq:pw-tmbmb}). To be more specific,
let us consider the case of $E=1.88$GeV. As discussed in the previous section,
the partial-wave  matrix elements
of the driving terms
$Z^{(E)}_{\pi\Delta,\pi\Delta}$ and
$Z^{(E)}_{\rho N,\pi\Delta}$ of Eq.(\ref{eq:pw-tmbmb}) diverge logarithmically 
in the moon-shape regions shown in Fig\ref{fig:moon-aos}.
To choose the grid points for applying the spline function expansion method,
we first select the following momenta 
\begin{eqnarray}
p_0 &=& 0 \ ,\\
p_1 &=& \frac{m_\pi}{m_N+m_\pi} \ p_6 \ ,\\
p_2 &=& \frac12 \ p_7 \ ,\\
p_3 &=& \left[ \frac14(E-m_N)^2 - m_\pi^2 \right]^{1/2} \ ,\\
p_4 &=& \frac{m_N}{m_N+m_\pi} \ p_6 \ ,\\
p_5 &=& \left[ \frac14 \left(E-m_\pi+\frac{m_\pi^2-m_N^2}
{E-m_\pi}\right)^2-m_\pi^2 \right]^{1/2}\ ,\\
p_6 &=& \left[ \frac14\left(E+\frac{m_\pi^2-(m_N+m_\pi)^2}{E}\right)^2 
- m_\pi^2 \right]^{1/2}\ ,\\
p_7 &=& \left[ \frac14\left(E+\frac{m_N^2-4m_\pi^2}{E}\right)^2
-m_N^2 \right]^{1/2}\ ,\\
p_8 &=& \left[ \frac14\left(E+\frac{m_N^2-m_\pi^2}{E}\right)^2
-m_N^2 \right]^{1/2}\,. 
\end{eqnarray}
The momentum $p_8$ is the on-shell momentum of the $\pi N$ state.
$p_6(p_7)$ corresponds to the  momentum at which
the invariant mass of the $\pi N$ ($\pi\pi)$  subsystem of the
$\pi\pi N$ state is $m_{12}=m_N+m_\pi$ ($2m_\pi$).
This momentum can be considered as the "breakup" threshold of the
unstable particle channels $\pi\Delta$ ($\rho N$ and $\sigma N$).
Specifically, we take $\{p_0,\ p_1,\ p_2,\ p_3,\ p_5,\ p_6,\ p_7,\ p_8,\ p_{max}\}$
for $\pi$-spectator channel $(\pi N,\pi\Delta)$, and
$\{p_0,\ p_1,\ p_2,\ p_4,\ p_5,\ p_6,\ p_7,\ p_8,\ p_{max}\}$
for $N$-spectator channel $(\rho N,\sigma N)$.
For example, numerical values at $E=1.88$ GeV are : 
$ p_1=  80.29, \  p_2=  334.8, \  p_3=  448.7, \ 
 p_4=  539.1, \  p_5=  605.8, \  p_6=  619.4, \  p_7=  669.7, \ 
 p_8=  696.3 $
and $p_{max}=6000$.
For 8 regions $R_1=[p_0,\ p_1],\ R_2=[p_1,\ p_2],\ldots,R_8=[p_8,\ p_{max}]$,
we prepare $\{3,\ 3,\ 3,\ 3,\ 3,\ 3,\ 3,\ 8\}$ grid points.
The distribution of the mesh points and the
integration over each region
are the same as those for the Amado model.

It is a rather complex numerical task to get accurate solutions of 
Eq.(\ref{eq:pw-tmbmb}).
We check our numerical accuracy by reproducing the
following optical theorem
 within $1\%$
\begin{eqnarray}
\frac{4\pi}{k}Im [ \hat{t}_{MB,MB}(\theta=0)]
=\sum_{M'B'= \pi N, \eta N, \gamma N} \hat{\sigma}_{MB,M'B'}
+ \hat{\sigma}_{MB,\pi\pi N} 
\end{eqnarray}
where $MB= \pi N, \eta N, \gamma N$ are stable particle channels,
the cross sections
$\hat{\sigma}_{a,b}$ are calculated from the non-resonant amplitudes
$\hat{t}_{MB,M'B'}$ by solving Eq.(\ref{eq:pw-tmbmb}). The two-pion production
cross sections $\hat{\sigma}_{MB,\pi\pi N}$ are calculated from 
the amplitudes Eqs.(\ref{eq:hat-t1})-(\ref{eq:hat-tsn})
 with resonant amplitude
$t^{R}_{MB,M'B'}=0$.

\section{Results}

\begin{figure}
\centering
\includegraphics[width=7cm,angle=-90]{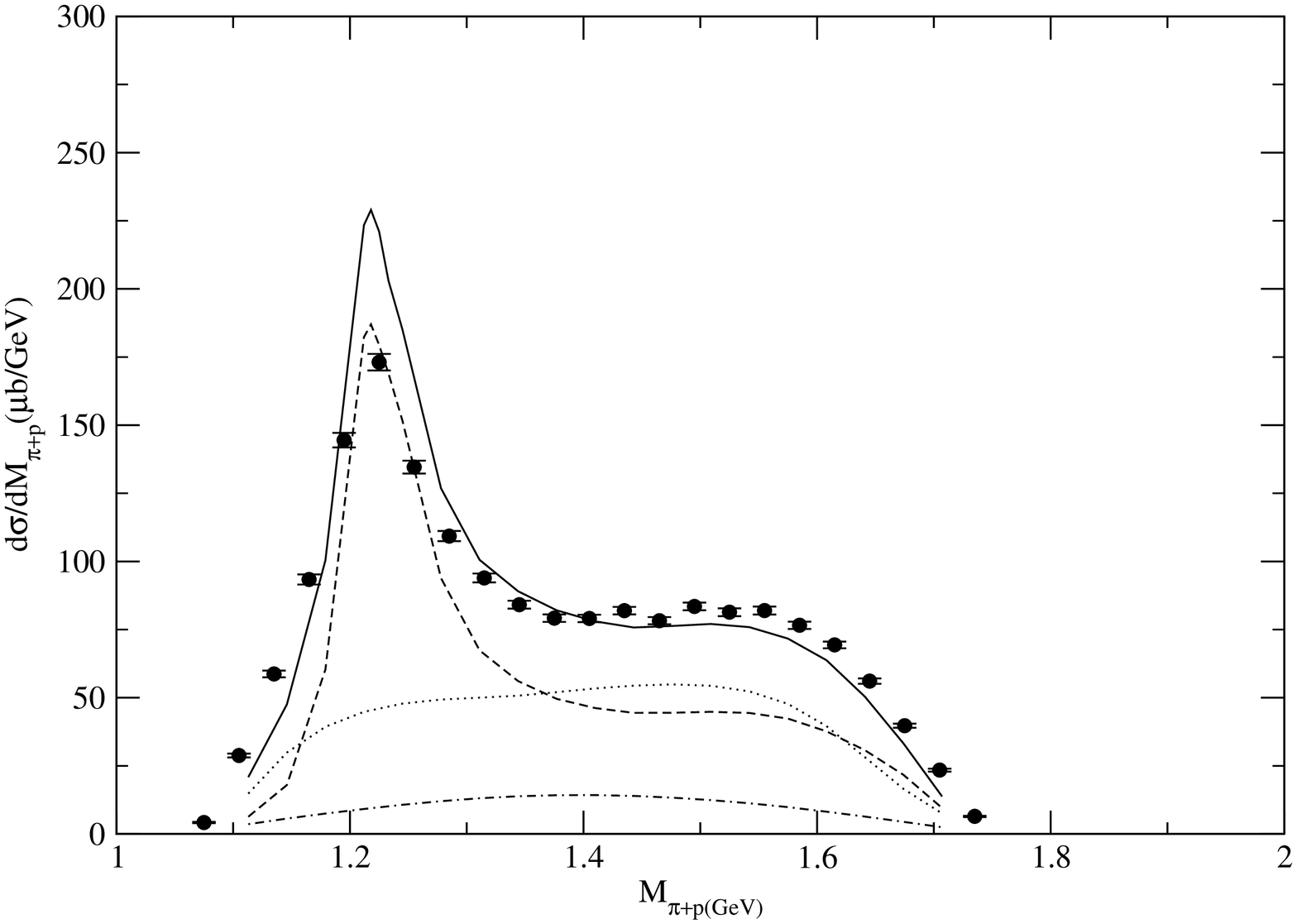}
\includegraphics[width=7cm,angle=-90]{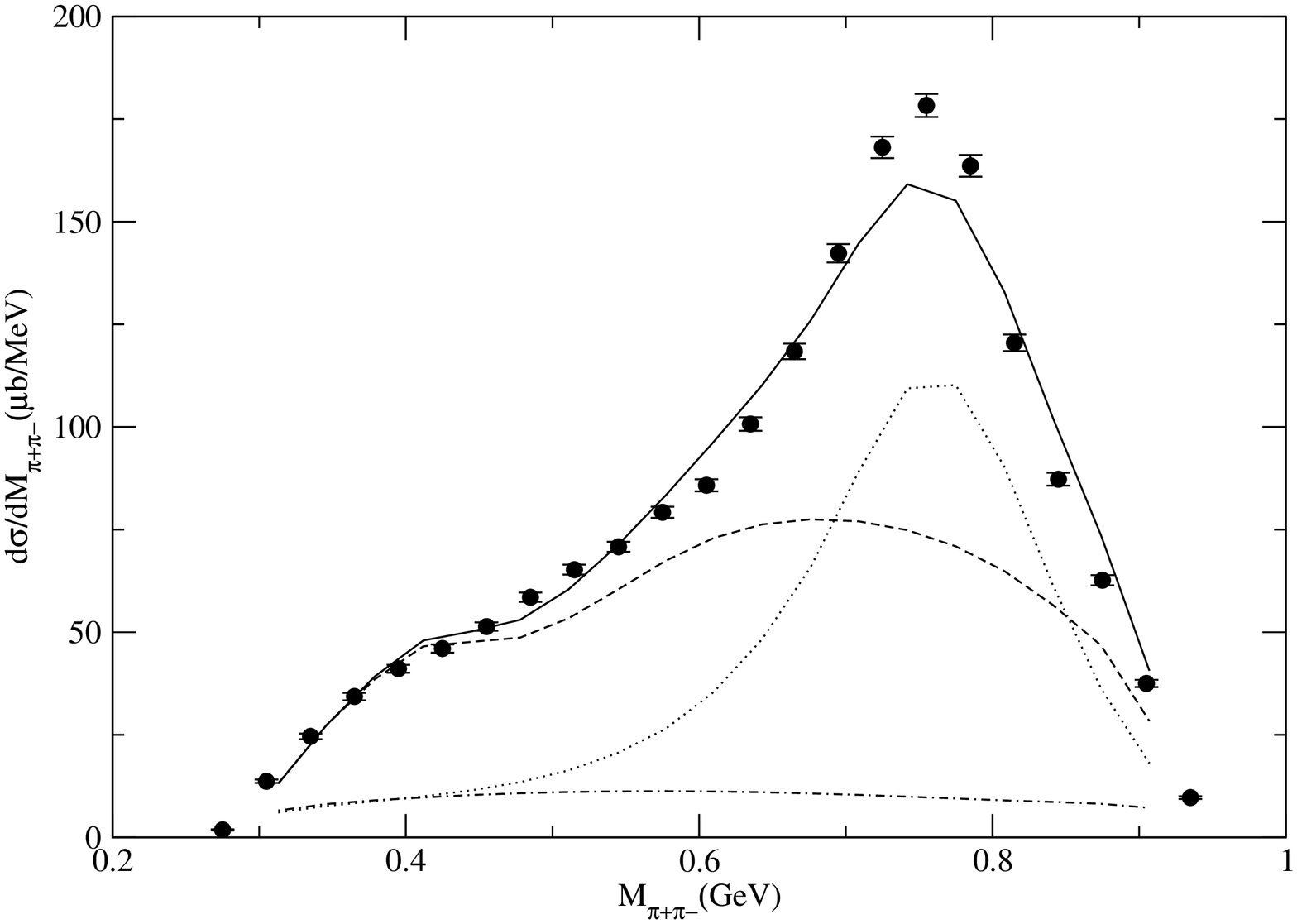}
\includegraphics[width=7cm,angle=-90]{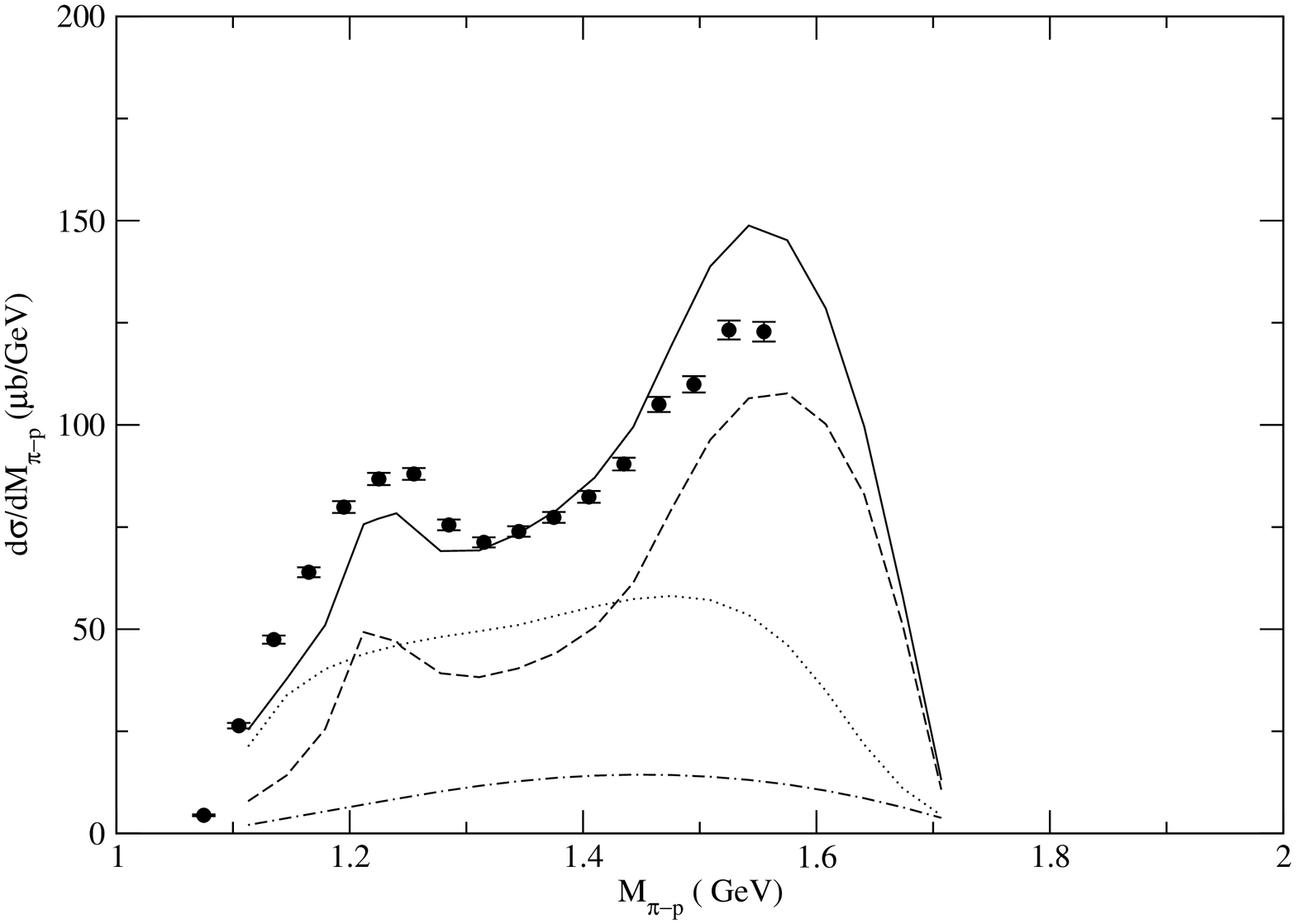}
\caption{The differential cross sections of
  $\gamma p \rightarrow \pi^+ \pi^- p$ reaction with respect to
the invariant mass $M_{\pi^+ p}$ (top), $M_{\pi^+\pi^-}$ (middle), and
$M_{\pi^- p}$ (bottom) at W=1.880 GeV. The data are from Ref.\cite{mokeev-1}.   
The solid curves are from full calculations,
The contributions from $\pi \Delta$ (dashed), $\rho N$ (dotted)
and $\sigma N$ (dot-dashed) to the invariant mass
distributions of $\gamma p \rightarrow \pi^+\pi^- p$ are also shown.}
\label{fig:mech-1880}
\end{figure}

Our main interest in this paper is to use the numerical methods described 
in section V to examine the dynamical consequences of 
the one-particle-exchange interaction $Z^{(E)}_{\pi\Delta,\pi\Delta}$,
$Z^{(E)}_{\rho N,\pi\Delta}$, 
and $Z^{(E)}_{\sigma N,\pi\Delta}$ (Fig.\ref{fig:z}). 
 As illustrated in Fig.\ref{fig:zpdpd} and discussed in section IV, 
the matrix elements of these interactions
have logarithmically divergent
structure due to the $\pi\pi N$ unitarity
cuts which are not accounted for in all of the
recent calculations of two-pion production. 
The parameters needed to evaluate the
partial-wave matrix elements of $Z^{(E)}_{\pi \Delta,\pi \Delta}$,
$Z^{(E)}_{\rho N,\pi \Delta}$,
and $Z^{(E)}_{\sigma N,\pi \Delta}$ are fixed by the
fitting the  low-energy
$\pi N$ and $\pi\pi$ scattering partial-wave amplitudes, 
as given in Appendices D and E.
With the resonant amplitudes also fixed by using the information
of PDG to evaluate Eqs.(\ref{eq:pdgres})-(\ref{eq:pdgres-1}), 
our first task is to choose the parameters of 
starting Lagrangians, given in Appendix A,
to evaluate the partial-wave matrix elements
${\it v}^{JT}_{L' S' M'B', L S MB}(k',k)$ defined in Appendix C
and ${\it v}^{JT}_{L' S' M'B', \lambda_\gamma \lambda_N}(k',q)$ in
Appendix F,  with $MB, M'B' =\pi N, \eta N, \pi\Delta, \rho N, \sigma N$. 
Here we are guided by the previous works on
meson-exchange models of $\pi N$ and $NN$ interactions, as
discussed in  Appendix A. We also need to regularize the resulting
matrix elements of all of the non-resonant interactions
given explicitly in Appendices C and F.
This is done by multiplying  each strong interaction vertex in the
considered non-resonant mechanisms,
illustrated in Figs.\ref{fig:mbmb}-\ref{fig:mbpipin},
 by a
form factor $[\Lambda^2/(\Lambda^2+\vec{k}^2)]^2$ with $\vec{k}$ being the
momentum associated with the meson at the $MBB$ vertex or the meson
being-exchanged. 
We adjust the cutoff
parameters $\Lambda$ 
as well as some of the less well determined 
coupling constants to get a reasonable description of
the Jlab data of invariant mass distributions of 
$\gamma p \rightarrow \pi^+ \pi^- p$ reactions. 
With the parameters listed in Tables I-II of Appendix A, 
our results (solid curves)
of the invariant mass
distributions are compared with the data at $W=1.88$ GeV
in Fig.\ref{fig:mech-1880}.
While the improvements are clearly needed,
the chosen parameters are sufficient for our present
very limited purposes of investigating the effects due to
$\pi\pi N$ cut. No attempt is made here  to adjust the parameters to
fit all
of the available data of $\gamma p \rightarrow \pi^+\pi^- p,
\pi^0\pi^0 p, \pi^+\pi^0 n$. This can be meaningfully pursued in
a coupled-channel approach only when
the data of $\pi N \rightarrow \pi N, \eta N, \pi\pi N$ and
$\gamma N \rightarrow \pi N, \eta N$ are also considered.
Here we focus on the effects
due to the $\pi\pi N$ cut
which are neglected in all recent two-pion production
calculations..

To see the dynamical content of our calculations, we also show
in Fig.\ref{fig:mech-1880} the contributions from each of the
unstable $\pi\Delta, \rho N, \sigma N$ channels. 
 The $M_{\pi^+ p}$ distribution (top panel) is
clearly dominated by the process 
$\gamma p \rightarrow \pi \Delta
\rightarrow \pi\pi N$ (dashed). The peak near $M_{\pi^+p} \sim 1.23$ GeV
is dominated by the 
$\gamma p \rightarrow \pi^- (\Delta^{++}\rightarrow \pi^+ p)$
process,
while the shoulder in the $M_{\pi^+p} \sim 1.4-1.6$ region is due to
the $\gamma p \rightarrow \pi^+ (\Delta^{0}\rightarrow \pi^- p)$ process.
The contributions from the $\rho N$ (dotted curve) and 
$\sigma N$ (dot-dashed curve) are sizable
and can change the shape and magnitude
of the cross sections through interference
effects.  The $M_{\pi^+\pi^-}$ distribution (middle panel) is dominated
by $\gamma p \rightarrow  p(\rho\rightarrow \pi^+\pi^-) $ (dotted)
and hence is peaked at $M_{\pi^+\pi^-} \sim 0.76 $ GeV. However
the contribution from $\pi\Delta$ channel (dashed) are clearly important
in getting the good description of the data. The situation
for $M_{\pi^-p}$ distribution (right) is similar to that for
$M_{\pi^+p}$ distribution (bottom panel), except that the relative strength
between two $\Delta$ peaks is changed.

\begin{figure}[h]
\centering
\includegraphics[width=6cm,angle=-0]{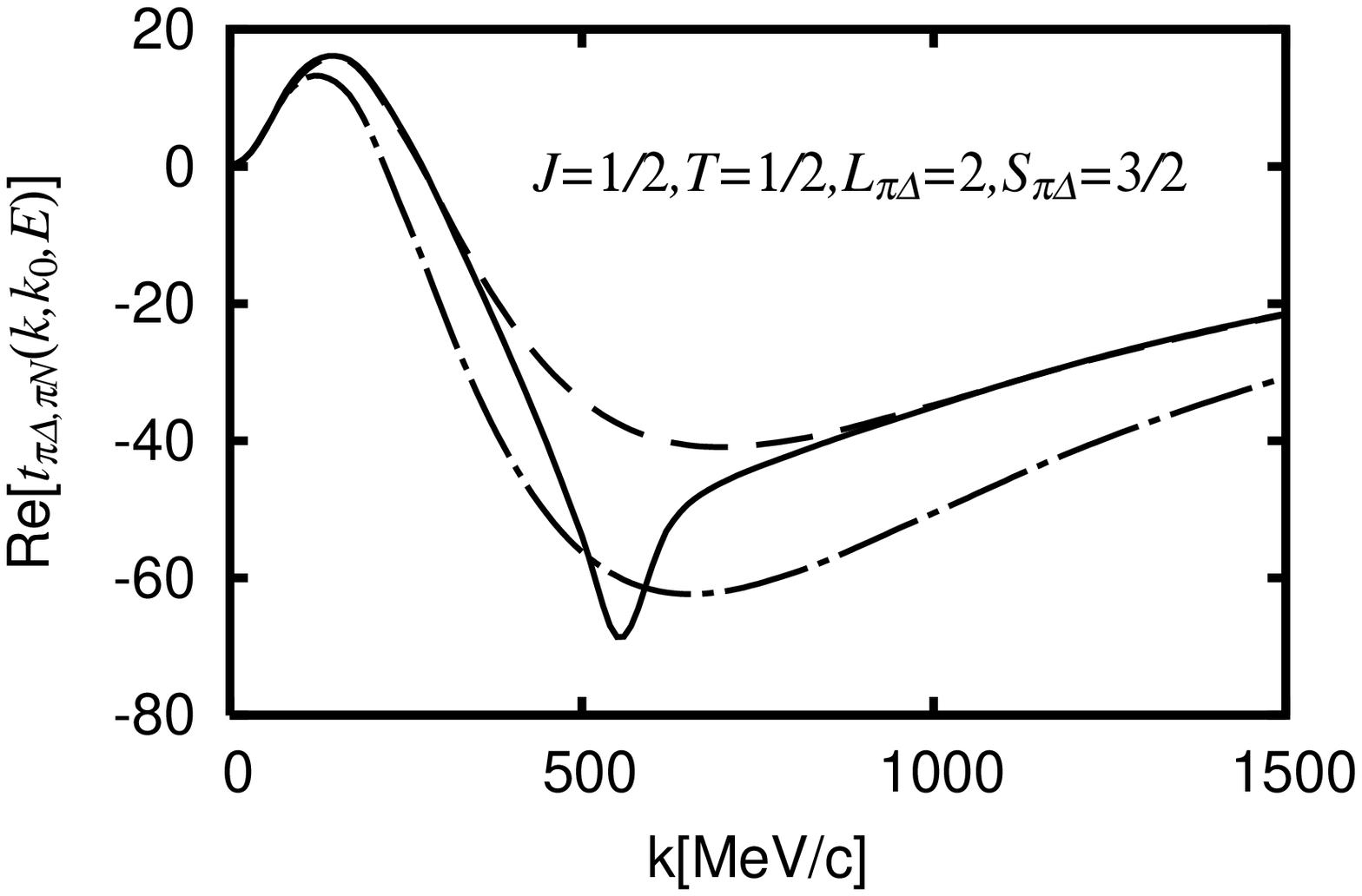}
\includegraphics[width=6cm,angle=-0]{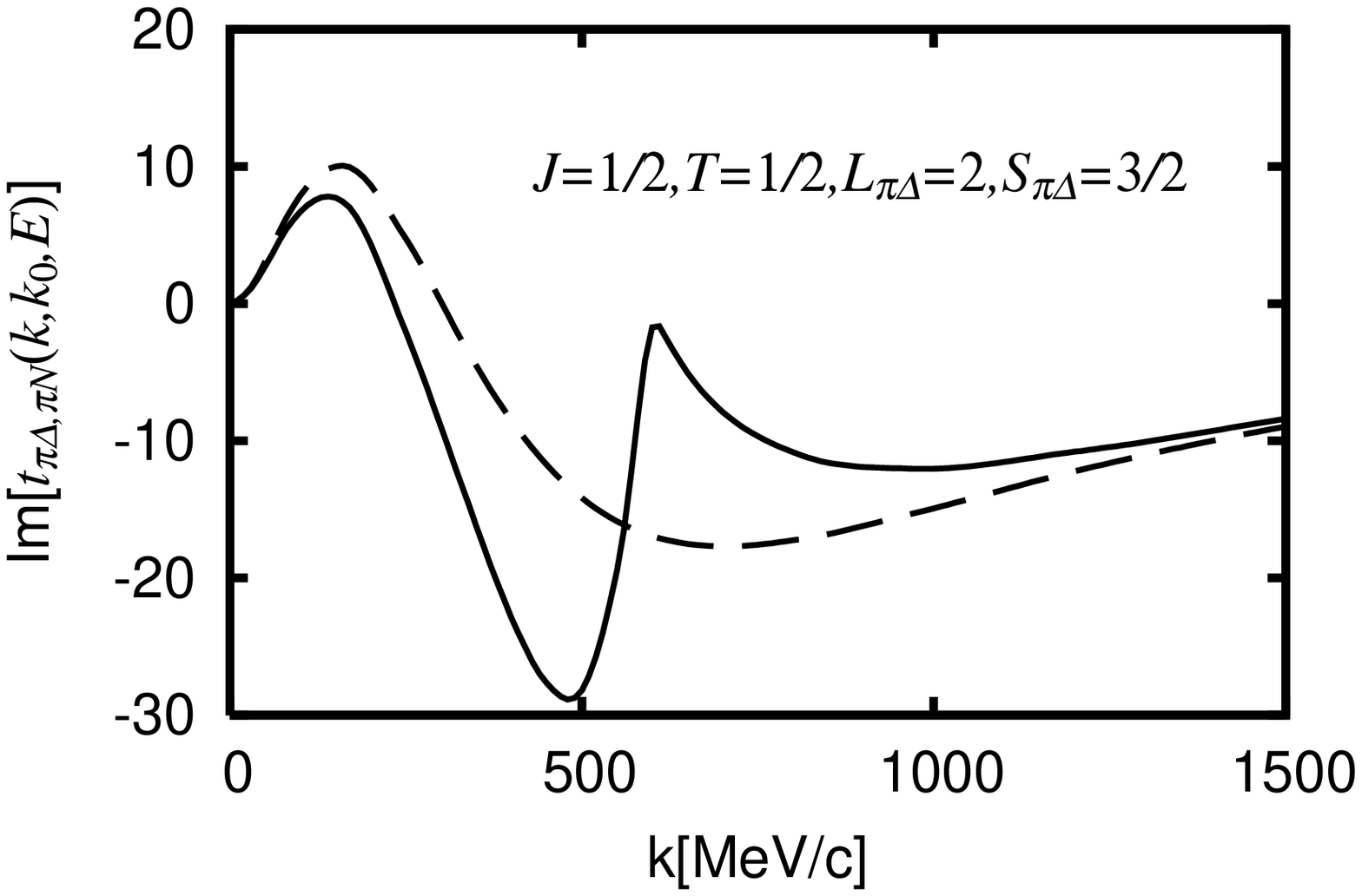}
\includegraphics[width=6cm,angle=-0]{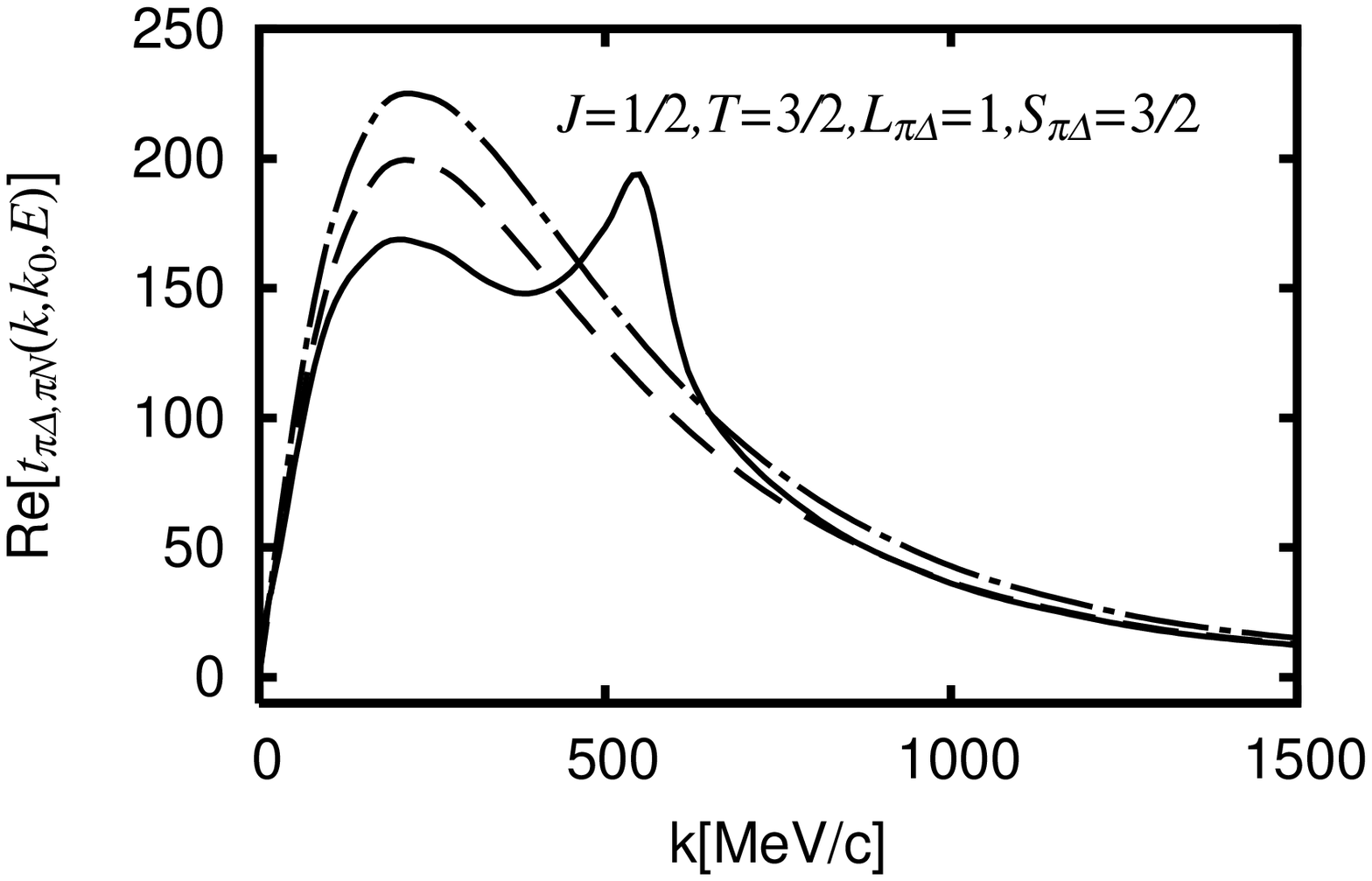}
\includegraphics[width=6cm,angle=-0]{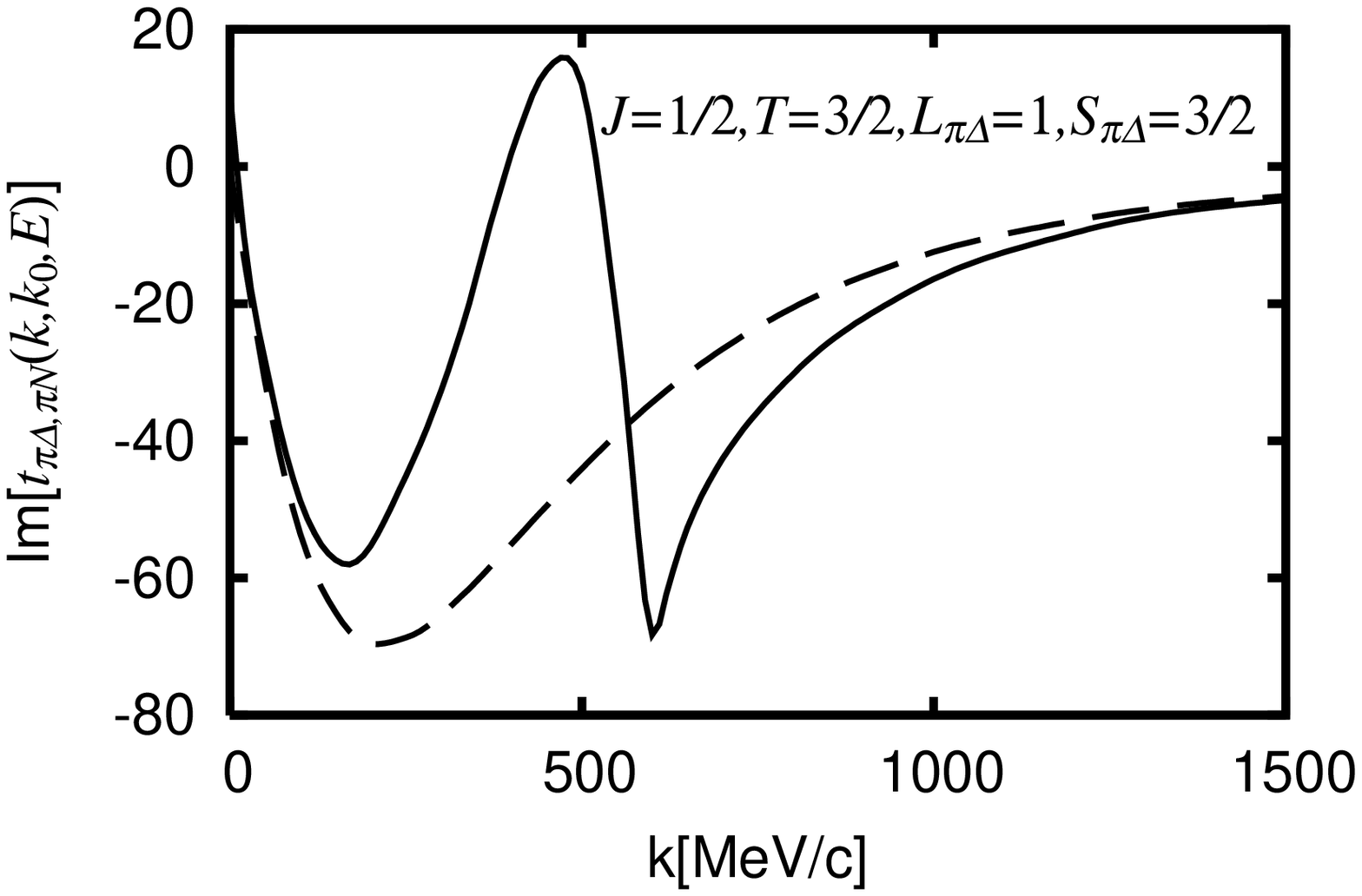}
\caption{ The half-off-shell amplitudes $\hat{t}_{\pi\Delta,\pi N} (k, k_0, E)$.
The invariant mass of the outgoing $\Delta$ is 1.232 GeV and the
total energy is E=1.880 GeV.
The left (right) hand sides are the real (imaginary) parts of the amplitudes
with $\pi N$ in $S_{11}$ (top) and $P_{31}$ (bottom). The
partial-wave quantum numbers for $\pi \Delta$ state are indicated
in each figure. The solid curves are from
full coupled-channel calculations. The dashed curves are from setting
$Z_{\pi\Delta,\pi\Delta}^{(E)}(E)=Z_{\rho N,\pi\Delta}^{(E)}(E)=
Z_{\sigma N,\pi\Delta}^{(E)}(E)=0$. The dot-dashed curves are from 
further setting multiple scattering terms of Eq.(55) to zero; i.e.
setting $\hat{t}^{JT}_{L' S' M'B', L S MB}(k',k,E)=
{v}^{JT}_{L' S' M'B', L S MB}(k',k,E)$. Note that the matrix elements
of $v_{\pi\Delta,\pi N}$ are real in our phase convention (see Appendix A)
and hence there is no dot-dashed curves in the
right hand side. }
\label{fig:pid-t}
\end{figure}

\begin{figure}[h]
\centering
\includegraphics[width=6cm,angle=-0]{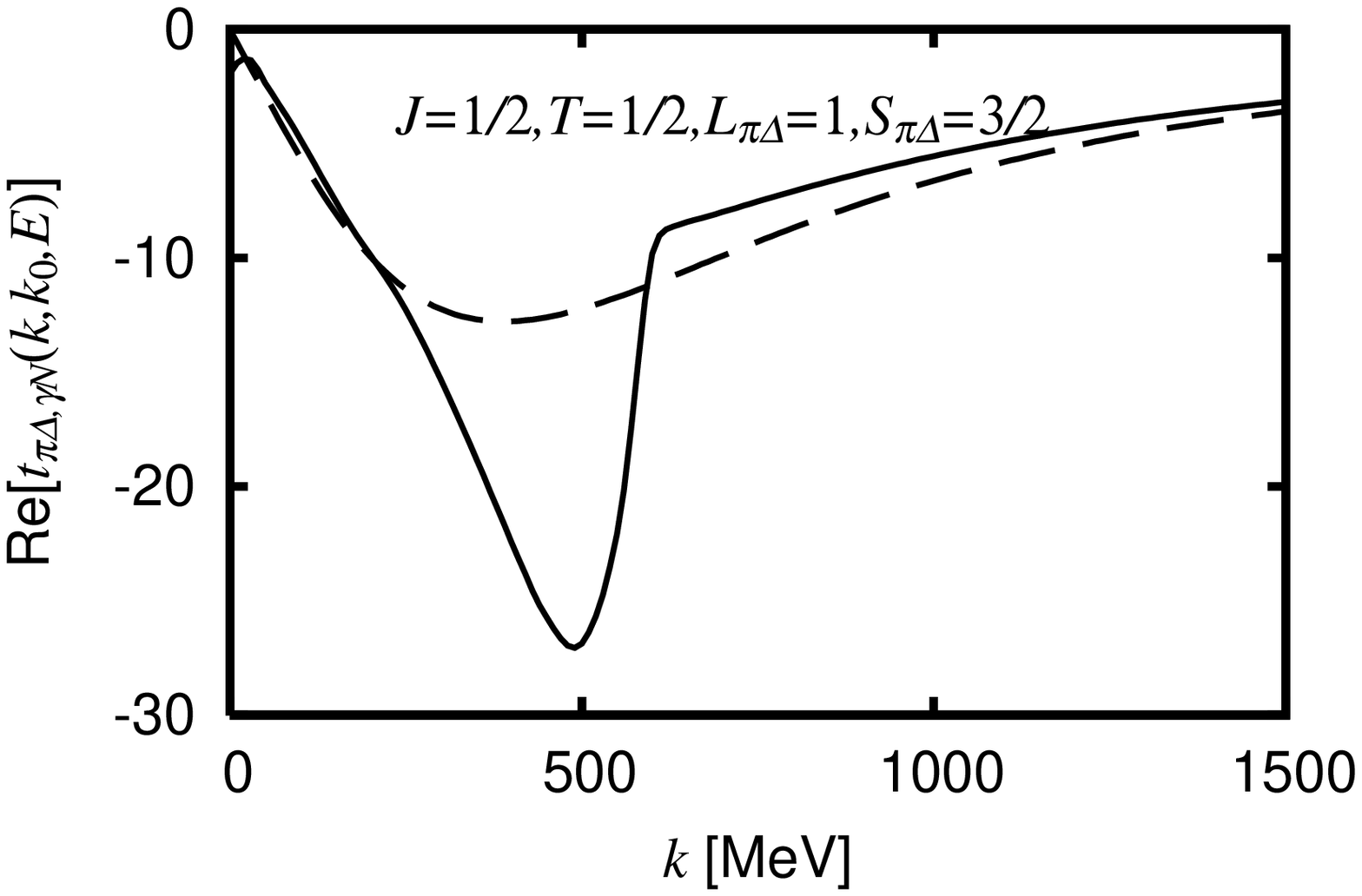}
\includegraphics[width=6cm,angle=-0]{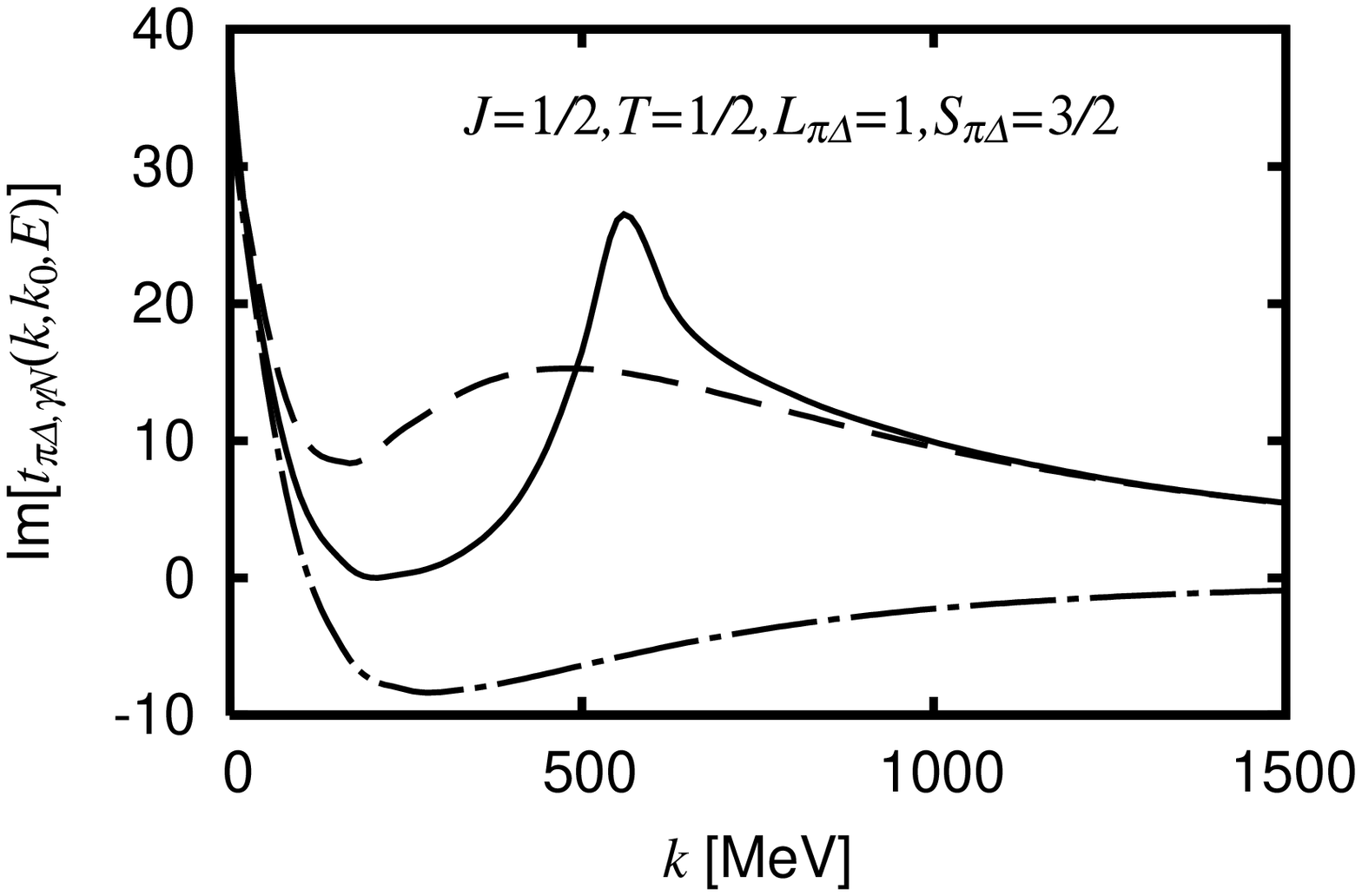}
\includegraphics[width=6cm,angle=-0]{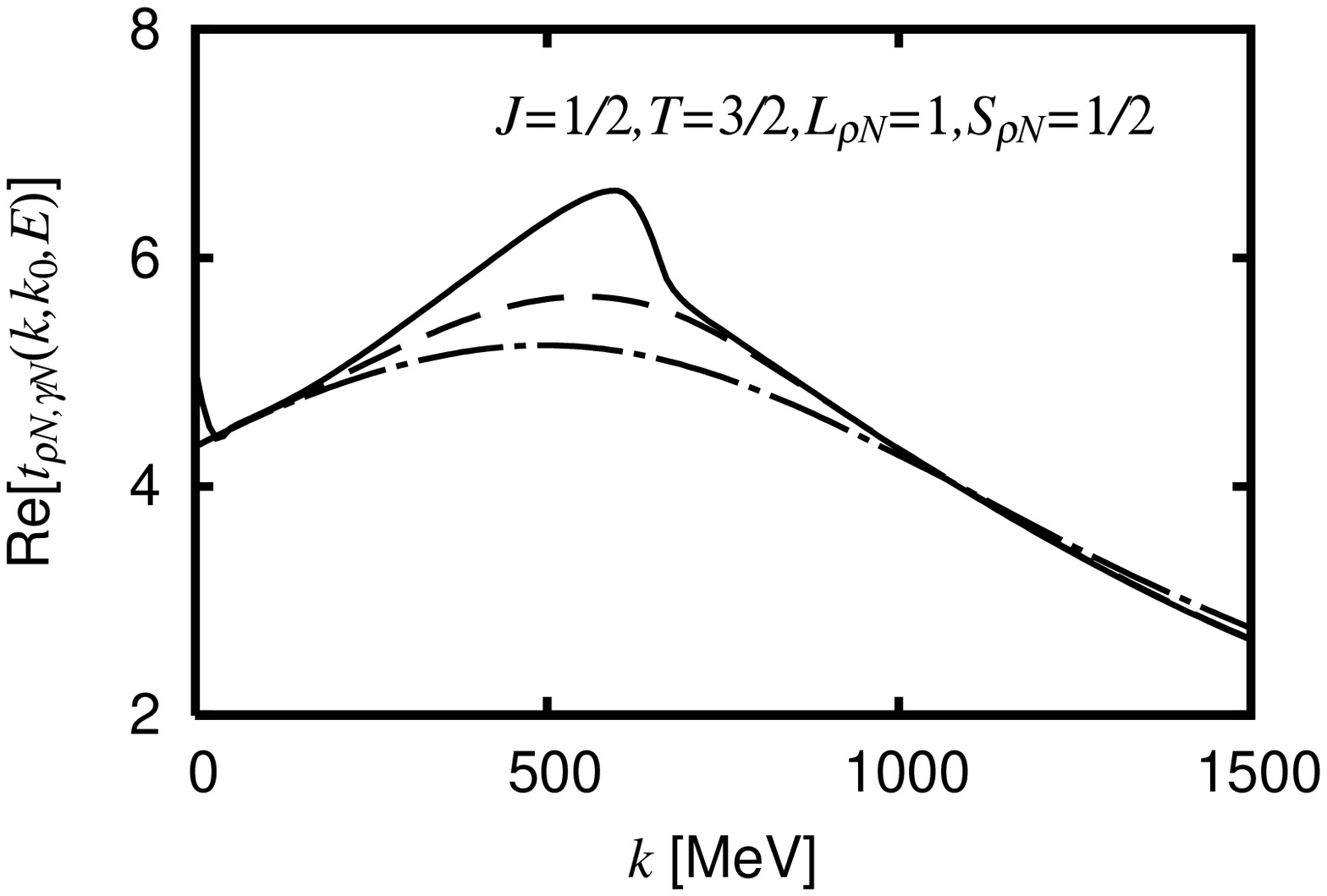}
\includegraphics[width=6cm,angle=-0]{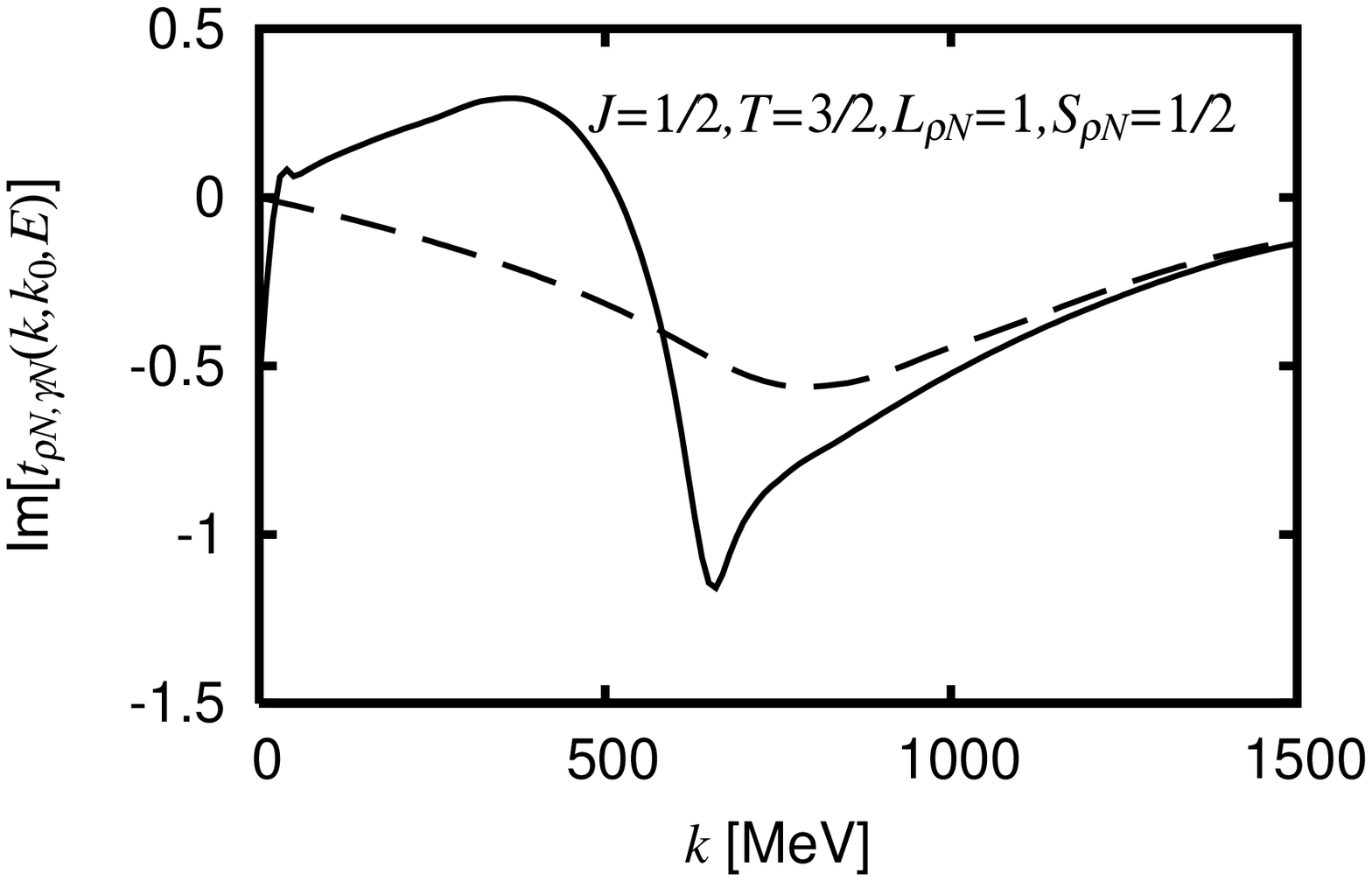}
\caption{The half-off-shell amplitudes 
$\hat{t}_{\pi \Delta, \gamma N} (k, q, E)$  (upper)
 and $\hat{t}_{\rho N, \gamma N} (k, q, E)$ (lower).
The invariant mass of the outgoing $\Delta$ ($\rho$) is 1.232 GeV (0.76 GeV)
and the total energy is E=1.880 GeV.
The partial-wave quantum numbers for the final
$\pi \Delta$ and $\rho N$ states are indicated
in each figure.  The solid curves are from
full coupled-channel calculations. The dashed curves are from setting
$Z_{\pi\Delta,\pi\Delta}^{(E)}(E)=Z_{\rho N,\pi\Delta}^{(E)}(E)=
Z_{\sigma N,\pi\Delta}^{(E)}(E)=0$. The dot-dashed curves are from
further setting multiple scattering terms of Eq.(65) to zero; i.e.
setting $\hat{t}^{JT}_{LSMB,\lambda_\gamma \lambda_N m_{\tau_N}}(k',k,E)=
{v}^{JT}_{LSMB,\lambda_\gamma \lambda_N m_{\tau_N}}(k',k,E)$. 
Note that the matrix elements
of $v_{\pi\Delta,\gamma N}$ ($v_{\rho N,\gamma N}$ ) are 
pure imaginary (real) in our phase convention (see Appendix A)
and hence there is no dot-dashed curves in the right (left) sides
of the upper (lower) parts.}
\label{fig:amp-1880b}
\end{figure}

We now turn to investigating the effects due to
the one-particle-exchange driving terms
$Z_{\pi\Delta,\pi\Delta}^{(E)}(E)$,$Z_{\rho N,\pi\Delta}^{(E)}(E)$,
and $Z_{\sigma N,\pi\Delta}^{(E)}(E)$  
which contain the effects due to the $\pi\pi N$ unitarity cut,
as discussed in section IV. Their singularity structure, 
illustrated in Fig.\ref{fig:moon-aos}, is similar to that shown in
Fig.\ref{fig:moon} of the three-boson case.
We thus expect that non-resonant   
 partial-wave amplitudes associated with $\pi\Delta$, $\rho N$,
and  $\sigma N$ states
 have similar momentum-dependent structure of
Fig.\ref{fig:3boson-x}. This is confirmed in our calculations.
Some of our typical results are shown in
Fig.\ref{fig:pid-t}
for $t_{\pi\Delta,\pi N}$ and Fig.\ref{fig:amp-1880b} for the
photo-production amplitudes $t_{\pi \Delta, \gamma N}$(upper panel)
and $t_{\rho N, \gamma N}$(lower panel).
The solid curves in these figures are from our full calculations, which show
rapid varying structure. 
When the driving terms
$Z_{\pi\Delta,\pi\Delta}^{(E)}(E)$, $Z_{\rho N,\pi\Delta}^{(E)}(E)$,
 and $Z_{\sigma N,\pi\Delta}^{(E)}(E)$
are turned off
in solving Eq.(\ref{eq:pw-tmbmb}), we obtain slow varying
dashed curves.
Here we note that
the momentum variable $k$ in Figs.\ref{fig:pid-t}-\ref{fig:amp-1880b}
is related to the sub-energy $\sigma(k,E)
=E - E_s(k)$ for the resonant particle ($\Delta$ or $\rho$)  to decay
in the presence of a spectator particle $s$ ($\pi$ or $N$) with energy $E_s(k)$.
Thus the full curves in
Figs.\ref{fig:pid-t}-\ref{fig:amp-1880b} also reflect the rapid dependence
on the sub-energy $\sigma(k,E)$.
We emphasize that the
 rapid dependence of these amplitudes 
on the sub-energy $\sigma(k,E)$ 
is a necessary consequence of $\pi\pi N$ unitarity condition, as
discussed by Aaron and Amado\cite{aa-76}, and
is similar to what
can be seen in the $\pi NN$ studies\cite{matsuyama,AM-1,AM-2}.
Our results clearly indicate that the usual tree-diagram approximation
should be used with cautions in interpreting the extracted $N^*$ parameters.
The rapidly varying structure associated with
an unstable particle channels must be
taken into account in any phenomenological extraction of the
partial-wave amplitudes. These were not taken into account in the early
partial-wave analyses\cite{manley-84} of the data of
$\pi N \rightarrow \pi\pi N$.

If we further turn off the multiple scattering mechanisms in solving
coupled-channel equation Eq.(\ref{eq:pw-tmbmb}), we get the dot-dashed curves
in Figs.\ref{fig:pid-t}-\ref{fig:amp-1880b}.
The large differences between the dash-dotted curves and the solid curves
indicate the difference between the dynamical coupled-channel
approaches and the recent tree-diagram models.
 
\begin{figure}[h]
\centering
\includegraphics[width=8cm,angle=-0]{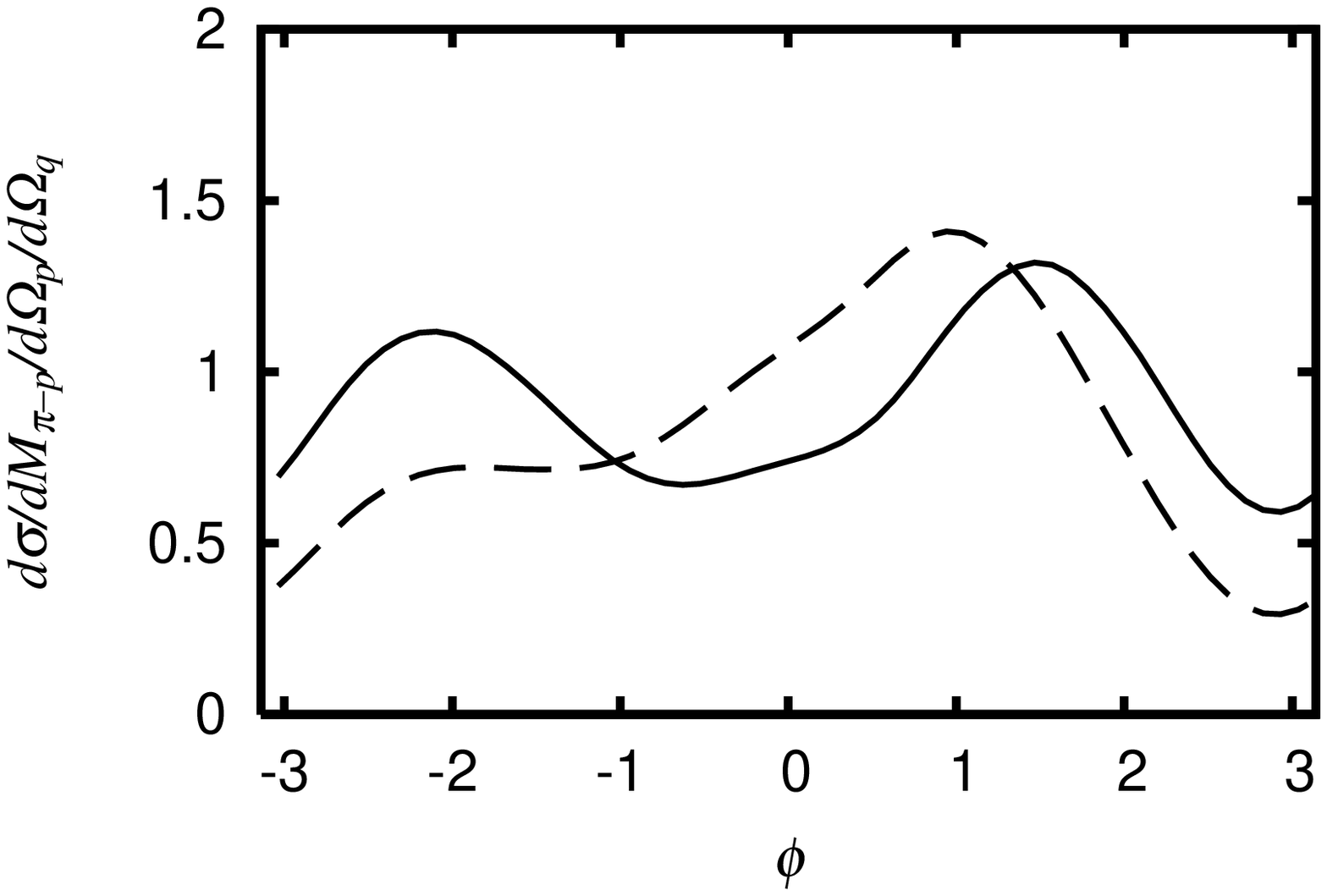}
\includegraphics[width=8cm,angle=-0]{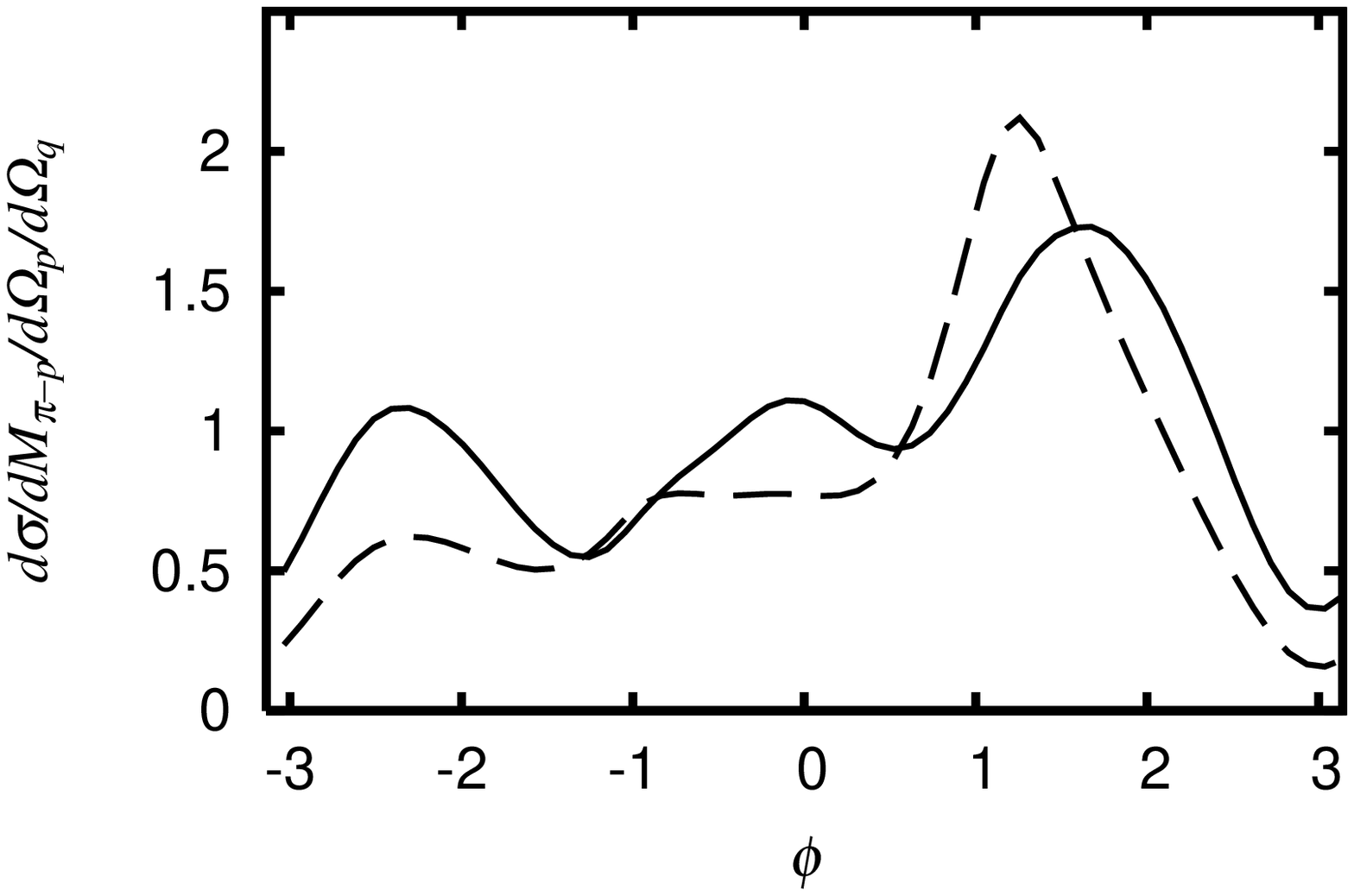}
\caption{ Differential cross sections of
$\gamma p \rightarrow \pi^+ \pi^- p$ at the $\gamma N$ invariant
mass W=1.880 GeV. 
The outgoing $\pi^+$ momentum
is $\vec{p}$ and the relative momentum between $\pi^-$ and
$p$ is $\vec{q}$.  $\phi$ is the azimuthal angle of
$\vec{q}$. The results are for the invariant mass
$M_{\pi^- p}=1.23$ GeV, $\cos\theta_p=0.183$, $\phi_p= -3.1$ rad.
The left (right) panel is for $\cos\theta_q=0.80 (0.183)$.
The dashed curves are obtained when $Z^{(E)}_{MB,M'B}$ term is turned off
in solving Eq.(65).  }
\label{fig:ds-1880a}
\end{figure}

We next examine the effects of the one-particle-exchange terms 
$Z_{\pi\Delta,\pi\Delta}^{(E)}(E)$, $Z_{\rho N,\pi\Delta}^{(E)}(E)$,
and $Z_{\sigma N,\pi\Delta}^{(E)}(E)$ on 
the differential cross sections of $\gamma p \rightarrow \pi^+\pi^-p$. 
Here we set $\vec{p}$ as
the outgoing $\pi^+$ momentum, 
$\vec{q}$ the relative momentum between $\pi^-$ and
$p$. Two of our typical 
results of the dependence of the differential cross sections  
$d\sigma/(dM_{\pi^- p}d\Omega_{p}d\Omega_{q})$  on the azimuthal
angle $\phi$ of $\vec{q}$ are shown in Fig.\ref{fig:ds-1880a} with the final
$\pi^+ \pi^- p$  kinematics fixed at
$M_{\pi^- p}=1.23$GeV, $\cos\theta_p=0.183$, $\phi_p= -3.1$ rad.,
and $\cos\theta_{q} = 0.80 (left), 0.183(right)$.
Our full results are the solid curves. The dotted curves are
obtained when
$Z^{(E)}_{\pi\Delta,\pi\Delta}$, $Z_{\rho N,\pi\Delta}^{(E)}(E)$, 
and $Z_{\sigma N,\pi\Delta}^{(E)}(E)$
are turned off
in solving the coupled-channel equation Eq.(\ref{eq:pw-tmbmb}).
Clearly, the effects due to these one-particle-exchange terms
 are very pronounced
in changing both the shapes and magnitudes of the differential
cross sections. Similar results are also seen in our calculations
for other values of $\vec{p}$ of the outgoing $\pi^+$ and
$\vec{q}$ of the relative momentum of the outgoing $\pi^-p$ system.
The results shown in  Fig.\ref{fig:ds-1880a}
further indicate that the rapid varying structure of
the amplitudes shown in Figs.\ref{fig:pid-t}-\ref{fig:amp-1880b}
must be accounted for in any analysis of two-pion production.

\begin{figure}[h]
\centering
\includegraphics[width=8cm,angle=-0]{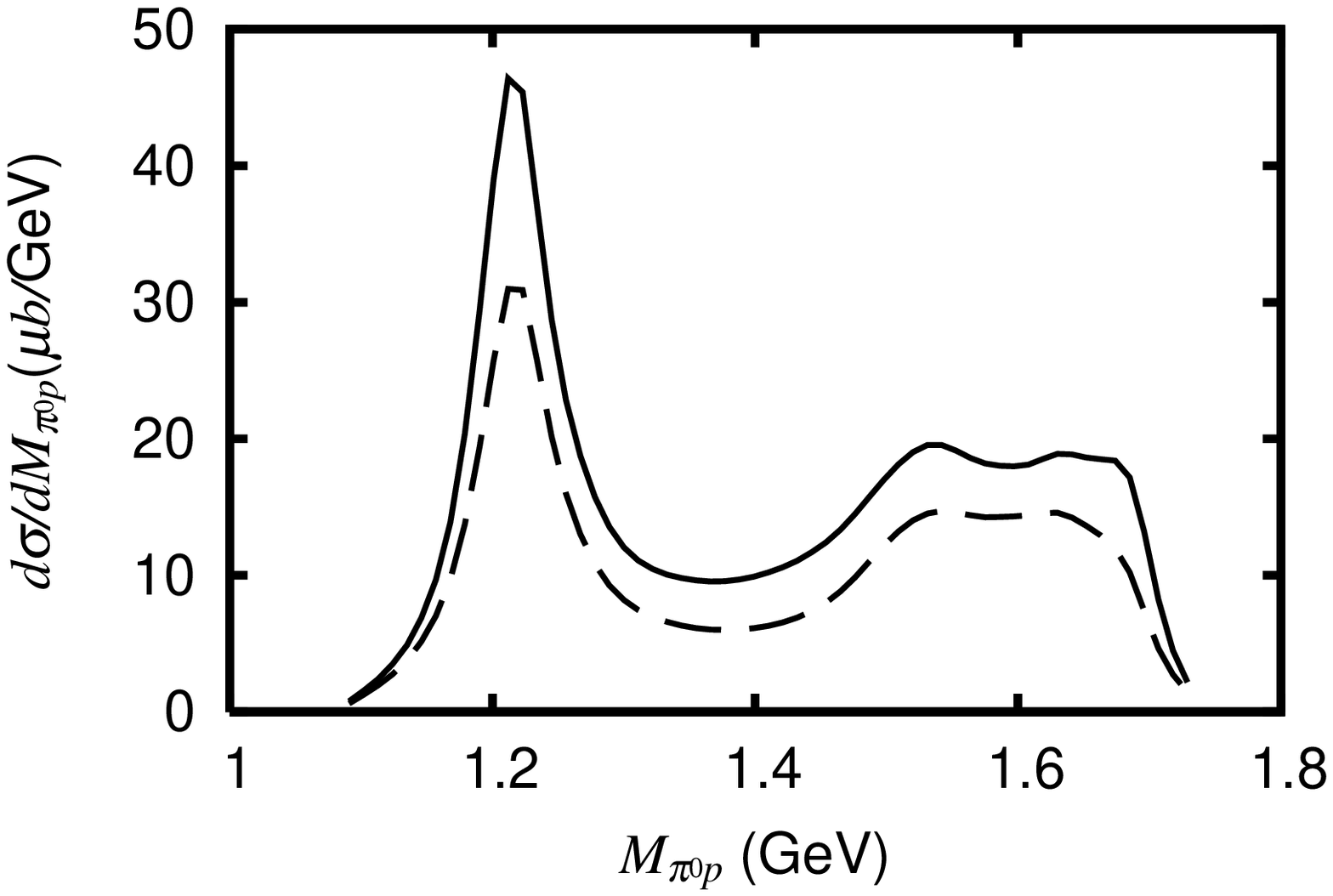}
\includegraphics[width=8cm,angle=-0]{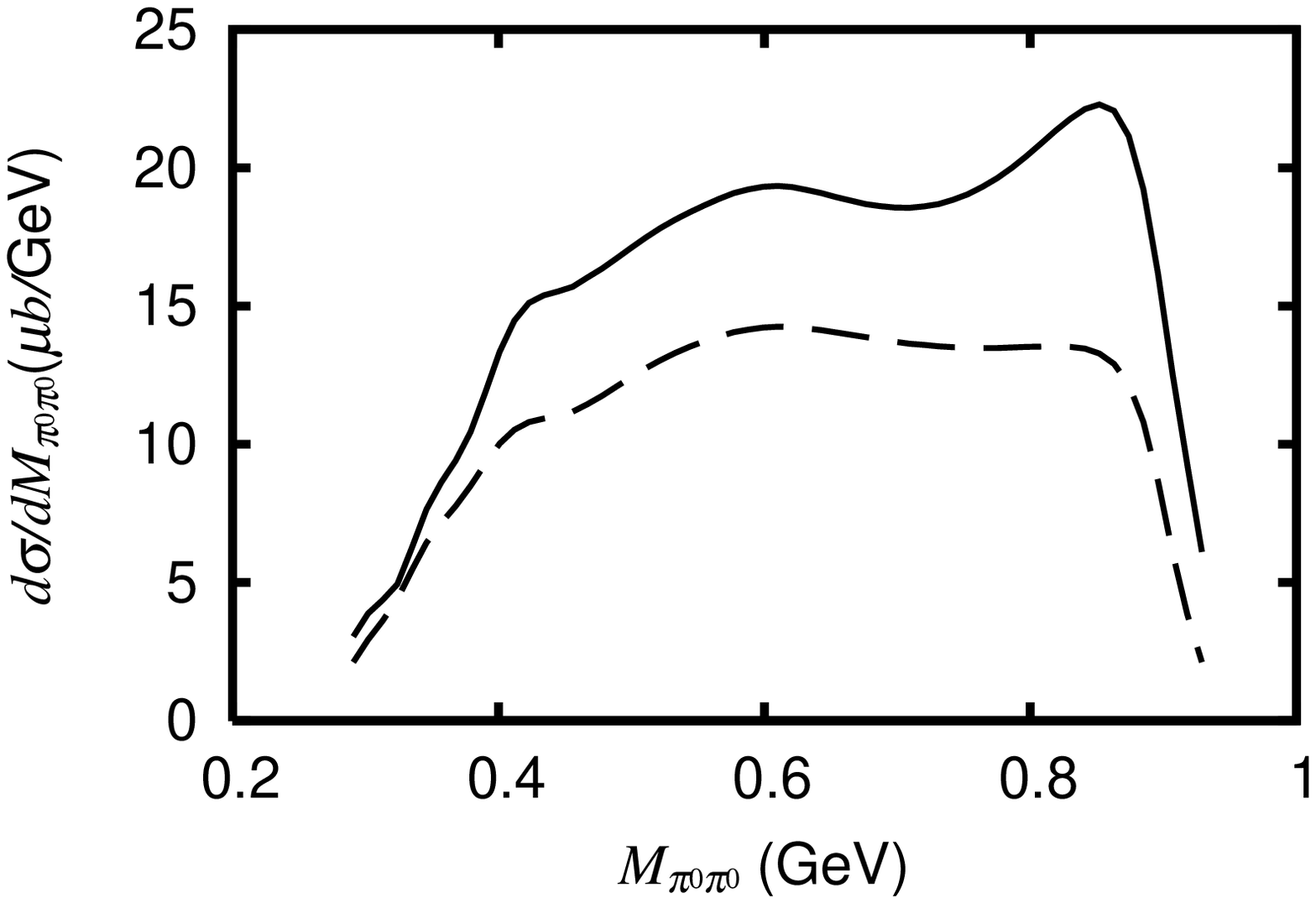}
\caption{ The invariant mass distributions of
$\gamma p \rightarrow \pi^0 \pi^0 p$ reaction at W=1.88 GeV.
The dashed curves are obtained when $Z^{(E)}_{MB,M'B}$ term is turned off
in solving Eq.(65).}
\label{fig:all-1880b}
\end{figure}

In the recent studies of two-pion production, the data of
invariant mass distributions $d\sigma/dM_{\pi N}$ and
$d\sigma/dM_{\pi\pi}$ of $\gamma N \rightarrow \pi\pi N$ are
most commonly used to extract $N^*$ parameters. Since these cross sections
involve integrations over angles of outgoing particles, 
as seen in Eq.(\ref{eq:crst:dcrstm12}),
the rapid varying structure of the partial-wave amplitudes
due to $\pi\pi N$ cut is washed out. 
We thus see the smooth distributions $d\sigma/dM_{\pi N}$ and
$d\sigma/dM_{\pi\pi}$, as shown in Figs.\ref{fig:mech-1880}.
However the one-particle-exchange terms 
$Z_{\pi\Delta,\pi\Delta}^{(E)}(E)$,
$Z_{\rho N,\pi\Delta}^{(E)}(E)$, and $Z_{\sigma N,\pi\Delta}^{(E)}(E)$
can change their magnitudes and shapes significantly.
One example is shown in
in Fig.\ref{fig:all-1880b} for
$\gamma p \rightarrow \pi^0\pi^0 p$. We see that when these
one-particle-exchange driving terms 
are turned off in solving coupled-channel equation Eq.(\ref{eq:pw-tmbmb}),
 the predicted invariant mass distributions are reduced
significantly. Such a large difference further indicate the importance of
including the $\pi\pi N$ cut effects in calculating
these particle-exchange terms for analyzing the two-pion production data.

\section{Summary and Future Developments}

For analyzing the meson production data in the nucleon resonance ($N^*$) region,
we have developed a dynamical coupled-channel reaction model.
With the assumption that the basic degrees of freedom of the considered
reactions  
are mesons ($M$) and baryons ($B$), our starting point is an
energy-independent effective Hamiltonian which is
derived from a set of Lagrangians by using a unitary transformation method.
Within the constructed Hamiltonian, the $N^*$ excitations are defined by 
bare $N^* \rightarrow MB, \pi\pi N$ vertex interactions
and the non-resonant meson-baryon interactions are defined by
 the tree-diagrams 
generated from the considered Lagrangians.  
We then apply the standard projection operator
techniques\cite{feshbach} to derive coupled-channel equations for calculating
the amplitudes of meson-baryon reactions. 
The model satisfies the unitary conditions within
the channel space spanned by the considered two-particle
meson-baryon states and the three-particle $\pi\pi N$ state.
In this paper, we present explicit formulations within a Fock-space
spanned by the basis states $\gamma N$, $\pi N$,
$\eta N$, $\pi \Delta$, $\rho N$, $\sigma N$, and $\pi \pi N$. 
However, the formulation
can be straightforwardly extended to include other meson-baryon states 
such as Kaon-Hyperon (KY) and $\omega N$, and other two meson production
channels such as $ \eta \pi N$ and $K\bar{K} N$.

To facilitate the interpretations of the extracted $N^*$ parameters,
we cast the reaction amplitudes  into a form such that
the meson-baryon scattering effects on $N^*$ excitations can be
explicitly calculated. These effects, called the meson cloud effects, 
are due to the mechanisms that
the incident meson interacts with the baryons through all possible
non-resonant scattering before the $N^*$ is excited by the
bare $N^* \rightarrow MB$ vertex interaction of the model Hamiltonian.
The determination of the meson cloud effects from the meson production data
could be useful for
interpreting the extracted $N^*$ parameters in 
terms of hadron structure calculations.
For example, it was found in
Refs.\cite{sl-1,sl-2} that the meson cloud effects can account for
the  main  differences  between 
the  extracted $\gamma N \rightarrow \Delta$ (1232) resonance transition
form factors and the 
constituent quark model predictions.
It will be interesting to explore how the meson cloud effects, as defined
in our formulation, can be related to the
$current$ Lattice QCD calculations.

In addition to giving a complete presentation of our theoretical framework,
we also present  in this paper a numerical method based on
a spline-function expansion 
for solving the resulting 
coupled-channel equations which contain logarithmically divergent
one-particle-exchange driving terms.
These driving terms contain the effects due to 
the $\pi\pi N$ unitarity cuts which must be included accurately
in calculating the two-pion production observables. We explain
how this method can be applied in practice for  a simple
three-boson Amado model, and 
then for our realistic model with $\gamma N, \pi N,
\pi\Delta, \rho N, \sigma N$, and $\pi\pi N$  channels. 

An another important step in carrying out numerical calculations is
to find an efficient way to calculate a large number of
 partial-wave matrix elements of  
the considered non-resonant meson-baryon interacting terms which are needed for
solving the coupled-channel equations.
Here we make use of the helicity representation of Jacob and Wick and
also introduce a helicity-LSJ mixed-representation which is most convenient for
calculating the electromagnetic matrix elements. While these are rather
technical details, but are also presented explicitly in this paper
for the completeness in explaining our numerics.

With the parameters of the model
chosen appropriately to fit JLab's two-pion photo-production data, we apply 
the developed numerical
methods to show that  the logarithmically divergent
one-particle-exchange driving terms in the
constructed coupled-channel equations generate rapid varying 
structure in the matrix elements of
reaction amplitudes associated 
with unstable particle channels $\pi \Delta$, $\rho N$, and $\sigma N$.
Our results confirm the analysis by Aaron and Amado\cite{aa-76}.
We further show that 
these one-particle-exchange terms have 
 large effects in
determining the two-pion production differential cross sections
both in shapes and magnitudes.
Our findings suggest that one needs to be cautious in
interpreting the $N^*$ parameters extracted from
the approaches which
do not account for the effects due to the $\pi\pi N$ unitarity cuts.

The  calculations presented in this paper are far from complete within
our formulation, while they are sufficient for testing the accuracy of
our numerical methods and illustrating  the importance of 
$\pi\pi N$ unitarity cut.  The $N^*$ parameters can be convincingly extracted
and properly interpreted only when we apply our full formulation
to analyze all available data of meson production reactions.
Obviously this is a rather complex process. We now discuss
how we will accomplish this rather ambitious research project in practice.

Our first task is to fit the $\pi N$ elastic scattering data to fix the
parameters defining the strong interaction parts 
of the model Hamiltonian Eqs.(\ref{eq:H-2})-(\ref{eq:v23}).
This must be done by extending the coupled-channel
calculations described in
section IV in two aspects, First, we must include the driving 
term $Z^{(I)}_{MB,M'B'}$ defined by Eq.(\ref{eq:z-mbmb-i}).
As discussed in section III, this term
 contains the $\pi\pi N$ cut effects
originated from the $\pi NN$ vertex. 
Our second main task is to develop appropriate parameterizations of
the bare $N^*\rightarrow MB$ form factors for
calculating the resonant amplitudes rigorously according to
Eqs.(\ref{eq:tmbmb-r})-(\ref{eq:nstar-mb}). Here we 
need to make use of the predictions from
hadron structure calculations. For example, we at least can
 fix the relative phases between different $N^*\rightarrow MB$ transitions
by using the naive SU(6) quark model with meson-quark coupling.
Predictions from more sophisticated models, such as the
$^3P_0$ model of Ref.\cite{capstick-roberts} and the model based
on Dyson-Schwinger Equation\cite{roberts}, could provide useful
information to our investigation. 
In fitting the $\pi N$ elastic scattering data, we
should also fit the available 
$\pi N \rightarrow \eta N$ reaction data and  use the
optical theorem to make sure that the predicted
$\pi N$ total cross sections are also in agreement with the data.

Once the $\pi N$ data are fitted by the above procedures, most of the strong
interaction vertexes in the non-resonant electromagnetic interactions 
$v_{\gamma N, MB}$ and $v_{\gamma N \rightarrow \pi\pi N}$ of our model
Hamiltonian have also been determined. We thus can focus on
the determination of $\gamma N \rightarrow N^*$ form factors.
From Eq.(\ref{eq:nstar-mb}), one can use the operator relations 
Eqs.(B32)-(B33) of Appendix B to write the dressed $N^* \rightarrow \gamma N$
vertex of the resonant amplitude (Eq.(\ref{eq:tmbmb-r})) as
\begin{eqnarray}
\bar{\Gamma}_{N^* \rightarrow \gamma N}(E)
 &=&  \Gamma_{N^* \rightarrow \gamma N} +
\sum_{MB} \Gamma_{N^*\rightarrow M B}
G_{M B }(E)t_{M B,\gamma N}(E) \nonumber \\
 &\equiv&  \Gamma_{N^* \rightarrow \gamma N} +
\sum_{MB} \bar{\Gamma}_{N^*\rightarrow M B}
G_{M B }(E)v_{M B,\gamma N}.
\end{eqnarray}
Since $\bar{\Gamma}_{N^*\rightarrow M B}$ in the second line of the above 
equation has been determined in
the fit to the $\pi N$ reaction data, 
the bare $N^*\rightarrow \gamma N $ vertex
 $\Gamma_{N^* \rightarrow \gamma N}$  is the main
unknown and can be determined by fitting the
data of photo-production and 
electro-production of $\pi, \eta$ and two pions.
 Of course some less well-determined parameters
in the non-resonant interaction $v_{\gamma N, M^\prime B^\prime}$ should 
also be adjusted in the fits.
In practice, one can extract
bare $N^*\rightarrow \gamma N$ form factor at each $Q^2$.
It of course will be more interesting if the parameterization of the bare
form factor $\Gamma_{N^* \rightarrow \gamma N}$
can be guided by some theoretical
calculations. 

We now turn to discussing the extension of the model to include
$KY$ and $\omega N$ channels which are also useful in probing
the structure of $N^*$.
In particular, we note that
$\eta N, K\Lambda$, and $\omega N$ channels are of isospin $T=1/2$.
The properties of $T=1/2$ $N^*$ states can therefore
be more selectively
extracted from analyzing the production data of these three channels.
Thus, an extension of the formulation
presented in this paper to include $KY$ 
and $\omega N$ channels is highly desirable and
technically straightforward. However, it will increase the needed
computation effort enormously.
Nevertheless, we can make use of the results from
fitting the $\pi N$, $\eta N$ and $\pi\pi N$ data
to perform simplified coupled-channel analyses of the
 $KY$ and $\omega N$ production data.
This can be done by following the approach of Ref.\cite{bruno}.

Considering the $KY$ production, we assume that it can be described by 
a coupled-channel model including  $\gamma N$, $KY$, $\pi N$,
and a dummy channel $QQ$ which represent all of the neglected channels.
If we further assume that $KY$ does not couple
directly with the $QQ$ state (mainly because
there is no information about how
$KY$ couples with $\pi\pi N$ channels), one can 
cast the coupled-channel equation Eq.(\ref{eq:cc-mbmb}) into 
the following form
\begin{eqnarray}
t_{\gamma N,KY}(E)&=&v_{\gamma N,KY}[1+G_{KY}(E)t_{KY,KY}(E)]
+v_{\gamma N,\pi N}G_{\pi N}(E)t_{\pi N,KY}(E)
\label{eq:ky-tmbmb}
\end{eqnarray}
with
\begin{eqnarray}
t_{KY,KY}(E)&=&v^{eff}_{KY,KY}(E) [1+G_{KY}(E)t_{KY,KY}(E)]
\label{eq:ky-1} \,, \\
t_{KY,\pi N}(E) &=&[1+t_{KY,KY}(E)G_{KY}(E)]v_{KY,\pi N}
[1+G_{\pi N}(E)\hat t_{\pi N,\pi N}(E)] \,.
\label{eq:ky-2}
\end{eqnarray}
Here the effective $KY$ interaction is defined by
\begin{eqnarray}
v^{eff}_{KY,KY}(E)=v_{KY,KY}+v_{KY\pi N}G_{\pi N}(E)[1+\hat{t}_{\pi N,\pi N}(E)
G_{\pi N}(E)]v_{\pi N, KY}\,,
\label{eq:ky-3}
\end{eqnarray}
and $\hat{t}_{\pi N, \pi N}$ is from solving the coupled-channel equation
Eq.(\ref{eq:cc-mbmb}) in the $\pi N\oplus QQ$ space.

If we assume that the dummy channel $QQ= \eta N\oplus \pi\Delta\oplus 
\rho N\oplus \sigma N$, the scattering amplitude
$\hat{t}_{\pi N,\pi N}$ in the above equations 
is just the solution of Eq.(\ref{eq:cc-mbmb}) of the model determined
 in the fit to $\pi N$ data described above.
 We therefore can 
use this information to solve Eqs.(\ref{eq:ky-1})-(\ref{eq:ky-3}) and 
determine the parameters associated with the non-resonant interaction
$v_{KY,\pi N}$ and $v_{KY,KY}$ by fitting the available
data of $\pi N \rightarrow KY$ reactions. This will then allow us to generate
$t_{KY,KY}$ and $t_{\pi N,KY}$ to evaluate Eq.(\ref{eq:ky-tmbmb}) and
also fix the strong vertexes in the non-resonant $v_{\gamma N, KY}$.
The $K\Lambda $ photo-production and electro-production data can then be used
to extract the $\gamma N \rightarrow N^*$ form factors for $T=1/2$ $N^*$
states.
The same procedure can be used to analyze the $\omega N$ production data.

To end this paper, we would like to emphasize
here that the objective of performing dynamical coupled-channel analyses
of meson production data is not only to extract the $N^*$ parameters,
but also to provide information on reaction mechanisms for interpreting
the extracted $N^*$ parameters in terms of the quark-gluon
substructure of hadrons.
In particular, we account for the
dynamical consequences of the $\pi\pi N$ unitarity condition
which is very difficult, if not impossible, to be treated rigorously
in the existing approaches for calculating the hadron structure
or the Lattice QCD calculations.
An another important point to note is that
our approach accounts for
the off-shell scattering
effects which describe the meson-baryon scattering wavefunctions in the
short range region where we want to explore the structure of $N^*$.
These  essential quantum-mechanical effects are absorbed in the
parameters of the approaches based
on tree-diagram models or K-matrix models. Thus
our dynamical approach perhaps has a better chance than these two
approaches  in  revealing the
quark-gluon substructure of baryons. 
Our progress in this direction will be published\cite{jlms} elsewhere.

\vspace{1.0cm}
We would like to thank B. Julia-Diaz and K. Tsushima for their help
in checking our calculations of the matrix elements of
non-resonant interactions. 
This work is supported by 
the U.S. Department of Energy, Nuclear Physics Division, under
Contract No. W-31-109-ENG-38 and
 the Japan Society for the Promotion of Science
Grant-in-Aid for Scientific Research (C) 15540275.

\clearpage
\appendix

\section{ Lagrangian }

In this appendix, we specify a set of Lagrangians for deriving the non-resonant
interactions $v_{MB,M'B'}$ which is the input to the coupled-channel equations
Eq.(\ref{eq:cc-mbmb}). Here we are guided by the previous works on 
meson-exchange models of $\pi N$ and $NN$ interactions.
The coupling with pseudo-scalar mesons $\pi$ and $\eta$ are consistent with
chiral symmetry. The vector meson couplings are less known and are mainly
constructed phenomenologically.
In the convention of Bjoken and Drell\cite{bj}), the Lagrangian 
with $\pi$, $\eta$, $N$, and $\Delta$ fields are
\begin{eqnarray}
L_{\pi NN} &=& -\frac{f_{\pi NN}}{m_\pi}\bar{\psi}_N\gamma_\mu \gamma_5 
\vec{\tau}\psi_N\cdot \partial^\mu\vec{\phi_\pi} \,, \label{eq:L-pinn} \\
& & \nonumber \\
L_{\pi N\Delta} &=& -\frac{f_{\pi N\Delta}}{m_\pi}\bar{\psi}_\Delta^\mu
\vec{T}\psi_N\cdot \partial_\mu\vec{\phi_\pi} \label{eq:L-pind}\,, \\
& & \nonumber \\
L_{\pi\Delta\Delta}&=&\frac{f_{\Delta\Delta\pi}}{m_\pi}\bar{\psi}_{\Delta\mu}
\gamma^\nu \gamma_5 \vec{T}_\Delta \psi_\Delta^\mu
\cdot \partial_\nu\vec{\phi}_\pi \,, \\
& &  \nonumber \\
L_{\eta NN} &=& -\frac{f_{\eta NN}}{m_\eta}\bar{\psi}_N\gamma_\mu \gamma_5
\psi_N \partial^\mu\phi_\eta \,.
\end{eqnarray}
The interactions involving $\rho$ meson are
\begin{eqnarray}
& & \nonumber \\
L_{\rho NN} &=& g_{\rho NN}\bar{\psi}_N[\gamma_\mu -
\frac{\kappa_\rho}{2m_N}\sigma_{\mu\nu}\partial^\nu]
\vec{\rho^\mu} \cdot \frac{\vec{\tau}}{2}\psi_N \label{eq:L-rnn} \,, \\
& & \nonumber \\
L_{\rho N\Delta} &=& -i\frac{f_{\rho N\Delta}}{m_\rho}\bar{\psi}_\Delta^\mu
\gamma^\nu \gamma_5 \vec{T} \cdot
[\partial_\mu \vec{\rho_\nu}-\partial_\nu\vec{\rho_\mu}]\psi_N
+[h.c.] \,, \\
& & \nonumber \\
L_{\rho\Delta\Delta}&=&g_{\rho \Delta\Delta}\bar{\psi}_{\Delta\alpha}
[\gamma^\mu- \frac{\kappa_{\Delta\Delta\rho}}{2m_\Delta}
\sigma^{\mu\nu}\partial_\nu]
\vec{\rho_\mu}\cdot \vec{T}_\Delta \psi_\Delta^\alpha \,, \\
& & \nonumber \\
L_{\rho\pi\pi}&=& g_{\rho\pi\pi}[\vec{\phi_\pi}\times
\partial_\mu \vec{\phi_\pi}] \cdot \vec{\rho}^\mu\,, \\
& & \nonumber \\
L_{NN\rho\pi} &=& \frac{f_{\pi NN}}{m_\pi}g_{\rho NN}
\bar{\psi}_N\gamma_\mu \gamma_5
\vec{\tau}\psi_N\cdot \vec{\rho^\mu} \times \vec{\phi_\pi}
\label{eq:rho-ct1} \,, \\
& & \nonumber \\
L_{NN\rho\rho}&=&-\frac{\kappa_\rho g_{\rho NN}^2}{8m_N}
\bar{\psi}_N\sigma^{\mu\nu}\vec{\tau}
\psi_N \cdot \vec{\rho_\mu}\times \vec{\rho_\nu} \,. \label{eq:rho-ct2}
\end{eqnarray}
Note that the contact terms Eqs.(\ref{eq:rho-ct1})-(\ref{eq:rho-ct2}) are 
from applying $[\partial^\mu \rightarrow \partial^\mu - g_{\rho NN} 
\vec{\rho}^\mu \times ]$ on 
$L_{\pi NN}$ Eq.(\ref{eq:L-pinn})
and $L_{\rho NN}$ Eq.(\ref{eq:L-rnn}). 

The interactions involving $\omega$ meson are
\begin{eqnarray}
L_{\omega NN} &=& g_{\omega NN}\bar{\psi}_N[\gamma_\mu -
\frac{\kappa_\omega}{2m_N}\sigma_{\mu\nu}\partial^\nu]
{\omega^\mu}  \psi_N  \,, \\
& & \nonumber \\
L_{\omega\pi\rho}&=&-\frac{g_{\omega\pi\rho}}{m_\omega}
\epsilon_{\mu\alpha\lambda\nu}\partial^\alpha\vec{\rho^\mu}\partial^\lambda
\vec{\phi_\pi}\omega^\nu \,. 
\end{eqnarray}
We also consider interaction involving a scalar isoscalar $\sigma$ meson
\begin{eqnarray}
L_{\sigma NN} &=& g_{\sigma NN}\bar{\psi}_N \psi_N\phi_\sigma \\
%& & \nonumber \\
L_{\sigma\pi\pi}&=& -\frac{g_{\sigma\pi\pi}}{2m_\pi}
 \partial^\mu \vec{\phi}_\pi\partial_\mu \vec{\phi}_\pi \phi_\sigma \,. 
\end{eqnarray}

\begin{table}[t]
\centering
\begin{tabular}{ccccccc} \hline\hline
$\frac{f_{\pi NN}^2}{4\pi}$ & $f_{\pi N\Delta}$ & 
$ \kappa_\rho$ &   $g_{\omega NN}$ &$\kappa_\omega $
& &$g_{\rho NN}g_{\rho\pi\pi}$ \\ \hline
0.08 & 2.049 &  1.825& 11.5 & 0 & & 38.4329
\\ \hline\hline
\end{tabular}
\caption{Coupling constants determined in Ref.\cite{sl-1}.}
\label{tab:L-coup}
\end{table}

\begin{table}[t]
\centering
\begin{tabular}{cccccccccc} \hline\hline
$g_{\rho \pi\pi}$ & $g_{\sigma \pi\pi}$ & $ g_{\omega\pi\rho} $
& $g_{\rho NN}$ & $ f_{\eta NN}$ & $g_{\sigma NN} $ & $ f_{\pi\Delta\Delta}$  
&$ f_{\rho N\Delta}$
&$f_{\rho\Delta\Delta}$ & $\kappa_{\rho\Delta\Delta}  $  \\ \hline
   6.1994 & 1.77 & 11.2 & 6.1994 & 1.77 &12.8 &1.78  
& -6.08  & -4.30 & 6.1 
\\ \hline\hline
\end{tabular}
\caption{Coupling constants used in the calculations in this paper.}
\label{tab:L-coupa}
\end{table}

To proceed, we need to know the coupling constants of the above Lagrangians.
The parameters determined from fitting the $\pi N$ data within the
SL model\cite{sl-1} are given in Table I. 
The $\rho, \sigma \rightarrow \pi\pi$ coupling constants can be estimated 
from fitting $\pi\pi$ phase shifts in the isobar model\cite{johnstone}, 
as described in Appendix C. The decay width of $\omega \rightarrow \pi \rho$
can be used to estimate the coupling constant $g_{\omega\pi\rho}$.
The $\eta NN$
coupling constant $f_{\eta NN}$ has been estimated in recent studies of
$\eta$ production from $\pi N$ and $\gamma N$ reactions. The
$\sigma NN$ coupling can be estimated from the
 previous works on $NN$ scattering.
These parameters
are adjusted around the values from
 these estimates  to fit
 the JLab data of 
$\gamma p \rightarrow \pi^+\pi^- p$ reactions, as described in 
section VII. They are listed
in Table II.
  
We have very little information on the coupling constants 
$f_{\pi \Delta \Delta}$, $f_{ N\Delta\rho}$ and $f_{\rho \Delta\Delta}$.
We simply follow the previous works and
use the simple SU(6) quark model to determine them
from the empirical values of the coupling constants
$f_{\pi NN}$ and $g_{\rho NN}$. 
To be more informative, we here also describe how this procedure is used in
practice.

First step is take the static-baryon limit of 
the matrix elements $<B'| L_{MBB'}|B M(q)>$
to define the effective $MBB'$ Hamiltonian operators
in the spin-isospin space of baryons. They are
\begin{eqnarray}
H_{\pi NN}&=&
  i \frac{f_{\pi NN}}{m_\pi}
 \vec{\sigma}\cdot\vec{q}\tau^\alpha \,,
\label{eq:hmdl-pipin} \\
H_{\pi N\Delta}&=&
  i \frac{f_{\pi N\Delta}}{m_\pi}
 \vec{S}\cdot\vec{q}T^\alpha \,,
\label{eq:hmdl-pind} \\
H_{\pi \Delta\Delta}&=&
i\frac{f_{\pi \Delta\Delta}}{m_\pi}\frac{2}{3}\vec{S}_\Delta\cdot\vec{q}T^\alpha_\Delta
\label{eq:hmdl-pidd} \,, \\
H_{\rho NN}&=&
 i \frac{g_{\rho NN}(1 + \kappa_\rho)}{4 m_N}
 \vec{\sigma}\times \vec{q} \cdot \vec{\epsilon}(\rho) \tau^\alpha \,,
\label{eq:hmdl-rnn} \\
H_{\rho N\Delta}&=&
  -i \frac{f_{\rho N\Delta}}{m_\rho}
            \vec{S}\times \vec{q}\cdot\vec{\epsilon}(\rho)T^\alpha 
\label{eq:hmdl-rnd} \,, \\
H_{\rho \Delta\Delta}&=&
   -i g_{\rho\Delta\Delta}\frac{1+\kappa_{\rho\Delta\Delta}}{2m_\Delta}
      \frac{2}{3}\vec{S}_\Delta\times
                 \vec{q}\cdot\vec{\epsilon_\rho}T^\alpha_\Delta \,.
\label{eq:hmdl-rdd}
\end{eqnarray}
Here, $\alpha$ is the isospin component of the considered meson, 
$\vec{S}$ and $\vec{T}$ are the spin and isospin operators of
the $N$-$\Delta$ transition, $\vec{S_\Delta}$ and $\vec{T_\Delta}$ are
the spin and isospin operators of the $\Delta$.
Along with the usual Pauli operators
 $\vec{\sigma}$ and $\vec{\tau}$, they are defined
by the following  reduced matrix elements
\begin{eqnarray}
<N||\sigma||N> & = & <N||\tau||N>= \sqrt{6} \,,
\label{eq:reduce-nn}\\
<\Delta||S||N> & = &<\Delta||T||N>  =  2 \,,
\label{eq:reduce-dn}\\
<\Delta||S_\Delta||\Delta> & = & <\Delta ||T_\Delta||\Delta> =\sqrt{15} 
\label{eq:reduce-dd}
\end{eqnarray}
with the convention that
\begin{eqnarray}
<j_f m_f | O^{I}_M|j_i m_i>
=\frac{1}{\sqrt{2j_f+1}}<j_f m_f | j_i I m_i M>
<j_f || O^{I} ||j_i> \,.
\label{eq:wigner}
\end{eqnarray}

We next consider a simple meson-quark interaction Hamiltonian
\begin{eqnarray}
H_{\pi qq} &=& \frac{f_{\pi qq}}{m_\pi }\sum_{i=1,3} 
i \vec{\sigma}_i\cdot \vec{q} \tau_i \,,
\label{eq:ccqm-piqq}\\
H_{\rho qq} &=& \frac{f_{\rho qq}}{m_\rho }\sum_{i=1,3} 
i \vec{\sigma}_i\times \vec{q} \cdot \vec{\epsilon}_\rho \tau_i \,,
\label{eq:ccqm-rqq}
\end{eqnarray}
where $\sigma_i, \tau_i$ are the spin and isospin operators of the constituent
quarks. By using the 0s constituent quark wavefunctions 
$\psi_{N,m_{s_N}m_{\tau_N}}$ and
$\psi_{\Delta,m_{s_\Delta}m_{\tau_\Delta}}$
for
the nucleon and $\Delta$ and the 
relations Eq.(\ref{eq:reduce-nn})-(\ref{eq:wigner}),
we have the following relations between the matrix elements in
the spin-isospin space
\begin{eqnarray}
<\psi_{N,m'_{s_N}m'_{\tau_N}} |\sum_{i=1,3}\vec{\sigma}_i \tau_i
|\psi_{N,m_{s_N}m_{\tau_N}}> & = &
 \frac{5}{3}<m'_{s_N}m'_{\tau_N}| \vec{\sigma}\tau|m'_{s_N}m'_{\tau_N}>\,, \\
<\psi_{\Delta,m'_{s_\Delta}m'_{\tau_\Delta}} |\sum_{i=1,3}\vec{\sigma}_i \tau_i
|<\psi_{N,m_{s_N}m_{\tau_N}} > & = &
 2\sqrt{2} <m'_{s_\Delta}m'_{\tau_\Delta} |\vec{S} T|m'_{s_N}m'_{\tau_N}>\,, \\
<\psi_{\Delta,m'_{s_\Delta}m'_{\tau_\Delta}} |\sum_{i=1,3}\vec{\sigma}_i \tau_i
|\psi_{\Delta,m_{s_\Delta}m_{\tau_\Delta}} > & = &
 \frac{4}{3} <m'_{s_\Delta}m'_{\tau_\Delta}|\vec{S}_\Delta  T_\Delta
|m_{s_\Delta}m_{\tau_\Delta} >\,.
\end{eqnarray}
Using the above formula and assume 
that the matrix elements of the hadron  Hamiltonians
Eqs.(\ref{eq:hmdl-pipin})-(\ref{eq:hmdl-rdd}) are equal
to the matrix elements of the quark-meson Hamiltonian 
Eqs.(\ref{eq:ccqm-piqq})-(\ref{eq:ccqm-rqq}) within
the SU(6) chiral constituent quark model
\begin{eqnarray} 
<\psi_{B,m'_{s_B},m'_{\tau_B}}|H_{Mqq}|\psi_{B,m_{s_B},m_{\tau_B}}> 
=<m'_{s_B},m'_{\tau_B}|H_{MBB'}|m_{s_B},m_{\tau_B}> \,,
\end{eqnarray}
we then obtain
\begin{eqnarray}
f_{\pi NN} & =& \frac{5}{3}f_{\pi qq} \,,  \\
f_{\pi N\Delta} & =& 2\sqrt{2}f_{\pi qq} \,, \\
\frac{2}{3}f_{\pi \Delta\Delta} & =& \frac{4}{3}f_{\pi qq}
   \,, \\
f_{\rho NN} & =& \frac{5}{3}f_{\rho qq} \,,  \\
f_{\rho N\Delta} & = &-2\sqrt{2}f_{\rho qq} \,, 
  \\
\frac{2}{3}f_{\rho \Delta\Delta} & =-& \frac{4}{3}f_{\rho qq} \,,
\end{eqnarray}
where we defined
\begin{eqnarray}
f_{\rho NN} & = & \frac{g_{\rho NN}(1 + \kappa_\rho)}{4 m_N} m_\rho
\label{eq:frnn0} \,, \\
f_{\rho\Delta\Delta} & = &
 g_{\rho\Delta\Delta}\frac{1+\kappa_{\rho\Delta\Delta}}{2m_\Delta}m_\rho \,.
\label{eq:frdd0}
\end{eqnarray}

From the above relations we finally have
\begin{eqnarray}
f_{\pi N \Delta} & = & \sqrt{\frac{72}{25}} f_{\pi NN} \,,
\label{eq:fpnd}\\
f_{\pi \Delta \Delta} & = & \frac{6}{5}f_{\pi NN} \,,
\label{eq:fpdd}\\
f_{\rho N\Delta} & = &- \sqrt{\frac{72}{25}} 
f_{\rho NN} \,, \label{eq:frnd} \\
f_{\rho\Delta\Delta} &=&-\frac{6}{5}f_{\rho NN} \,.
\label{eq:frdd}
\end{eqnarray}
By using the vector meson dominance assumption and  the recently 
determined $\Delta$ magnetic moment, we can set
\begin{eqnarray}
\kappa_{\rho\Delta\Delta} = 6.1 \,.
\label{eq:krdd}
\end{eqnarray}
With the values in Eq.(\ref{eq:krdd}) and values listed in Table I,
we can use Eqs.(\ref{eq:frnn0})-(\ref{eq:frdd}) to get
$f_{\pi \Delta\Delta}$, $f_{\rho N\Delta}$ and $f_{\rho \Delta\Delta}$.
The resulting values are also listed in Table II.

The electromagnetic interactions are obtained from the 
usual non-interacting Lagrangian and the above
interaction Lagrangian by using the minimum substitution
$\partial_\mu \rightarrow \partial_\mu - ie A_\mu$.
The resulting Lagrangian are given below :

\begin{eqnarray}
L_{\gamma NN} & = 
& \bar{\psi}_N[\hat{e}_N\gamma^\mu -\frac{\hat{\kappa}_N}{2m_N}\sigma^{\mu\nu}
\partial_\nu ]\psi_N A_\mu  \label{eq-Lgnn} \,, \\
L_{\gamma \pi\pi} &=& 
[\vec{\phi}_\pi\times \partial^\mu \vec{\phi}_\pi]_3A_\mu \,, \\
L_{\gamma N \pi N} &=&
 \frac{f_{\pi NN}}{m_\pi}[\bar{\psi}_N\gamma^\mu \gamma_5 
\vec{\tau} \psi_N)\times \vec{\phi}_\pi]_3A_\mu \,, \\
L_{\gamma \rho\rho} &=& [(\partial^\mu\vec{\rho^\nu}
-\partial^\nu\vec{\rho^\mu}) \times \vec{\rho}_\nu]_3A_\mu \,, \\
L_{\gamma \rho\pi\pi} &=& -g_{\rho\pi\pi}[(\vec{\rho^\mu}
 \times \vec{\phi}_\pi)\times \vec{\phi}_\pi]_3A_\mu \,, \\
L_{\gamma N\pi \Delta} &=&
 \frac{f_{\pi N\Delta}}{m_\pi}[(\bar{\psi}_{\Delta}^\mu \vec{T}\psi_{N})
\times \vec{\phi}_\pi]_3A_\mu  \label{eq-Lgnd} \,, \\
L_{\gamma N \rho N} &=& g_{\rho NN}[\frac{\kappa_\rho}{2m_N}
(\bar{\psi}_N \frac{\vec{\tau}}{2}\sigma^{\nu\mu}\psi_N)
\times \vec{\rho_\nu}]_3 A_\mu \,, \\
L_{\gamma N \Delta} &=& -i \bar{\psi}_{\Delta}^\mu \Gamma^{em,\Delta}_{\mu\nu} 
T_3 \psi_NA^\nu + (h.c.) \,,  \\
L_{\gamma\rho\pi} &=& \frac{g_{\rho\pi\gamma}}{m_\pi}
\epsilon_{\alpha\beta\gamma\delta}\vec{\phi}_\pi\cdot
(\partial^\gamma \vec{\rho^\delta})(\partial^\alpha A^\beta)  \,, \\
L_{\gamma\omega\pi} &=&\frac{g_{\omega\pi\gamma}}{m_\pi}
\epsilon_{\alpha\beta\gamma\delta}(\partial^\alpha A^\beta) \phi_\pi^3
(\partial^\gamma \omega^\delta) \,, \\
L_{\gamma\rho\eta} &=& \frac{g_{\rho\eta\gamma}}{m_\rho}
\epsilon^{\mu\nu\alpha\beta}
\partial_\mu\rho_\nu^3 \partial_\alpha A_\beta \phi_\eta \,, \\
L_{\gamma\rho\sigma} &=& -\frac{g_{\rho\sigma\gamma}}{m_\rho}
(\partial_\mu\rho_\nu^3)(\partial^\mu A^\nu-\partial^\nu A^\mu)\sigma \,,
\\
L_{\gamma\Delta\Delta} &=& \bar{\psi}_{\Delta}^\eta
(T^3_\Delta+\frac{1}{2})[
- \gamma^\mu g_{\eta \nu} + (g^\mu_\eta \gamma^\nu + g^\mu_\nu
\gamma^\eta) + \frac{1}{3}\gamma_\eta \gamma^\mu \gamma_\nu
] \psi_\Delta^\nu A_\mu \,.
\end{eqnarray}
For Eq. (\ref{eq-Lgnn}), we have defined
\begin{eqnarray}
\hat{e} & = & \frac{F_{1S}+ F_{1V}\tau^3}{2} \,, \\
\hat{\kappa}& =&  \frac{F_{2S} + F_{2V}\tau^3}{2} \,,
\end{eqnarray}
where $F_{1S}(0)= F_{1V}(0)=1$, $F_{2S}(0)=\mu_p + \mu_n -1 \sim -0.12 $
and $F_{2V}(0)=\mu_p - \mu_n -1 \sim 3.7 $
 The  matrix element of $\gamma N\Delta$ vertex of Eq. (\ref{eq-Lgnd})
 between an $N$ with momentum $p$ and a $\Delta$ with momentum
 $p_{\Delta}$ can be written explicitly as
\begin{eqnarray}
<\Delta(p_{\Delta}^\prime)|\Gamma^{em \Delta}_{\mu\nu}|N(p_N)>
  = & & \frac{m_\Delta+m_N}{2m_N}\frac{1}{(m_\Delta+m_N)^2 - q^2}
 \nonumber \\
& & \times [(G_M-G_E)3\epsilon_{\mu\nu\alpha\beta}P^\alpha q^\beta
 \nonumber \\
& & + G_E i\gamma_5 \frac{12}{(m_\Delta - m_N)^2 - q^2}
  \epsilon_{\mu\lambda\alpha\beta}P^\alpha q^\beta
  \epsilon^{\lambda}_{\ \ \nu\alpha\delta}p_\Delta^\gamma q^\delta
 \nonumber \\
& & + G_C i\gamma_5 \frac{6}{(m_\Delta-m_N)^2 - q^2}
    q_\mu (q^2 P_\nu - q\cdot P q_\nu)],
\end{eqnarray}
with $P = (p_\Delta^\prime +p_N)/2$ and $p_\Delta^\prime = p_N + q$.
Note that the  index $\mu$ of $\Gamma^{em \Delta}_{\mu\nu}$
contracts with the $\Delta$ field and $\nu$
 with the photon field.
The coupling strength $G_M=1.85$, $G_E=0.025$, and
$G_C=-0.238$ are taken from  the SL model\cite{sl-1,sl-2}.

\section{Derivation of Coupled-channel Equations}

In this appendix, we give the derivation of coupled-channel equations
from the model Hamiltonian $H_{eff}=H_0+V$ defined by 
Eqs.(\ref{eq:H-1})-(\ref{eq:v23}).
We apply the standard projection operator techniques\cite{feshbach}.
The procedure is similar to that used in  the derivation of 
$\pi NN$ equations\cite{LM85}.
We start with Eq.(\ref{eq:loweq})
\begin{eqnarray}
T(E) = V + V \frac{1}{E-H_0}T(E) \,.
\label{eq:appb-T}
\end{eqnarray}
The propagator in the above equation
is understood to include $+ i\epsilon$ for defining
the boundary condition, but is omitted to simplify the
presentation in this appendix.
The interaction $V$, defined in
Eqs.(\ref{eq:H-2})-(\ref{eq:v23}), can be more clearly written as
\begin{eqnarray}
V = v_{22}+ v_{33}+ (\Gamma_{12}+u_{23}+\gamma_{13}) 
+ (\Gamma_{21}+u_{32}+\gamma_{31}) \,,
\label{eq:appb-V}
\end{eqnarray}
where $v_{22}=v_{MB,M'B'}+v_{\pi\pi}$,
 $\Gamma_{12}=\Gamma_{N^*\rightarrow MB} + h_{M^*\rightarrow \pi\pi}$ with
$M^* = \rho, \sigma$, $\gamma_{13}=\Gamma_{N^*\rightarrow \pi\pi N}$,
 $u_{23}=v_{MB,\pi\pi N}$, and
$v_{33}=v_{\pi\pi N\pi\pi N}$. Here we restrict $MB=\gamma N,\pi N,\eta N,
\pi\Delta, \rho N, \sigma N$.
 In Eq.(\ref{eq:appb-V}), we have also introduced more
transparent notations
$\Gamma_{21}=\Gamma^\dagger_{12}$, $u_{23}=u^\dagger_{23}$, and $\gamma_{31}=\gamma^\dagger_{13}$.

We next introduce projection operators
\begin{eqnarray}
P+Q=1
\end{eqnarray}
with
\begin{eqnarray}
Q&=& |\pi\pi N><\pi\pi N| \,, \\
P&=& P_1 + P_2 + P_{2*} \,,
\end{eqnarray}
where
\begin{eqnarray}
P_1 &=& \sum_{N^*} |N^*><N^*| \label{eq:appb-p1} \,, \\
P_2 &=& |\gamma N><\gamma N| + |\pi N><\pi N| +|\eta N><\eta N|\,,
 \label{eq:appb-p2} \\
P_{2*} &=& |\pi \Delta><\pi\Delta| +|\rho N><\rho N| + |\sigma N><\sigma N| \,.
\label{eq:appb-p2s}
\end{eqnarray}
We then obtain the equations for the projected operators $T_{PP}=PTP$ and
$T_{QP}= QTP$
\begin{eqnarray}
T_{PP}&=&\bar{V}_{PP} + \bar{V}_{PP}\frac{1}{E-H_0}{T}_{PP} \,,
\label{eq:appb-tpp} \\
T_{QP}&=&\frac{1}{1-V_{QQ}\frac{Q}{E-H_0-V_{QQ}}}V_{QP}
[1+\frac{P}{E-H_0}T_{PP}]\,,
\label{eq:appb-tqp}
\end{eqnarray}
where
\begin{eqnarray}
\bar{V}_{PP}=V_{PP} + V_{PQ}\frac{Q}{E-H_0-V_{QQ}}V_{QP} \,,
\label{eq:appb-vpp}
\end{eqnarray}
with
\begin{eqnarray}
V_{PP} &=& PVP = v_{22}+\Gamma_{12} + \Gamma_{21} \,, \\
V_{QP} &=& QVP = u_{32}+\Gamma_{21} + \gamma_{31} \,, \\
V_{QQ}&=&QVQ = v_{22} + v_{33} \,.
\end{eqnarray}

Eq.(\ref{eq:appb-vpp}) can be written explicitly as
\begin{eqnarray}
\bar{V}_{PP} = P[(v_{22}+\Gamma_{12} +\Gamma_{21})
+(u_{23}+\gamma_{13}+\Gamma_{12})G_Q(u_{32}+\gamma_{31}+\Gamma_{21}) ]P \,,
\label{eq:appb-vppa}
\end{eqnarray}
with
\begin{eqnarray}
G_Q=\frac{Q}{E-H_0-Q(v_{22}+v_{33})Q} \,.
\end{eqnarray}
From the definitions Eqs.(\ref{eq:appb-p1})-(\ref{eq:appb-p2s}) for the projection operators, 
we have the following conditions
\begin{eqnarray}
P_2\Gamma_{12}Q &=&Q\Gamma_{12}P_2 =0 \,, \nonumber \\
P_2\gamma_{31}Q&=&P_{2*}\gamma_{31}Q = Q\gamma_{31}P_2=Q\gamma_{31}P_{2*} =0 \,.
\end{eqnarray}
With the above "doorway" conditions, we can decompose $\bar{V}_{PP}$ as
\begin{eqnarray}
\bar{V}_{PP}=P[\Sigma+ \bar{v} ]P \,,
\label{eq:appb-vppc}
\end{eqnarray}
where
\begin{eqnarray}
\Sigma&=& [\Gamma_{12}G_Q\Gamma_{21}]_{un-connected} \,.
\label{eq:appb-sigma} 
\end{eqnarray}
Here ${un-connected}$ in Eq.(\ref{eq:appb-sigma}) means that the
pion emitted from one baryon is also absorbed by the same baryon.
Obviously this is the self-energy of
the unstable particles in the $\pi\Delta$,
$\rho N$ and $\sigma N$ states of $P_{2*}$ space. We thus have
\begin{eqnarray}
P\Sigma P = P_{2*}\Sigma P_{2^*} \,.
\end{eqnarray}
All other interactions within the P-space are in $\bar{v}$ of 
Eq.(\ref{eq:appb-vppc})
\begin{eqnarray}
\bar{v}&=&V_E +\hat{\Gamma}_{12}+\hat{\Gamma}_{21} +\hat{\Sigma}
\label{eq:appb-barv}
\end{eqnarray}
with
\begin{eqnarray}
V_E &=&v_{22}+(u_{23} +\Gamma_{12})G_Q (u_{32}+\Gamma_{21}) - \Sigma \,, \\
\hat{\Gamma}_{21} &= &\Gamma_{21}+ +\Delta \hat{\Gamma}_{21} \,, \\
\hat{\Gamma}_{12}& =& \Gamma_{12}+ +\Delta \hat{\Gamma}_{12} \,, \\
\hat{\Sigma} &=& \gamma_{13}G_Q\gamma_{31} \,,
\end{eqnarray}
where $\Delta \hat{\Gamma}_{21}$ and $\Delta \hat{\Gamma}_{12}$ contain
interactions due to $N^*\leftrightarrow \pi\pi N$ transitions
\begin{eqnarray}
\Delta \hat{\Gamma}_{21} =[u_{23}+\Gamma_{21}]G_Q\gamma_{31} \,, \\
\Delta \hat{\Gamma}_{12} =\gamma_{13}G_Q[u_{32}+\Gamma_{12}] \,.
\end{eqnarray}

To follow the derivations given below, we note that 
the  well known
operator relations
\begin{eqnarray}
t&=&v+v\frac{1}{E-H_0}t \nonumber \\
&=& v + t\frac{1}{E-H_0}v
\label{eq:op-0}
\end{eqnarray}
lead to
\begin{eqnarray}
t &=&[1-v\frac{1}{E-H_0}]^{-1}v \nonumber \\
 &=&v[1-\frac{1}{E-H_0}v]^{-1} \,.
\label{eq:op-1}
\end{eqnarray}
Eqs.(\ref{eq:op-0}) and (\ref{eq:op-1}) then  lead to
\begin{eqnarray}
[1-v\frac{1}{E-H_0}]^{-1} &=& 1+t\frac{1}{E-H_0} \,, \nonumber 
\end{eqnarray}
\begin{eqnarray}
[1-\frac{1}{E-H_0}v]^{-1}&=& 1+\frac{1}{E-H_0}t \,.
\label{eq:op-2}
\end{eqnarray}
Eq.(\ref{eq:op-0}) also leads to
\begin{eqnarray}
t = v + v\frac{1}{E-H_0-v}v \,.
\label{eq:op-3}
\end{eqnarray} 
Comparing Eqs.(\ref{eq:op-0}) and (\ref{eq:op-3}), we have
\begin{eqnarray}
\frac{1}{E-H_0-v}v &=& \frac{1}{E-H_0}t \,, \nonumber \\
v\frac{1}{E-H_0-v} &=&t \frac{1}{E-H_0} \,.
\label{eq:op-4}
\end{eqnarray}
It can also easily be seen that
\begin{eqnarray}
\frac{1}{E-H_0-v} =\frac{1}{E-H_0}+\frac{1}{E-H_0}t\frac{1}{E-H_0} \,.
\label{eq:op-5}
\end{eqnarray}
In the following derivations, the above relations Eqs.(\ref{eq:op-0})
-(\ref{eq:op-5}) will be often used without mentioned them again.

By using Eqs.(\ref{eq:op-0}),  
(\ref{eq:op-3}) and (\ref{eq:op-5}), we can write $T_{PP}$ defined by 
Eq.(\ref{eq:appb-tpp}) as
\begin{eqnarray}
T_{PP}&=&P[ (\Sigma + v) +
(\Sigma + v)\frac{1}{E-H_0-\Sigma + v}(\Sigma + v)]P \nonumber \\
&=&P[ \Sigma + \Sigma \frac{P_{2*}}{E-H_0-\Sigma} \Sigma
+(1+\Sigma\frac{P_{2*}}{E-H_0-\Sigma})T_{\bar{v}}
(1+\frac{P_{2*}}{E-H_0-\Sigma}\Sigma)]P
\label{eq:appb-tppf}
\end{eqnarray}
with
\begin{eqnarray}
T_{\bar{v}} = \bar{v} + \bar{v}\frac{P}{E-H_0-\Sigma}T_{\bar{v}} \,.
\label{eq:apprb-btv}
\end{eqnarray}
By using Eq.(B13) and relation (B32), we
 can write Eq.(\ref{eq:appb-tqp}) as
\begin{eqnarray}
T_{QP}=Q[(1+ t_Q\frac{Q}{E-H_0})(u_{32}+\gamma_{31} +\Gamma_{21})
[1+\frac{P}{E-H_0}T_{PP}]P \,,
\label{eq:tqp}
\end{eqnarray}
where
\begin{eqnarray}
t_Q = V_{QQ} + V_{QQ}\frac{Q}{E-H_0}t_Q 
\label{eq:tq}
\end{eqnarray}
describes $\pi\pi N \rightarrow \pi\pi N$ scattering through 
$V_{QQ}=Q[v_{22}+ v_{33}]Q = v_{\pi\pi}+v_{\pi N,\pi N} 
+ v_{\pi\pi N,\pi\pi N}$ interactions.

We now derive equations for calculating the
 scattering amplitudes between two particle 
channels in $P^\prime_2 = P_2 + P_{2*}$ space. 
We first note that
\begin{eqnarray}
P^\prime_2 \bar{v} P^\prime_2 &=& V_E \,, \\
P^\prime_2 \bar{v} P_1 &=& \hat{\Gamma}_{21} \,, \\
P_1 \bar{v} P^\prime_2 &=& \hat{\Gamma}_{12} \,,\\
P_1 \bar{v} P_1 &=& \hat{\Sigma} \,.
\end{eqnarray}
The above relations and Eq.(B35) lead to
\begin{eqnarray}
P^\prime_2 T_{\bar{v}} P^\prime_2
&=& V_E +V_E\frac{P^\prime_2}{E-H_0-\Sigma}P^\prime_2 T_{\bar{v}}P^\prime_2
+\hat{\Gamma}_{21}\frac{P_1}{E-m^0_{N^*}} P_1T_{\bar{v}}P^\prime_2 
\label{eq:appb-tv1} \,, \\
P_1 T_{\bar{v}} P^\prime_2
&=& \hat{\Gamma}_{12} + \hat{\Gamma}_{12}
\frac{P^\prime_2}{E-H_0-\Sigma}P^\prime_2 T_{\bar{v}}P^\prime_2
+\hat{\Sigma}\frac{P_1}{E-m^0_{N^*}} P_1T_{\bar{v}}P^\prime_2 \,.
\label{eq:appb-tv2}
\end{eqnarray} 
Eq.(\ref{eq:appb-tv2}) can be written as
\begin{eqnarray}
P_1T_{\bar{v}}P^\prime_2&=&[1-\hat{\Sigma}\frac{1}{E-m^0_{N^*}}]^{-1}
[\hat{\Gamma}_{12} + \hat{\Gamma}_{12}
\frac{P^\prime_2}{E-H_0-\Sigma}P^\prime_2 T_{\bar{v}}P^\prime_2]
\nonumber \\
&=& \frac{E-m^0_{N^*}}{E-m^0_{N^*}-\hat{\Sigma}}
[\hat{\Gamma}_{12} + \hat{\Gamma}_{12}
\frac{P^\prime_2}{E-H_0-\Sigma}P^\prime_2 T_{\bar{v}}P^\prime_2] \,.
\label{eq:appb-tv2a}
\end{eqnarray}
Substituting Eq.(\ref{eq:appb-tv2a}) into Eq.(\ref{eq:appb-tv1}), we have
\begin{eqnarray}
P^\prime_2 T_{\bar{v}} P^\prime_2 = X + X\frac{1}{E-H_0-\Sigma}
P^\prime_2 T_{\bar{v}}P^\prime_2 \,,
\label{eq:appb-tp2pp2pa} 
\end{eqnarray}
where
\begin{eqnarray}
X= V_E+\hat{\Gamma}_{21}\frac{1}{E-M^0_{N^*}-\hat{\Sigma}} \hat{\Gamma}_{12} \,.
\end{eqnarray}

Eq.(\ref{eq:appb-tp2pp2pa}) can be written as 
\begin{eqnarray}
P^\prime_2 T_{\bar{v}} P^\prime_2
&=& [1-V_E\frac{1}{E-H_0-\Sigma}]^{-1}[V_E \nonumber \\
& & +[1-V_E\frac{1}{E-H_0-\Sigma}]^{-1}
\hat{\Gamma}_{21}\frac{1}{E-M^0_{N^*}-\hat{\Sigma}} \hat{\Gamma}_{12}
[1+\frac{1}{E-H_0-\Sigma}P^\prime_2 T_{\bar{v}} P^\prime_2] \nonumber \\
&=& t_E + [1+ t_E\frac{1}{E-H_0-\Sigma}]
\hat{\Gamma}_{21}\frac{1}{E-M^0_{N^*}-\hat{\Sigma}} \hat{\Gamma}_{12}
[1+\frac{1}{E-H_0-\Sigma}P^\prime_2 T_{\bar{v}} P^\prime_2] \nonumber \\
&=& t_E +\bar{\Gamma}_{21}\frac{1}{E-M^0_{N^*}-\hat{\Sigma}} \hat{\Gamma}_{12}
[1+\frac{1}{E-H_0-\Sigma}P^\prime_2 T_{\bar{v}} P^\prime_2] \nonumber \\
  &=& t_E
+\bar{\Gamma}_{21}\frac{1}{E-M^0_{N^*}-\hat{\Sigma}} \hat{\Gamma}_{12}
[1+ \frac{1}{E-H_0-\Sigma-X}X] \nonumber \\
&=&  t_E
+\bar{\Gamma}_{21}\frac{1}{E-M^0_{N^*}-\hat{\Sigma}} \hat{\Gamma}_{12}
\frac{1}{E-H_0-\Sigma-X}[E-H_0-\Sigma] \,,
\label{eq:b-1}
\end{eqnarray}
where
\begin{eqnarray}
t_E &=& V_E + V_E\frac{1}{E-H_0-\Sigma} t_E 
\label{eq:appb-te} \,, \\
\bar{\Gamma}_{21} &=& [1 + t_E\frac{1}{E-H_0-\Sigma}]\hat{\Gamma}_{21} \,.
\label{eq:hat}
\end{eqnarray}

We further note that
\begin{eqnarray}
\frac{1}{E-H_0-\Sigma-X} &=& \frac{1}{E-H_0-\Sigma-V_E -
\hat{\Gamma}_{12}\frac{1}{E-M^0_{N^*}-\hat{\Sigma}}\hat{\Gamma}_{21}}
\nonumber \\
&=&\frac{1}{E-H_0-\Sigma-V_E}+\frac{1}{E-H_0-\Sigma-V_E} 
t_s\frac{1}{E-H_0-\Sigma-V_E}
\label{eq:c1}
\end{eqnarray}
with
\begin{eqnarray}
t_s &=& \hat{\Gamma}_{12}\frac{1}{E-M^0_{N^*}-\hat{\Sigma}}\hat{\Gamma}_{21}
[1 + \frac{1}{E-H_0-\Sigma-V_E}t_s] \nonumber \\
&=& \hat{\Gamma}_{12}\frac{1}{E-M^0_{N^*}-\hat{\Sigma}-\bar{\Sigma}}
\hat{\Gamma}_{21} \,,
\label{eq:c2}
\end{eqnarray}
where
\begin{eqnarray}
\bar{\Sigma} &=& \hat{\Gamma}_{12}\frac{1}
{E-H_0-\Sigma - V_E}\hat{\Gamma}_{21} \nonumber \\
&=& \hat{\Gamma}_{12}\frac{1}
{E-H_0-\Sigma }\bar{\Gamma}_{21} \,.
\label{eq:hatsigma}
\end{eqnarray}
Here $\bar{\Gamma}_{21}$ has been defined in Eq.(\ref{eq:hat}). 

By using Eqs.(\ref{eq:c1}) and (\ref{eq:c2}), Eq.(\ref{eq:b-1}) can be 
written as
\begin{eqnarray}
 P^\prime_2 T_{\bar{v}} P^\prime_2 &=& t_E
+\bar{\Gamma}_{21}\frac{1}{E-M^0_{N^*}-\hat{\Sigma}}[1+
(\hat{\Gamma}_{21}\frac{1}{E-H_0-\Sigma -V_E}
\hat{\Gamma}_{12})\frac{1}
{E-M^0_{N^*}-\hat{\Sigma}-\bar {\Sigma} }] \nonumber \\
& &\times\hat{\Gamma}_{12}
\frac{1}{E-H_0-\Sigma -V_E}[E-H_0-\Sigma] \nonumber \\
 &=& t_E+
\bar{\Gamma}_{21}\frac{1}{E-M^0_{N^*}-\hat{\Sigma}}[1+
\bar{\Sigma}\frac{1}
{E-M^0_{N^*}-\hat{\Sigma}-\bar{\Sigma} }]\hat{\Gamma}_{12}
\frac{1}{E-H_0-\Sigma -V_E}[E-H_0-\Sigma] \nonumber \\
 &=& t_E
+\bar{\Gamma}_{21}\frac{1}
{E-M^0_{N^*}-\hat{\Sigma}-\bar {\Sigma} }\hat{\Gamma}_{12}
\frac{1}{E-H_0-\Sigma -V_E}[E-H_0-\Sigma] \nonumber \\
&=& t_E
+\bar{\Gamma}_{21}\frac{1}
{E-M^0_{N^*}-\hat{\Sigma}-\bar{\Sigma} }\hat{\Gamma}_{12}
[1+\frac{1}{E-H_0-\Sigma -V_E}V_E] \nonumber \\
&=& t_E
+\bar{\Gamma}_{21}\frac{1}
{E-M^0_{N^*}-\hat{\Sigma}-\bar{\Sigma} }\hat{\Gamma}_{12}
[1+\frac{1}{E-H_0-\Sigma}t_E] \,. \nonumber 
\end{eqnarray}
The above then gives
\begin{eqnarray}
 P^\prime_2 T_{\bar{v}} P^\prime_2
 &=& t_E +\bar{\Gamma}_{21}\frac{1}
{E-M^0_{N^*}-\hat{\Sigma}-\bar{\Sigma} }\bar{\Gamma}_{12} \,,
\label{eq:appb-t22f-1}
\end{eqnarray}
where (also recalling Eq.(\ref{eq:hat}))
\begin{eqnarray}
\bar{\Gamma}_{12}&=&\hat{\Gamma}_{12}[1+\frac{1}{E-H_0-\Sigma}t_E] \,, \\
\bar{\Gamma}_{21}&=&[1+t_E\frac{1}{E-H_0-\Sigma }]\hat{\Gamma}_{21} \,.
\label{eq:appb-t22f-2}
\end{eqnarray}

We now turn to deriving equations for calculating two-pion production.
For initial $\pi N$ or $\gamma N$ of $P_2$-space,
Eq.(\ref{eq:tqp}) can be written explicitly as
\begin{eqnarray}
T_{QP_2} &=& Q(1+t_Q\frac{Q}{E-H_0})[u_{32}+u_{32}\frac{P_2}{E-H_0}
T_{PP}P_2 \nonumber \\
& & +(u_{32}+\Gamma_{21})P_{2*}\frac{P_{2*}}
{E-H_0} T_{PP}P_2 +\gamma_{31}\frac{P_1}{E-H_0}T_{PP} P_2] \,.
\label{eq:qp1}
\end{eqnarray}
From definition Eq.(\ref{eq:appb-tppf}), we have
\begin{eqnarray}
P_2T_{PP}P_2 &=& P_2T_{\bar{v}}P_2 \,, \\
P_1T_{PP}P_2 &=& P_1T_{\bar{v}}P_2 \,, \\
P_{2*}T_{PP}P_2 &=& P_{2*}
(1+\Sigma\frac{1}{E-H_0-\Sigma}P_{2*}]T_{\bar{v}}P_2 \nonumber \\
& =& P_{2*}
[E-H_0]\frac{1}{E-H_0-\Sigma}P_{2*}T_{\bar{v}}P_2 \,.
\end{eqnarray}
By using the above relations and Eq.(\ref{eq:appb-t22f-1}) and
$P^\prime_2=P_2+P_{2*}$, 
the 3rd in the bracket of Eq.(\ref{eq:qp1})
can be written as
\begin{eqnarray}
3rd &=& (u_{32}+\Gamma_{21})\frac{P_{2^*}}{E-H_0-\Sigma} 
[t_E +\bar{\Gamma}_{21}\frac{1}
{E-M^0_{N^*}-\hat{\Sigma}-\bar{\Sigma} }\bar{\Gamma}_{12}]P_2 \,.
\end{eqnarray} 
By using Eq.(B45), the 4th term in the bracket of Eq.(\ref{eq:qp1})
becomes
\begin{eqnarray}
4th &=&\gamma_{31}\frac{1}{E-M^0_{N^*}-\hat{\Sigma}}
\hat{\Gamma}_{12}[1+\frac{1}{E-H_0-\Sigma}P^\prime_2T_{\bar{v}}P^\prime_2]
\nonumber \\
&=& 
\gamma_{31}\frac{1}{E-M^0_{N^*}-\hat{\Sigma}}
\hat{\Gamma}_{12}[1+\frac{1}{E-H_0-\Sigma}
(t_E +\bar{\Gamma}_{21}\frac{1}
{E-M^0_{N^*}-\hat{\Sigma}-\bar{\Sigma} }\bar{\Gamma}_{12})] \nonumber \\
&=&
\gamma_{31}\frac{1}{E-M^0_{N^*}-\hat{\Sigma}}
[\hat{\Gamma}_{12}(1+\frac{1}{E-H_0-\Sigma}t_E)
+(\hat{\Gamma}_{12}\frac{1}{E-H_0-\Sigma}\bar{\Gamma}_{21})
\frac{1}{E-M^0_{N^*}-\hat{\Sigma}-\bar{\Sigma} }\bar{\Gamma}_{12}]
\nonumber \\
&=&\gamma_{31}\frac{1}{E-M^0_{N^*}-\hat{\Sigma}}
[1+\bar{\Sigma}\frac{1}{E-M^0_{N^*}-\hat{\Sigma}-\bar{\Sigma}}]
\bar{\Gamma}_{12} \nonumber \\
&=&\gamma_{31}
\frac{1}{E-M^0_{N^*}-\hat{\Sigma}-\bar{\Sigma}}
\bar{\Gamma}_{12} \,.
\end{eqnarray}

We finally obtain
\begin{eqnarray}
T_{QP_2} &=& Q\Omega^{(-)\dagger}_{\pi\pi N}[
u_{23} + \{ u_{32}\frac{P_2}{E-H_0} +(u_{32}
+\Gamma_{21})\frac{P_{2*}}{E-H_0-\Sigma}\}
\{ t_E +\bar{\Gamma}_{21}\frac{1}
{E-M^0_{N^*}-\hat{\Sigma}-\bar{\Sigma} }\bar{\Gamma}_{12}\} 
\nonumber \\
& & +\gamma_{31}
\frac{1}{E-M^0_{N^*}-\hat{\Sigma}-\bar{\Sigma}}
\bar{\Gamma}_{12}]P_2 \,,
\end{eqnarray}
where
\begin{eqnarray}
\Omega^{(-)\dagger}_{\pi\pi N} = (1+t_Q\frac{Q}{E-H_0}) \,.
\end{eqnarray}
Here $t_Q$ is defined by Eq.(\ref{eq:tq}) and hence 
$\Omega^{(-)\dagger}_{\pi\pi N}$
is the $\pi\pi N$ scattering operator.

In the above rather detailed derivations, Eqs.(B53) and (B62) are
what we need to investigate meson-baryon scattering and two-pion production.
In practice,
the interaction $\gamma_{31}=\Gamma_{N^*\rightarrow \pi\pi N}$ will be neglected
in first calculations. If we set $\gamma_{31}=0$, we then find
from Eqs.(B23)-(B27) that
\begin{eqnarray}
\hat{\Gamma}_{21} &\rightarrow& \Gamma_{21} \,, \\
\hat{\Gamma}_{12} &\rightarrow& \Gamma_{12} \,,\\
\hat{\Sigma} &\rightarrow& 0 \,.
\end{eqnarray}
Eqs.(B52), and (B54)-(B55) lead to
\begin{eqnarray}
\bar{\Gamma}_{21} &\rightarrow& 
[1+t_E\frac{1}{E-H_0-\Sigma}]\Gamma_{21} \,,  \\
\bar{\Gamma}_{12} &\rightarrow&  
\Gamma_{12}[1+\frac{1}{E-H_0-\Sigma}t_E] \,, \\
\bar{\Sigma} &\rightarrow& \Gamma_{12}\frac{1}{E-H_0-\Sigma}
[1+t_E\frac{1}{E-H_0-\Sigma}]\Gamma_{21}  \nonumber \\
 &=& \Gamma_{12}\frac{1}{E-H_0-\Sigma}\bar{\Gamma}_{21} \,.
\end{eqnarray}

Recalling Eq.(B7)-(B8) for the projection operators $P_2$ and $P_{2*}$,
we can write
\begin{eqnarray}
P^\prime_2 = P_2+P_{2*} = \sum_{MB} |MB><MB| \,,
\end{eqnarray}
where $MB =\gamma N, \pi N, \eta N, \pi\Delta, \rho N,\sigma N$ include
all meson-baryon states in the considered model space.
Defining
\begin{eqnarray}
T_{MB,M'B'}(E)&=&<MB|P^\prime_2 T_{\bar{v}}P^\prime_2|M'B'> \,, \\
t_{MB,M'B'}(E)&=&<MB|t_E|M'B'> \,, \\
V_{MB,M'B'}(E) &=& <MB|V_E|M'B'> \,, \nonumber
\end{eqnarray}
and
\begin{eqnarray}
G_{MB}(E) &=& <MB|\frac{1}{E-H_0-\Sigma}|MB> \,, 
\end{eqnarray}
 using the simplifications Eqs.(B64)-(B69), Eq.(B22) for $V_E$, and
Eq.(B48) for $t_E$, the matrix element of Eq.(B53)
between two $MB$ states then become
\begin{eqnarray}
T_{MB,M'B'}(E)
&=& t_{MB,M'B'}(E)+t^R_{MB,M'B'}(E) \,, 
\label{eq:ap-totmbmb}
\end{eqnarray}
where
\begin{eqnarray}
t_{MB,M'B'}(E)&=& V_{MB,M'B'}(E) + \sum_{M''B''}V_{MB,M''B''}(E)G_{M''B''}(E)
t_{M''B'',M'B'}(E) \,,
\label{eq:ap-tmbmb}
\end{eqnarray}
with 
\begin{eqnarray}
V_{MB,M'B'}(E) = <MB|v_{22}+
(u_{23}+\Gamma_{12})G_Q(u_{32}+\Gamma_{21})-\Sigma |M'B'> \,.
\label{eq:ap-vmbmb}
\end{eqnarray}
As defined in the beginning of this appendix, we have 
 $v_{22}=v_{MB,M'B'}+v_{\pi\pi}$,
 $\Gamma_{12}=\Gamma_{N^*\rightarrow MB} + h_{M^*\rightarrow \pi\pi}$ with
$M^* = \rho, \sigma$, $\Gamma_{21}=\Gamma^\dagger_{12}$ and
$u_{23}=v_{MB,\pi\pi N}$, $u_{32}=u^\dagger_{23}$. Eq.(\ref{eq:ap-vmbmb})
can be written explicitly as
\begin{eqnarray}
V_{MB,M^\prime B^\prime}(E)= v_{MB,M^\prime B^\prime}
+Z_{{M}{B},{M}^\prime {B}^\prime}(E) \,.
\label{eq:veff-mbmb-app}
\end{eqnarray}
Here $Z_{{M}{B},{M}^\prime {B}^\prime}(E)$
contains the effects due to the coupling with $\pi\pi N$ states.
It has the following form
\begin{eqnarray}
Z_{{M}{B},{M}^\prime {B}^\prime}(E)
&=& < {M}{B} \mid F
\frac{ P_{\pi \pi N}}
{E- H_0 - \hat{v}_{\pi \pi N}+ i\epsilon}
F^\dagger  \mid {M}^\prime {B}^\prime > \nonumber \\
& & \nonumber \\
& & -[ \delta_{MB,M'B'} \Sigma_{MB}(E) ] \,,
\label{eq:z-mbmb}
\end{eqnarray}
where
\begin{eqnarray}
\Sigma_{MB}(E)&=&<MB|\Sigma|MB> \,, \\
F &=& g_V + v_{MB,\pi\pi N} \nonumber \\
&=& [\Gamma_{ \Delta\rightarrow \pi N}+
h_{\rho\rightarrow \pi\pi} + h_{\sigma \rightarrow \pi\pi}]
+v_{MB,\pi\pi N} 
\label{eq:vertex-f } \,, \\
\hat{v}_{\pi\pi N} &=& v_{\pi N,\pi N} + v_{\pi\pi} + v_{\pi\pi N,\pi\pi N} \,.
\end{eqnarray}

 The resonant term in Eq.(\ref{eq:ap-totmbmb}) is
\begin{eqnarray}
t^R_{MB,M'B'}(E)&=&<MB|\bar{\Gamma}_{21}\frac{1}{E-H_0-\bar{\Sigma}}
\bar{\Gamma}_{12} |MB>\nonumber  \\
&=& \sum_{N^*_i,N^*_j}\bar{\Gamma}_{MB\rightarrow N^*_i}
<N^*_i|\frac{1}{E-H_0 - \bar{\Sigma}}|N^*_j>
\bar{\Gamma}_{N^*_j\rightarrow M'B'} \,.
\end{eqnarray}
Note that $\Sigma$ in Eqs.(B73) and (B79) is defined by Eq.(B19). If we neglect
the non-resonant interactions in $\pi\pi N$ $Q$-space, we then have
\begin{eqnarray}
\Sigma &\rightarrow&[ \Gamma_{12}\frac{Q}{E-H_0}\Gamma_{21}]_{un-connected}\,.
\end{eqnarray}
Since $\Gamma_{12}$ does not have a $N \rightarrow \pi N$, we obviously have
$<\pi N|\Sigma|\pi N>=0$ and hence
\begin{eqnarray}
G_{\pi N}(E) &=& \frac{1}{E-K_\pi (k) - K_N(p) + i\epsilon} \,, \\
G_{\pi\Delta}(E)&=&\frac{1}{E-K_\pi(k)-K_\Delta(p)
 - \Sigma_{\pi\Delta}(E-K_\pi(k))} \,, \\
G_{\rho N}(E)&=&\frac{1}{E-K_\rho(k)-K_N(p)-\Sigma_{\rho N}(E-K_N(p))}\,, \\
G_{\sigma N}(E)&=&\frac{1}{E-K_\sigma-K_N(p)-\Sigma_{\sigma N}(E-K_N(p))}\,,
\end{eqnarray}
where
\begin{eqnarray}
\Sigma_{\pi\Delta}(\omega ) &=& <\pi \Delta|
\Gamma_{\Delta\rightarrow \pi N}\frac{1}{\omega - K_\pi(k)-K_N(p)+i\epsilon}
\Gamma_{\pi N \rightarrow \Delta} |\pi \Delta > \,, \\
\Sigma_{\rho N}(\omega ) &=&<\rho N|
\Gamma_{\rho \rightarrow \pi \pi}\frac{1}{\omega - E_\pi(k_1)-E_\pi(k_2)
+i\epsilon}\Gamma_{\pi\pi \rightarrow \rho}|\rho N> \,, \\
\Sigma_{\sigma N}(\omega ) &=&<\sigma N|
\Gamma_{\sigma \rightarrow \pi \pi}\frac{1}{\omega - E_\pi(k_1)-E_\pi(k_2)
+i\epsilon}\Gamma_{\pi\pi \rightarrow \sigma}|\sigma N> \,.
\end{eqnarray}
In the above equations, $K_\alpha(p)=\sqrt{m_\alpha^2+\vec{p}^2}$ 
is the free energy operator defined by  momentum operator $\vec{p}$.

When $N^* \rightarrow \pi\pi N$ is neglected, the two-pion production
operator $T_{QP}$ defined in Eq.(B62) also becomes simpler,
since its last term in the right-hand side  does not contribute.
By using Eqs.(B66), (B72) and (B77), the matrix element of Eq.(B62)
$T_{\pi\pi N,MB}(E)=<\pi\pi N |T_{QP_2} |MB> $ can be written as
\begin{eqnarray}
T_{\pi\pi N,MB}(E)&=&
<\psi^{(-)}_{\pi\pi N}| u_{32}|MB> \nonumber \\
& & + \sum_{M'B'}[<\psi^{(-)}_{\pi\pi N}u_{32}\frac{P^\prime_2}{E-H_0-\Sigma}
|M'B'>T_{M'B',MB} \nonumber \\
& &+ <\psi^{(-)}_{\pi\pi N}|\Gamma_{21}
\frac{P_{2*}}{E-H_0-\Sigma}|M'B'>T_{M'B',MB} ] \,.
\label{eq:ap-pipinmb}
\end{eqnarray}
Recalling that $u_{32}=v_{\pi\pi N, MB}$, 
$\Gamma_{21}=\Gamma_{\pi N\rightarrow \Delta} 
+ \Gamma_{\pi\pi \rightarrow \rho}+\Gamma_{\pi\pi \rightarrow \sigma}$,
we can write Eq.(\ref{eq:ap-pipinmb}) explicitly as
\begin{eqnarray}
T_{\pi\pi N,MB}(E) = T^{dir}_{\pi\pi N,MB}(E)+
T^{\pi\Delta}_{\pi\pi N,MB}(E)+T^{\rho N}_{\pi\pi N,MB}(E)
+T^{\sigma N}_{\pi\pi N,MB}(E)
\label{eq:ap-tpipin-1}
\end{eqnarray}
with
\begin{eqnarray}
T^{dir}_{\pi\pi N,MB}(E)
&=&
<\psi^{(-)}_{\pi\pi N}(E)|
 \sum_{M'B'}v_{\pi\pi N,M'B'}|M'B'>[\delta_{M'B',MB} \nonumber \\
& & +G_{M'B'}(E)T_{M'B',MB}(E) 
\label{eq:ap-tpipin-dir} \,, \\
& & \nonumber \\
T^{\pi \Delta}_{\pi\pi N,MB}(E)
&=&
 <\psi^{(-)}_{\pi\pi N}(E)|
\Gamma_{\pi N\rightarrow \Delta}|\pi \Delta> 
G_{\pi\Delta}(E) T_{\pi\Delta, MB}(E) \,,
\label{eq:ap-tpipin-pid}
 \\
& & \nonumber \\
T^{\rho N}_{\pi\pi N,MB}(E)
&=&
 <\psi^{(-)}_{\pi\pi N}(E)|
h_{\pi\pi \rightarrow \rho}|\rho N>
G_{\rho N}(E)T_{\rho N, MB}(E)
\label{eq:ap-tpipin-rhon} \,,  \\
& & \nonumber \\
T^{\sigma N}_{\pi\pi N,MB}(E)
&=&
 <\psi^{(-)}_{\pi\pi N}(E)|
h_{\pi\pi \rightarrow \sigma}|\sigma N>
G_{\sigma N}(E)T_{\sigma N, MB}(E) \,.
\label{eq:ap-tpipin-sigma}
\end{eqnarray}

\section{Matrix elements of meson-baryon potentials}

To solve Eq.(\ref{eq:pw-tmbmb}) for generating the non-resonant amplitudes, 
we need to first calculate the partial-wave matrix elements
of meson-baryon non-resonant interactions $v_{MB,M'B'}$ 
generated from the Lagrangians specified in Appendix A,
and the 
one-particle-exchange interaction $Z^{(E)}_{MB,M'B'}(E)$ defined by
Eq.(\ref{eq:z-mbmb-e}) and illustrated in Fig.\ref{fig:z}. 
In this appendix, we present formula for calculating 
the partial-wave matrix elements of $v_{MB,M'B'}$ with
$MB, M'B' = \pi N, \eta N, \sigma N, \rho N, \pi\Delta$.
The partial-wave matrix elements  of $Z^{(E)}_{MB,M'B'}(E)$
will be given in Appendix D.

In general, each of 
the constructed $v_{MB,M'B'}$ consists of various combinations
of tree-diagram mechanisms illustrated in Fig.\ref{fig:mbmb}. They can be
computed by the usual Feynman rules, except that the time components of
the propagators of the intermediate states are specified by the
unitarity transformation method, such that the resulting matrix elements
are independent  of the collision energy $E$ of Eq.(\ref{eq:pw-tmbmb})
and free of any singularity on the real momentum
axis. We will explain this feature of our model at the end of this appendix.
 
It is convenient to get the partial matrix elements by first evaluating
the matrix elements of $v_{MB,M'B'}$ in helicity representation and then
transforming them into the usual $|(LS)J T >$ representation with
$J$, $T$, $L$, and $S$  denoting the total angular momentum,
isospin, orbital   
angular momentum, and spin quantum numbers, respectively.
For each meson-baryon ($MB$) state, we use $k$($p$) to denote
the momentum of $M$($B$). In the center of mass frame,
we thus have $\vec{p}=-\vec{k}$.
Following the Jacob-Wick formulation\cite{jw},
the partial-wave matrix elements of the
 non-resonant interaction $v_{MB,M'B'}$ can be written as 
\begin{eqnarray}
 {\it v}^{JT}_{L' S' M'B', L S MB}(k',k,E) 
 & =& \sum_{ \lambda'_M\lambda'_B\lambda_M\lambda_B } [
\frac{\sqrt{(2L+1)(2L'+1)}}{2J+1} \nonumber \\
& & \times <j'_M j'_B \lambda'_M  -\lambda'_B \mid S' S'_z>
<L'S'0S'_z\mid J S'_z> \nonumber \\
& & \times <j_M j_B \lambda_M  -\lambda_B \mid S S_z>
<LS0S_z\mid J S_z> \nonumber \\
& & \times <J,k' \lambda'_M -\lambda'_B \mid v_{M'B',MB}\mid
 J,k \lambda_M -\lambda_B >] \,,
\label{eq:c-1}
\end{eqnarray}
where $j_M$ and $j_B$ are the spins of the meson  and baryon, 
respectively, and $\lambda_M$ and $\lambda_B$ are their helicities,
 and
\begin{eqnarray}
& &<J,k' \lambda'_M -\lambda'_B \mid v_{M'B',MB}\mid
 J,k \lambda_M -\lambda_B > \nonumber \\
& & = 2\pi \int_{-1}^{+1} d(\cos\theta)
d^{J}_{\lambda'_M-\lambda'_B,\lambda_M-\lambda_B}(\theta)
\nonumber \\
& & \times <(\vec{k}', s'_M \lambda'_M), (-\vec{k}', s'_B,-\lambda'_B)
\mid v_{M'B',MB}
\mid (\vec{k}, s_M \lambda_M), (-\vec{k}, s_B, -\lambda_B) >\,.
\label{eq:c-2} \nonumber \\
& &
\end{eqnarray}
Here we have chosen the coordinates such that
\begin{eqnarray}
\vec{k}' &=& (k'\sin\theta ,0, k'\cos\theta )\,, \\
\vec{k} &=& (0 ,0, k )\,,
\end{eqnarray}
and the helicity eigenstates are defined by
\begin{eqnarray}
\hat{k}\cdot \vec{s}_M \mid M( \vec{k}, s_M \lambda_M) > &=& \lambda_M
\mid M( \vec{k}, s_M \lambda_M) >  \,,
\end{eqnarray}
\begin{eqnarray}
[-\hat{k}\cdot \vec{s}_B] \mid B( -\vec{k}, s_B \lambda_B) > 
&=& \lambda_B \mid B( -\vec{k}, s_B \lambda_B) > \,.
\label{eq:helicity-2}
\end{eqnarray}
Note the $-$ sign in Eq.(\ref{eq:helicity-2}).

\begin{table}
\centering
\begin{tabular}{c|cccccc} \hline\hline
Channels & $\pi N$ & $\eta N$ & $\sigma N$ & $\rho N$ & $\pi\Delta$ & \\ \hline
$\pi N$ & 1& 2& 4& 7& 11& \\
$\eta N$& & 3& 5& 8& 12& \\
$\sigma N$ & & & 6& 9& 13& \\
$\rho N$  & & &  & 10& 14& \\
$\pi\Delta$ & & & & & 15& \\
\hline\hline
\end{tabular}
\caption{ Labels for $v_{a,b}$ with $a,b$ = $\pi N$, $\eta N$,
$\sigma N$, $\rho N$, and $\pi\Delta$}
\label{tab:pot}
\end{table}

To evaluate the matrix elements in the right hand side of
  Eq.(\ref{eq:c-2}) with the normalization defined by Eq.(\ref{eq:crst}), 
we define (suppress the helicity and isospin indices)
\begin{eqnarray}
<k^\prime(j),p^\prime \mid v_{M'B',MB}\mid k(i), p> &=& \frac{1}{(2\pi)^3}
\sqrt{\frac{m_B^\prime}{E_{B^\prime}(p^\prime)}}
\frac{1}{\sqrt{2E_{M^\prime}(k^\prime)}}
\sqrt{\frac{m_B}{E_{B}(p^\prime)}}
\frac{1}{\sqrt{2E_{M}(k)}} \nonumber \\
& & \times \bar{u}_{B'}(\vec{p'}) \bar{V}(n) u_B(\vec{p})
\end{eqnarray}
where $n$ defined the $MB \rightarrow M'B'$ transitions 
as specified in Table~\ref{tab:pot},
and $i,j$ are the isospin indices of the mesons.
We also have defined $q= k^\prime-k$ or $q=p-p^\prime$.
The expressions of each term in Table \ref{tab:pot} are given in the following 
subsections.

\subsection{$\pi(k,i)+ N(p) \rightarrow \pi(k^\prime,j)+ N(p^\prime)$} 

\begin{eqnarray}
\bar{V}(1) = \bar{V}^1_a +\bar{V}^1_b+\bar{V}^1_c+\bar{V}^1_d+\bar{V}^1_e
\end{eqnarray}
with
\begin{eqnarray}
\bar{V}^1_a &=& [\frac{f_{\pi NN}}{m_\pi}]^2 \slas{k}^\prime
\gamma_5 \tau^j S_N(p+k) \slas{k}\gamma_5 \tau^i \\
\bar{V}^1_b &=& [\frac{f_{\pi NN}}{m_\pi}]^2 \slas{k}
\gamma_5 \tau^i S_N(p-k^\prime) \slas{k}^\prime\gamma_5 \tau^j \\
\bar{V}^1_c &=& [\frac{f_{\pi N\Delta}}{m_\pi}]^2 {k}_\alpha
(T^\dagger)^i S^{\alpha\beta}_\Delta(p-k^\prime){k}^\prime_{\beta} T^j \\
\bar{V}^1_d &=&i g_{\rho NN}g_{\rho\pi\pi}\frac{\tau^{\it l}}{2}
\epsilon_{ji{\it l}}\frac{1}{q^2-m^2_\rho} \nonumber \\
& & \times [(\slas{k}+\slas{k}^\prime)+
\frac{\kappa_\rho}{4m_N}\{(\slas{k}+\slas{k}^\prime)\slas{q}
-\slas{q}(\slas{k}+\slas{k}^\prime)\}] \\
\bar{V}^1_e &=& - g_{\sigma NN}\frac{g_{\sigma \pi\pi}}{m_\pi}
\delta_{i,j} \frac{k\cdot k^\prime}{q^2-m_\sigma^2}
\end{eqnarray}

\subsection{$\pi(k,i)+ N (p)\rightarrow \eta(k^\prime)+ N (p^\prime)$ }
\begin{eqnarray}
\bar{V}(2) = \bar{V}^2_a +\bar{V}^2_b
\end{eqnarray}
with 
\begin{eqnarray}
\bar{V}^2_a &=& \frac{f_{\pi NN}f_{\eta NN}}{m_\pi m_\eta} \slas{k}^\prime
\gamma_5  S_N(p+k) \slas{k}\gamma_5 \tau^i \\
\bar{V}^2_b &=& \frac{f_{\pi NN}f_{\eta NN}}{m_\pi m_\eta} \slas{k}
\gamma_5 \tau^i S_N(p-k^\prime) \slas{k}^\prime\gamma_5 
\end{eqnarray}

\subsection{$\eta(k)+ N(p) \rightarrow \eta(k^\prime)+ N(p^\prime)$ }
\begin{eqnarray}
\bar{V}(3) = \bar{V}^3_a +\bar{V}^3_b
\end{eqnarray}
with
\begin{eqnarray}
\bar{V}^3_a&=&[\frac{f_{\eta NN}}{m_\eta}]^2\slas{k}^\prime
\gamma_5 S_N(p+k) \slas{k}\gamma_5 \\
\bar{V}^3_b&=&[\frac{f_{\eta NN}}{m_\eta}]^2\slas{k}
\gamma_5 S_N(p-k^\prime) \slas{k}^\prime\gamma_5 
\end{eqnarray}

\subsection{$\pi(k,i)+ N (p)\rightarrow \sigma(k^\prime)+ N (p^\prime)$ }
\begin{eqnarray}
\bar{V}(4) = \bar{V}^4_a +\bar{V}^4_b+\bar{V}^4_c
\end{eqnarray}
with
\begin{eqnarray}
\bar{V}^4_a&=& i g_{\sigma NN}\frac{f_{\pi NN}}{m_\pi}S_N(p+k)\slas{k}
\gamma_5\tau^i \\
\bar{V}^4_b&=& i g_{\sigma NN}\frac{f_{\pi NN}}{m_\pi}\slas{k}
\gamma_5 S_N(p-k^\prime) \tau^i \\
\bar{V}^4_c &=& i\frac{f_{\pi NN}g_{\sigma\pi\pi}}{m_\pi^2}\slas{q}\gamma_5
\tau^i\frac{q\cdot k}{q^2-m^2_\pi} % ; \,\,\,\, with \,\,\, q = p-p'
\end{eqnarray}

\subsection{$\eta(k)+ N (p)\rightarrow \sigma(k^\prime)+ N (p^\prime)$ }
\begin{eqnarray}
\bar{V}(5) = \bar{V}^5_a +\bar{V}^5_b
\end{eqnarray}
with
\begin{eqnarray}
\bar{V}^5_a&=&i g_{\sigma NN}\frac{f_{\eta NN}}{m_\eta}S_N(p+k)\slas{k}
\gamma_5 \\
\bar{V}^5_b&=&i g_{\sigma NN}\frac{f_{\eta NN}}{m_\eta}\slas{k}\gamma_5
S_N(p-k^\prime) 
\end{eqnarray}

\subsection{$\sigma(k)+ N(p) \rightarrow \sigma(k^\prime)+ N(p^\prime)$ }
\begin{eqnarray}
\bar{V}(6) = \bar{V}^6_a +\bar{V}^6_b    %+\bar{V}^6_c
\end{eqnarray}
with
\begin{eqnarray}
\bar{V}^6_a&=&g_{\sigma NN}^2 S_N(p+k)  \\
\bar{V}^6_b&=&g_{\sigma NN}^2 S_N(p-k^\prime) %\\
\end{eqnarray}

\subsection{$\pi(k,i)+ N(p) \rightarrow \rho(k^\prime,j)+ N(p^\prime)$}

\begin{eqnarray}
\bar{V}(7) = \bar{V}^7_a +\bar{V}^7_b+\bar{V}^7_c+\bar{V}^7_d+\bar{V}^7_e
\end{eqnarray}
with
\begin{eqnarray}
\bar{V}^7_a &=& i\frac{f_{\pi NN}}{m_\pi}g_{\rho NN}\Gamma_{\rho^\prime}
 S_N(p+k) \slas{k}\gamma_5 \tau^i \\
\bar{V}^7_b &=& i\frac{f_{\pi NN}}{m_\pi}g_{\rho NN}
\slas{k}\gamma_5 \tau^i S_N(p-k^\prime)\Gamma_{\rho^\prime}
\\
\bar{V}^7_c &=&\frac{f_{\pi NN}}{m_\pi} g_{\rho\pi\pi}
\epsilon_{ij{\it l}}\tau^{\it l}\frac{(q-k)\cdot\epsilon_{\rho^\prime}^*
\slas{q}\gamma_5}{q^2-m_\pi^2} \\   %; \,\,\, with \,\,\, q=p-p'\\
\bar{V}^7_d&=&-\frac{f_{\pi NN}}{m_\pi}g_{\rho NN}
\slas{\epsilon_{\rho^\prime}}^*\gamma_5 \epsilon_{ji{\it l}}\tau^{\it l}
\\
\bar{V}^7_e&=& \frac{g_{\omega NN}g_{\omega \pi\rho}}{m_\omega}\delta_{ij}
\frac{\epsilon_{\alpha\beta\gamma\delta}\epsilon_{\rho^\prime}^{*\alpha}
 k^{\prime\beta}
k^\gamma}{q^2-m^2_\omega} % \nonumber \\
[\gamma^\delta+\frac{\kappa_\omega}{4m_N}
(\gamma^\delta\slas{q}-\slas{q}\gamma^\delta)]   %\\
\end{eqnarray}
where
\begin{eqnarray}
\Gamma_{\rho^\prime}& =&\frac{\tau^j}{2}[\slas{\epsilon_{\rho^\prime}}^*+
\frac{\kappa_\rho}{4m_N}
(\slas{\epsilon_{\rho^\prime}}^*
\slas{k}^\prime-\slas{k}^\prime\slas{\epsilon_{\rho^\prime}}^*)] 
\end{eqnarray}

\subsection{$\eta(k)+ N(p) \rightarrow \rho(k^\prime,j)+ N(p^\prime)$}

\begin{eqnarray}
\bar{V}(8) = \bar{V}^8_a +\bar{V}^8_b
\end{eqnarray}
with
\begin{eqnarray}
\bar{V}^8_a &=& i\frac{f_{\eta NN}}{m_\eta}g_{\rho NN}\Gamma_{\rho^\prime}
 S_N(p+k) \slas{k}\gamma_5  \\
\bar{V}^8_b &=& i\frac{f_{\eta NN}}{m_\eta}g_{\rho NN}
\slas{k}\gamma_5  S_N(p-k^\prime)\Gamma_{\rho^\prime}
\end{eqnarray}

\subsection{$\sigma(k)+ N(p) \rightarrow \rho(k^\prime,j)+ N(p^\prime)$}

\begin{eqnarray}
\bar{V}(9) = \bar{V}^9_a +\bar{V}^9_b
\end{eqnarray}
with
\begin{eqnarray}
\bar{V}^9_a&=& g_{\rho NN} g_{\sigma NN}\Gamma_{\rho^\prime}
S_N(p+k) \\
\bar{V}^9_b&=& g_{\rho NN} g_{\sigma NN} 
S_N(p-k^\prime) \Gamma_{\rho^\prime}
\end{eqnarray}

\subsection{$\rho(k,i)+ N(p) \rightarrow \rho^\prime(k^\prime,j)+ N(p^\prime)$}

\begin{eqnarray}
\bar{V}(10) = \bar{V}^{10}_a +\bar{V}^{10}_b+ \bar{V}^{10}_c
\end{eqnarray}
with
\begin{eqnarray}
\bar{V}^{10}_a +\bar{V}^{10}_b &=&
g_{\rho NN}^2[\Gamma_{\rho^\prime} S_N(p+k)\Gamma_\rho 
+\Gamma_{\rho} S_N(p-k^\prime)\Gamma_{\rho^\prime}] 
\end{eqnarray}
where
\begin{eqnarray}
\Gamma_{\rho}& =&\frac{\tau^i}{2}[\slas{\epsilon_{\rho}}-
\frac{\kappa_\rho}{4m_N}
(\slas{\epsilon_{\rho}}
\slas{k}-\slas{k}\slas{\epsilon_{\rho}})] 
\end{eqnarray}

\begin{eqnarray}
\bar{V}^{10}_c &=& i\frac{\kappa_\rho g^2_{\rho NN}}{8 m_N}
[\slas{\epsilon_\rho}\slas{\epsilon_{\rho^\prime}}^*-
\slas{\epsilon_{\rho^\prime}}^*\slas{\epsilon_{\rho}} ]
\epsilon_{i j {\it l}}\tau^{\it l} % \\
\end{eqnarray}

\subsection{$\pi(k,i)+ N(p) \rightarrow \pi^\prime(k^\prime,j)+ 
\Delta(p^\prime)$}
\begin{eqnarray}
\bar{V}(11) = \bar{V}^{11}_a +\bar{V}^{11}_b+ \bar{V}^{11}_c
+\bar{V}^{11}_d+ \bar{V}^{11}_e
\end{eqnarray}
with
\begin{eqnarray}
\bar{V}^{11}_a&=&\frac{f_{\pi NN}f_{\pi N\Delta}}{m^2_\pi}
T^j\epsilon_\Delta^*\cdot k^\prime S_N(p+k)\slas{k}\gamma_5\tau^i \\
\bar{V}^{11}_b&=&\frac{f_{\pi NN}f_{\pi N\Delta}}{m^2_\pi}
T^i\epsilon_\Delta^*\cdot k S_N(p-k^\prime)\slas{k}^\prime\gamma_5 
\tau^j \\
\bar{V}^{11}_c&=&i \frac{f_{\rho N\Delta}f_{\rho\pi\pi}}{m_\rho}
\frac{\epsilon_{ji{\it l}}T^{\it l}}{q^2-m^2_\rho}
[\epsilon_\Delta^*\cdot q (\slas{k}+\slas{k}^\prime)\gamma_5
-\epsilon_\Delta^*\cdot(k+k^\prime)\slas{q}\gamma_5] \\
\bar{V}^{11}_d&=& -\frac{f_{\pi\Delta\Delta}f_{\pi N\Delta}}{m^2_\pi}
[\epsilon_\Delta^*]_\mu \slas{k}^\prime\gamma_5 T^j_\Delta
 S^{\mu\nu}_{\Delta}(p' + k') T^i k_\nu \\
\bar{V}^{11}_e&=& -\frac{f_{\pi\Delta\Delta}f_{\pi N\Delta}}{m^2_\pi}
[\epsilon_\Delta^*]_\mu \slas{k}\gamma_5 T^i_\Delta
 S^{\mu\nu}_{\Delta}(p- k^\prime) T^j k_\nu^\prime %\\
\end{eqnarray}

\subsection{$\eta(k)+ N(p) \rightarrow \pi^\prime(k^\prime,j)+
\Delta(p^\prime)$}
\begin{eqnarray}
\bar{V}(12) 
 =\frac{f_{\eta NN}f_{\eta N\Delta}}{m_\pi m_\eta}
T^j\epsilon_\Delta^*\cdot
k^\prime S_N(p+k) \slas{k}\gamma_5
\end{eqnarray}

\subsection{$\sigma(k)+ N(p) \rightarrow \pi^\prime(k^\prime,j)+
\Delta(p^\prime)$}
\begin{eqnarray}
\bar{V}(13)
 =-i g_{\sigma NN}\frac{f_{\pi N\Delta}}{m_\pi}
T^j\epsilon_\Delta^*\cdot
k^\prime S_N(p+k) 
\end{eqnarray}

\subsection{$\rho(k,i)+ N(p) \rightarrow \pi^\prime(k^\prime,j)+
\Delta(p^\prime)$}
\begin{eqnarray}
\bar{V}(14) = \bar{V}^{14}_a +\bar{V}^{14}_b+ \bar{V}^{14}_c
+\bar{V}^{14}_d
\end{eqnarray}
with
\begin{eqnarray}
\bar{V}^{14}_a &=& -i\frac{f_{\pi N\Delta}g_{\rho NN}}{m_\pi}
T^j\epsilon_\Delta^*\cdot k^\prime S_N(p+k)\Gamma_\rho\\
%tsato june 23 2005
\bar{V}^{14}_b &=&i\frac{f_{\pi N\Delta}g_{\rho NN}}{m_\pi}
T^i[
  \epsilon_\Delta^*\cdot k \slas{\epsilon}_\rho \gamma_5
- \epsilon_\Delta^*\cdot \epsilon_\rho \slas{k} \gamma_5]
\nonumber \\
& & \times
 S_N(p-k^\prime)\slas{k}^\prime \gamma_5 \tau^j \\
\bar{V}^{14}_c &=& -i \frac{f_{\pi \Delta\Delta}f_{\rho N\Delta}}{m_\pi m_\rho}
[\epsilon_\Delta^*]_\alpha\slas{k}^\prime \gamma_5 T^j_\Delta 
S^{\alpha\beta}_{\Delta}(p'+k')
[k_\beta\slas{\epsilon_\rho}\gamma_5 -
[\epsilon_\rho]_\beta \slas{k}\gamma_5]T^i \\
\bar{V}^{14}_d &=&-i \frac{g_{\rho \Delta\Delta}f_{\pi N\Delta}}{m_\pi}
[\epsilon_\Delta^*]_\alpha [\slas{\epsilon_\rho}
-\frac{\kappa_{\rho\Delta\Delta}}{4m_\Delta}
(\slas{\epsilon_\rho}\slas{k}-\slas{k}\slas{\epsilon_\rho})]T_\Delta^i
S^{\alpha\beta}_\Delta(p-k^\prime) T^j k^{\prime}_\beta \nonumber
\\
\end{eqnarray}
\subsection{$\pi(k,i)+ \Delta(p) \rightarrow \pi^\prime(k^\prime,j)+
\Delta^\prime (p^\prime)$}
\begin{eqnarray}
\bar{V}(15) = \bar{V}^{15}_a +\bar{V}^{15}_b+ \bar{V}^{15}_c
+\bar{V}^{14}_d
\end{eqnarray}
with
\begin{eqnarray}
\bar{V}^{15}_a &=& [\frac{f_{\pi N\Delta}}{m_\pi}]^2
\epsilon_\Delta^{\prime *}\cdot k^\prime T^j S_N(p+k) \epsilon_\Delta \cdot 
k (T^i)^\dagger \\
\bar{V}^{15}_b &=& [\frac{f_{\pi \Delta\Delta}}{m_\pi}]^2
\slas{k}^\prime \gamma_5 T^j_\Delta [\epsilon_{\Delta^{\prime *}}]_\mu
S^{\mu\nu}_\Delta(p+k) [\epsilon_{\Delta}]_\nu \slas{k}\gamma_5 T^i_\Delta
\\
\bar{V}^{15}_c &=& [\frac{f_{\pi \Delta\Delta}}{m_\pi}]^2
\slas{k} \gamma_5 T^i_\Delta [\epsilon_{\Delta^{\prime *}}]_\mu
S^{\mu\nu}_\Delta(p-k^\prime) [\epsilon_{\Delta}]_\nu 
\slas{k}^\prime \gamma_5 T^j_\Delta
\\
\bar{V}^{15}_d&=&i g_{\rho \Delta\Delta}g_{\rho\pi\pi}
\frac{\epsilon_{ji{\it l}}T_\Delta^{\it l}}{q^2-m^2_\rho}
\{(\slas{k}+\slas{k}^\prime)+
\frac{\kappa_{\rho\Delta\Delta}}{4m_\Delta}((\slas{k}+\slas{k}^\prime)
\slas{q}-\slas{q}(\slas{k}+\slas{k}^\prime)\}
\epsilon_{\Delta^{\prime *}}\cdot\epsilon_{\Delta}
 \nonumber \\
\end{eqnarray}

The baryon propagators in Eqs.(C8)-(C64) are
\begin{eqnarray}
S_N(p) &=& \frac{1}{\slas{p} - m_N} \,,\\
S^{\mu\nu}_\Delta(p)&=&\frac{1}{3(\slas{p}-m_\Delta)}[2(-g^{\mu\nu}
+\frac{p^\mu p^\nu}{m_\Delta^2})
+\frac{\gamma^\mu\gamma^\nu-\gamma^\nu\gamma^\mu}{2}
-\frac{p^\mu\gamma^\nu-p^\nu\gamma^\mu}{m_\Delta}] \,.
\label{eq:delta-p}
\end{eqnarray}
Eq.(\ref{eq:delta-p}) is the simplest choice of many possible
definitions of the $\Delta$ propagator. It is part of our
phenomenology for this rather complex coupled-channel calculations.

Although the expressions Eqs.(C8)-(C64)
look like the usual Feynman amplitudes, the
unitary transformation method defines definite procedures in evaluating the
time component of each propagator. For each propagator, the vertex
interactions associated with its ends define  
either a "virtual" process or a "real" process.
The real process is the process that can occur in free space such as 
$\Delta \rightarrow \pi N$. The virtual processes,
such as the  $\pi N \rightarrow N$,
$\pi\Delta \rightarrow \Delta$, and $\pi \Delta \rightarrow N$
transitions, are not allowed by the energy-momentum conservation.
The consequences of the unitary transformation is the following.
When both vertex interactions  
are 'virtual', the propagator is the average of
the propagators calculated with two different momenta specified by
the initial and final external momenta. For example, the propagator of
$\bar{V}_a$ of Eq. (C9), which corresponds to 
$\pi (k) N (p) \rightarrow N \rightarrow
\pi (k') N(p')$, should be evaluated by
\begin{eqnarray}
S_N(p+k) &\rightarrow &\frac{1}{2}[S_N(p+k) + S_N(p'+k')] \nonumber \\
& & =\frac{1}{2}[ \frac{(E_N(p)+E_\pi(k))\gamma^0 
- \vec{\gamma}\cdot (\vec{p}+\vec{k})+ m_N}
{(E_N(p) +E_\pi(k))^2- (\vec{p}+\vec{k})^2 - m_N^2} \nonumber \\
& & +\frac{(E_N(p')+E_\pi(k'))\gamma^0 - \vec{\gamma}\cdot
 (\vec{p'}+\vec{k'}) + m_N}
{(E_N(p')+E_\pi(k'))^2- (\vec{p'}+\vec{k'})^2 - m_N^2} ] \,.
\end{eqnarray}
One see clearly that the denominators of the above expression
 are independent of
the collision energy $E$ of scattering equation Eq. (\ref{eq:pw-tmbmb}) and 
finite in all real momentum region. 
This is the essence of
the unitary transformation method in deriving the interactions from
Lagrangians.
When only one of the vertex interactions is 'real', the propagator
is evaluated by using the momenta associated with the 'virtual' vertex.
For example, the propagator of $V^{11}_d$  Eq.(C51), which corresponds to
$\pi(k)  N(p) \rightarrow \Delta \rightarrow \pi (k')  \Delta (p')$ is
$S^{\mu\nu}_\Delta (p'+k')$,  not $S^{\mu\nu}_\Delta (p+k)$ or 
$[S^{\mu\nu}_\Delta (p'+k')+S^{\mu\nu}_\Delta (p+k)]/2$. The structure
of its denominator is
similar to that of Eq.(C67) and hence the resulting
matrix elements are also independent of scattering energy $E$ and
finite in all momentum region.
The terms which have one 'real' and one 'virtual' vertex interactions are
$\bar{V}^4_c$, $\bar{V}^7_c$, $\bar{V}^{11}_b$.
Their corresponding intermediate momentum variables
have been correctly specified. The average, such as that of Eq.(C67), must
be used in all other terms of Eqs.(C8)-(C64).
We  note that there is no propagator in Eqs.(C8)-(C64)
which is attached by two 
real processes such as $\pi N \rightarrow \Delta \rightarrow \pi N$.
Such real processes are generated from $\Gamma_V$ of the Hamiltonian and
included in the resonant term $t^R_{MB,M'B'}$ of Eq.(\ref{eq:tmbmb}).

\section{Matrix elements of vertex interactions}

We need to have partial-wave matrix elements of vertex interactions
$\Gamma_{\Delta\rightarrow \pi N}$, $h_{\rho\rightarrow \pi\pi}$,
and $h_{\sigma \rightarrow \pi\pi}$ to evaluate the 
self-energy $\Sigma_{\alpha}$ with $\alpha=\pi\Delta, \rho N, \sigma N$
of Eqs.(\ref{eq:sig-pid})-(\ref{eq:sig-sn}), and the one-particle-exchange 
interactions $Z^{(E)}_{\pi\Delta,\pi\Delta}$, 
$Z^{(E)}_{\rho N, \pi\Delta}$, and $Z^{(E)}_{\sigma N, \pi\Delta}$,
illustrated in Fig.\ref{fig:z}

In consistent with the normalizations defined by Eqs.(\ref{eq:crst}), we write
the matrix element of the  
$\alpha (p_\alpha) \rightarrow \beta (p_\beta) + \gamma (p_\gamma) 
$ vertex interaction $f_{\alpha,\beta\gamma}$ 
as
\begin{eqnarray}
& & <\vec{p}_\alpha ; j_\alpha m_{j_\alpha} |f_{\alpha, \beta\gamma}
| \vec{p^\prime}_\alpha \vec{q^\prime}_{\alpha} ; j_\beta m_{j_\beta}j_\gamma
m_{j_\gamma} > 
= \delta(\vec{p}_\alpha - \vec{p^\prime}_{\alpha})
\nonumber \\
& &\times \sum_{all m's}<\bar{j}_\alpha m_{\bar{j}_\alpha} | l_\alpha s_\alpha m_{l_\alpha}
m_{s_\alpha} > <s_\alpha m_{s_\alpha} | j_\beta j_\gamma m_{j_\beta}
m_{j_\gamma} > f_{n_\alpha}(q'_\alpha)
Y_{l_\alpha m_{l_\alpha}}(\hat{q'_\alpha}) \,,
\end{eqnarray}
where $j_\alpha$ is the spin of the particle $\alpha$,  $l_\alpha$
is the relative orbital angular momentum of the pair $(\beta,\gamma)$,
$n_\alpha =[(l_\alpha (j_\beta j_\gamma)s_\alpha) ]\bar{j}_\alpha]$ denotes
collectively all quantum numbers specifying the interacting $(\beta,\gamma)$
pair, The momenta are related by relativistic kinematics
\begin{eqnarray}
\vec{p'}_\alpha &=&\vec{p'}_\beta +\vec{p'}_\gamma \,,\\ 
\vec{q'}_\alpha &=& \vec{p'}_\beta 
+ \rho_\alpha(p'_\alpha,p'_\beta,x) \vec{p'}_\alpha \,,
\\
\vec{q'}_\beta &=& -\vec{p'}_\alpha 
- \rho_\beta(p'_\alpha,p'_\beta,x) \vec{p'}_\beta \,,
\end{eqnarray}
where $x=\hat{p'_\alpha}\cdot \hat{p'_\beta}$, and
\begin{eqnarray}
\rho_\alpha(p_\alpha,p_\beta,x) &=&
\xi_{p_\alpha}^{-1/2} \left[ \varepsilon_{p_\beta}
+ \vec{p}_\alpha \cdot \vec{p}_\beta
(\varepsilon_{p_\beta}+\varepsilon_{\vec{p}_\alpha+\vec{p}_\beta}
+\xi_{p_\alpha}^{1/2})^{-1} \right] \,, \nonumber \\
\rho_\beta(p_\alpha,p_\beta,x) &=& \rho_\alpha(p_\beta,p_\alpha,x)\,,
 \nonumber \\
\xi_{p_\alpha} &=&
(\varepsilon_{p_\beta}+\varepsilon_{\vec{p}_\alpha+\vec{p}_\beta})^2-p_\alpha^2
\,.  \nonumber
\end{eqnarray}
The form factor $g_{n_\alpha}(q'_\alpha)$ for $\rho \rightarrow \pi\pi$
and $\sigma \rightarrow \pi\pi$ are  
taken from Ref.\cite{johnstone} and
$\Delta \rightarrow \pi N$ are from SL model. They are related to phase 
shifts by
\begin{eqnarray}
-[\sin\delta_\alpha (E)]e^{i\delta_\alpha}(E) = 
\frac{|f_{n_\alpha} (q_0)|^2}{E - M^0_\alpha - \Sigma_\alpha (E)}
\end{eqnarray}
with
\begin{eqnarray}
\Sigma_\alpha(E) =\int q^2 dq \frac{|f_{n_\alpha}(q)|^2}
{E-E_\beta(q) -E_\gamma(q) + i\epsilon}
\end{eqnarray}

Explicitly we have
\begin{eqnarray}
f_{\Delta,\pi N}(q)=-\frac{f_{\pi N\Delta}}{m_\pi}
\frac{1}{(2\pi)^{3/2}}\frac{1}{\sqrt{2E_\pi(q)}}
\sqrt{\frac{E_N(q)+m_N}{2E_N(q)}}\sqrt{\frac{4\pi}{3}}
q[\frac{q^2}{q^2+\Lambda_{\pi N\Delta}^2}]^2 
\end{eqnarray}
with $f_{\pi N\Delta}=2.049$, $\Lambda_{\pi N\Delta}=3.29$ (fm)$^{-1}$,
and $M^0_\Delta=1299.07$ MeV,
\begin{eqnarray}
f_{\rho, \pi\pi }(q) = g_{\rho,\pi\pi }\frac{qr}{(1+(qr)^2)^2}
\end{eqnarray}
with $g_{\rho,\pi\pi}= 0.6684$ $ m^{1/2}_\pi$, $r=0.428$ fm, and 
$M^0_{\rho}= 811.7$
MeV, 
\begin{eqnarray}
g_{\sigma,\pi\pi}(q) = g_{\sigma,\pi\pi}\frac{1}{1+(qr)^2}
\end{eqnarray}
with $g_{\sigma,\pi\pi}= 0.7550 m^{1/2}_\pi$, $r=0.522$ fm, and
$ M^0_{\sigma}= 896.8$ MeV.

\section{Matrix elements of $Z^{(E)}_{MB,M'B'}(E)$}

With the matrix elements of the vertex interactions defined by 
Eqs.(D1)-(D4) of Appendix D,
we can evaluate the partial-wave matrix elements of one-particle-exchange
interaction $Z^{(E)}_{MB,M'B'}$.
We use the cyclic notation $(\alpha,\beta,\gamma)$ to 
specify the particles involved in the 
vertex interaction $\alpha \rightarrow \beta +\gamma$.
By using the angular momentum quantum numbers defined in Appendix D,
we  then define  the basis state of a $MB$ system 
with a given total angular momentum $(J M)$ in the center of mass frame
as
\begin{eqnarray}
|N_\alpha; p_\alpha J M >
=|\{L_\alpha [(l_\alpha (j_\beta j_\gamma )s_\alpha )\bar{j}_\alpha 
j_\alpha ]S_\alpha \} J M; p_\alpha > \,,
\end{eqnarray}
where we have introduced a concise notation 
$N_\alpha = [\{L_\alpha [(l_\alpha (j_\beta j_\gamma )s_\alpha )
\bar{j}_\alpha j_\alpha ] S_\alpha\}]$.

Following the standard procedures of Ref.\cite{thomas-review}, one then
obtained
\begin{eqnarray}
Z^J_{N_\beta, N_\alpha} (p_\beta,p_\alpha)
&=&< N_\beta; p_\beta J M| G_{\pi\pi N}(E) |N_\alpha; p_\alpha J M >
\nonumber \\
 &=&\bar{\delta}_{\alpha,\beta}
[p^{l_\alpha}_\beta]\, [p^{l_\beta}_\alpha] \sum_{L}
\sum_{a=0}^{l_\alpha} \sum_{b=0}^{l_\beta}
F^{L,b,a}_{n_\beta,n_\alpha}(p_\beta,p_\alpha;E) 
A^{L,a,b}_{N_\beta,N_\alpha}
(p_\alpha/p_\beta)^{a-b} \,,
\end{eqnarray}
where $\bar{\delta}_{\alpha,\beta} = 1-{\delta}_{\alpha,\beta}$, and
\begin{eqnarray}
A^{L,a,b}_{N_\beta,N_\alpha}
 & = & (-1)^R \hat{l}_\alpha \hat{l}_\beta
              \hat{L}_\alpha \hat{L}_\beta
              \hat{S}_\alpha \hat{S}_\beta
              \hat{\bar{j}}_\alpha \hat{\bar{j}}_\beta
              \hat{s}_\alpha \hat{s}_\beta
              \hat{L}^2  \nonumber \\
 & \times &
  \{ \frac{(2l_\alpha + 1 )! (2l_\beta + 1 )!}
     {(2a)! (2b)! (2l_\alpha - 2a)! (2l_\beta - 2b)!}
  \}^{1/2} \nonumber \\
& \times &
 \sum_{f \Lambda \Lambda'} [\hat{f}\hat{\Lambda}\hat{\Lambda}' ]^2
\left \{ \begin{array}{ccc}
                 S_\alpha  & S_\beta &  f\\
                 L_\beta  & L_\alpha &  J
                \end{array}
                 \right\}
\left \{ \begin{array}{cccccccc}
j_\alpha &                & S_\alpha &   & S_\beta & & j_\beta & \\
         & \bar{j}_\alpha & & f & & \bar{j}_\beta & & j_\gamma \\
s_\alpha && l_\alpha && l_\beta && s_\beta &
                \end{array}
                 \right\} \nonumber \\
& \times &
\left \{ \begin{array}{ccc}
                 L_\alpha  & L_\beta &  f\\
                 \Lambda'  & \Lambda &  L
                \end{array}
                 \right\}
\left \{ \begin{array}{ccc}
l_\alpha &    l_\beta &   f \\
a        & l_\beta-b  & \Lambda \\
l_\alpha -a & b & \Lambda'
                \end{array}
                 \right\}  \nonumber \\
& \times & \left( \begin{array}{ccc}
\Lambda' &   L  &  L_\beta \\
0         &   0  & 0
                \end{array}
                 \right)
\left( \begin{array}{ccc}
\Lambda &   L  &  L_\alpha \\
0         &   0  & 0
                \end{array}
                 \right)
\left( \begin{array}{ccc}
l_\alpha-a  &   b  &  \Lambda' \\
0         &   0  & 0
                \end{array}
                 \right)
\left( \begin{array}{ccc}
a &  l_\beta - b  &  \Lambda \\
0         &   0  & 0
                \end{array}
                 \right)
\end{eqnarray}
with $\hat{a}=\sqrt{2a+1}$ and
\begin{eqnarray}
R = -J + L + L_\alpha +L_\beta + S_\alpha + S_\beta + \bar{j}_\alpha
 + \bar{j}_\beta + s_\beta
 + l_\beta - j_\alpha \,.
\end{eqnarray}
In the above equations, the usual 3-j, 6-j, 9-j, and 12-j symboles
have been used to define the angular momentum coupling.
The details can be found in Ref.\cite{thomas-review}.

The three-body cut effects are in $F^L_{n_\beta,n_\alpha}(p_\beta,p_\alpha;E)$
of Eq.(E2). They are calculated from the vertex functions 
$f_{n_\alpha}(q_\alpha)$ by
\begin{eqnarray}
F^{L,b,a}_{n_\beta,n_\alpha}(p_\beta,p_\alpha;E) = \frac{1}{2}
\int_{-1}^{+1} d x \frac{q^{-l_\beta}_\beta 
\rho_\beta^b (x) f^\dagger_{n_\beta}(q_\beta)
q^{-l_\alpha}_\alpha 
\rho_\alpha^a (x) f_{n_\alpha}(q_\alpha)P_L(x)}
{E - E_{\alpha '}(-\vec{p}_\alpha )- E_{\beta'} (-\vec{p}_\beta) - 
E_{\gamma}(\vec{p}_\alpha + \vec{p}_\beta) + i\epsilon} \,,
\end{eqnarray}
where $x=\hat{p_\alpha}\cdot \hat{p_\beta}$ and
the
vertex function  $f_{n_\alpha}$ defines
$\alpha \rightarrow \beta' + \gamma$ and
$f_{n_\beta}$ defines
$\gamma + \alpha' \rightarrow \beta$. Namely, $\alpha'$ and $\beta'$
are the spectators of the decay of particle $\alpha$ and $\beta$ respectively.
We obviously  have  $\alpha=\Delta$,
$\alpha'=\pi$, $\gamma=N$, $\beta = \Delta $ and $\beta'= \pi$ for
$Z^{(E)}_{\pi\Delta,\pi\Delta}(E)$, and
$\alpha=\rho,\sigma $,
$\alpha'= N$, $\gamma=\pi$, $\beta = \Delta $ and $\beta'= \pi$ for
$Z^{(E)}_{\rho N,\pi\Delta}(E)$.

In the actual calculations, 
the integration path $-1\le x \le 1$ of eq.(E5) is deformed into the complex
$x-$plane in order to avoid the singularity $x_0 \ (-1\le x_0 \le 1)$
where the denominator vanishes.
We have used a simple parabolic form, i.e., $x=t+i(t^2-1)$.

\setcounter{equation}{0}

\section{Matrix elements of $\gamma N \rightarrow MB$ transitions}

To include the final meson-baryon interactions in the
photo-production, it is only necessary to
perform the partial-wave decomposition of the final $MB$ state.
We thus introduce the following helicity-LSJ mixed-representation
\begin{eqnarray}
 {\it v}^{JT}_{L' S' M'B', \lambda_\gamma \lambda_N}(k',q)
 & =& \sum_{ \lambda'_M\lambda'_B } [
\sqrt{\frac{(2L'+1)}{2J+1}}
 <j'_M j'_B \lambda'_M  (-\lambda'_B) \mid S' S'_z> \nonumber \\
& &\times <L'S'0S'_z\mid J S'_z> \nonumber \\
& & \times <J,k' \lambda'_M (-\lambda'_B) \mid v_{M'B',\gamma N}\mid
J, q\lambda_\gamma (-\lambda_N) >]\,,
\label{eq:e-1}
\end{eqnarray}
where $<J,k' \lambda'_M (-\lambda'_B) \mid v_{M'B',\gamma N}\mid
 J,q\lambda_\gamma (-\lambda_N) >$ can be evaluated using the same
expression of Eq.(C2) using the heliclity matrix elements
of $v_{M'B',\gamma N}$. To evaluate these quantities with
 the normalization defined by Eq.(\ref{eq:crst}), we define
for a photon four momentum $q=(\omega,\vec{q})$
\begin{eqnarray}
 <(k'j),p'|v_{MB,\gamma N}|q,p> 
 &=&\frac{1}{\sqrt{2q_0}} 
 <(k'j),p'|\sum_{n}J^\mu(n) \epsilon_\mu|q,p> \nonumber \\
& =& \frac{1}{(2\pi)^3}\sum_{n}
\sqrt{\frac{m_B}{E'_B(k')}}\sqrt{\frac{1}{2 E'_M(k')}}
\bar{u}_B(\vec{p'}) I(n) u_N(\vec{p}) \sqrt{\frac{m_N}{E_N(q)}}
\frac{1}{\sqrt{2q_0}} \nonumber \\
\end{eqnarray}
where $\epsilon_\mu $ is the photon polarization vector, 
and $n$  denotes a given considered process
\begin{eqnarray}
I(n) = \epsilon \cdot \bar{j}(n)\,.
\end{eqnarray}
Here $\bar{j}(n)$ can be constructed by using the
Feynman rules. The resulting expressions for each of
$\gamma N \rightarrow \pi N, \eta N, \sigma N, \rho N, \pi\Delta$ are listed
below :

\subsection{$\gamma(q)+N(p) \rightarrow \pi(k',j)+ N(p')$ }
\begin{eqnarray}
I(1)=I^1_a+I^1_b+I^1_c+I^1_d+I^1_e+I^1_f+I^1_g + I^1_h
\end{eqnarray}
with 
\begin{eqnarray}
I^1_a &=&i\frac{f_{\pi NN}}{m_\pi}\slas{k'}\gamma_5\tau^j\frac{1}{\slas{p'}+\slas{k'}
-m_N }\Gamma_N \\
& & \nonumber \\
& & \mbox{where}\,\,\,\,
 \Gamma_N =\hat{e}_N\slas{\epsilon_\gamma}-\frac{\kappa_N}{4m_N}
[\slas{\epsilon_\gamma}\slas{q}-\slas{q}\slas{\epsilon_\gamma}] \\
& & \nonumber \\
I^1_b &=&i\frac{f_{\pi NN}}{m_\pi}\Gamma_N \frac{1}{\slas{p}-\slas{k'}
-m_N }\slas{k'}\gamma_5\tau^j \\
& & \nonumber \\
& & \nonumber \\
I^1_{c}&=& i \frac{f_{\pi N \Delta}}{m_\pi}{\Gamma^{em,\Delta}_\nu}^\dagger
S^{\nu\mu}_\Delta(p-k') k'_\mu T^j \\
& & \nonumber \\
I^1_d &=& \frac{f_{\pi NN}}{m_\pi}\epsilon_{ij3}\tau^i
\slas{\epsilon_\gamma}\gamma_5
 \\
& & \nonumber \\
I^1_e&=&-\frac{f_{\pi NN}}{m_\pi}\frac{\slas{\tilde{k}}
\gamma_5}{\tilde{k}^2
-m^2_\pi}\epsilon_{ij3}\tau^i(\tilde{k}+k')\cdot\epsilon_\gamma \\
& & \nonumber \\
& & \mbox{where} \,\,\,\, \tilde{k}=p-p' \nonumber \\
& & \nonumber \\
I^1_f&=&-\frac{g_{\rho NN}g_{\rho\pi\gamma}}{m_\pi}
\frac{\tau^j}{2}
 [\gamma^\delta+\frac{\kappa_\rho}{4m_N}(\gamma^\delta\slas{\tilde{k}}-
\slas{\tilde{k}} \gamma^\delta)] 
\nonumber \\
& &\times  \epsilon_{\alpha\beta\eta\delta}
{\tilde{k}}^\eta q^\alpha \epsilon^\beta_\gamma 
\frac{1}{\tilde{k}^2-m^2_\rho} \\
& & \nonumber \\
I^1_g&=&-\frac{g_{\omega NN}g_{\omega\pi\gamma}}{m_\pi}
 [\gamma^\delta+\frac{\kappa_\omega}{4m_N}(\gamma^\delta\slas{\tilde{k}}-
\slas{\tilde{k}} \gamma^\delta)] \nonumber \\
& &\times  \epsilon_{\alpha\beta\eta\delta}
{\tilde{k}}^\eta q^\alpha \epsilon^\beta_\gamma \delta_{j3}
\frac{1}{\tilde{k}^2-m^2_\omega} %\\
\end{eqnarray}
where $\Gamma^{em,\Delta}_\mu = \Gamma^{em,\Delta}_{\mu\nu}\epsilon^\nu_\gamma$.

\subsection{$\gamma(q)+ N(p) \rightarrow \eta(k')+ N(p')$ }
\begin{eqnarray}
I(2)=I^2_a+I^2_b+I^2_c
\end{eqnarray}
with
\begin{eqnarray}
I^2_a &=&i\frac{f_{\eta NN}}{m_\eta}\slas{k'}\gamma_5\frac{1}{\slas{p'}+\slas{k'}
-m_N }\Gamma_N \\
& & \nonumber \\
I^2_b &=&i\frac{f_{\eta NN}}{m_\eta}\Gamma_N \frac{1}{\slas{p}-\slas{k'}
-m_N }\slas{k'}\gamma_5 \\
& & \nonumber \\
I^2_c&=&-\frac{g_{\rho NN}g_{\rho\eta\gamma}}{m_\rho}\frac{\tau^3}{2}
 [\gamma^\nu+\frac{\kappa_\rho}{4m_N}(\gamma^\nu\slas{\tilde{k}}-
\slas{\tilde{k}} \gamma^\nu)] \nonumber \\
& &\times  \epsilon_{\mu\nu\alpha\beta}
{\tilde{k}}^\mu q^\alpha \epsilon^\beta_\gamma
\frac{1}{\tilde{k}^2-m^2_\rho} %\\
\end{eqnarray}

\subsection{$\gamma(q)+ N(p) \rightarrow \sigma(k')+ N(p')$ }
\begin{eqnarray}
I(3)=I^3_a+I^3_b
%I(3)=I^3_a+I^3_b+I^3_c
\end{eqnarray}
with
\begin{eqnarray}
I^3_a &=& -g_{\sigma NN}\frac{1}{\slas{p'}+\slas{k'}
-m_N }\Gamma_N \\
& & \nonumber \\
I^3_b &=&-g_{\sigma NN}\Gamma_N \frac{1}{\slas{p}-\slas{k'}
-m_N } 
%& & \nonumber \\
%I^3_c&=&\frac{g_{\rho NN}g_{\rho\sigma\gamma}}{m_\rho}\frac{\tau^0}{2}
% [\gamma_\mu+\frac{\kappa_\rho}{4m_N}(\gamma_\mu\slas{\tilde{k}}-
%\slas{\tilde{k}} \gamma_\mu)] \nonumber \\
%& &\times
%\frac{1}{\tilde{k}^2-m^2_\rho}[(q\cdot\tilde{k})\epsilon^\mu_\gamma
%-(q\cdot \epsilon_\gamma)k^\mu] %\\
%& & \nonumber \\
\end{eqnarray}

\subsection{$\gamma(q)+N(p) \rightarrow \rho(k',j,\lambda)+ N(p')$ }
\begin{eqnarray}
I(4)=I^4_a+I^4_b+I^4_c+I^4_d+I^4_e+I^4_f+I^4_g
\end{eqnarray}
with
\begin{eqnarray}
I^4_a &=& -g_{\rho NN}\Gamma_{\rho^\prime}\frac{1}{\slas{p'}+\slas{k'}
         -m_N }\Gamma_N \\
%& & \nonumber \\
%& & \mbox{where}\,\,\,\, 
%\Gamma_\rho =\slas{\epsilon_\rho}^* +\frac{\kappa_\rho}{4m_N}
%[\slas{\epsilon_\rho}^* \slas{k'}-\slas{k'} \slas{\epsilon_\rho}^*]
% \\
& & \nonumber \\
I^4_b &=&-g_{\rho NN}\Gamma_N \frac{1}{\slas{p}-\slas{k'}-m_N }
                     \Gamma_{\rho^\prime} \\
& & \nonumber \\
I^4_{c}&=& \frac{f_{\rho N\Delta}}{m_\rho}
          (k'_\mu \slas{\epsilon_{\rho^\prime}}^*
        -  \slas{k'}\epsilon_{\rho^\prime \mu}^*)\gamma_5
           {T^\dagger}^j S^{\mu\nu}_\Delta(p'+k')  
           \Gamma^{em,\Delta}_\nu  \\
& & \nonumber \\
I^4_{d}&=& -\frac{f_{\rho N\Delta}}{m_\rho}[\Gamma^{em,\Delta}_\mu]^\dagger
             S^{\mu\nu}_\Delta(p-k')  T^j
           (k'_\nu \slas{\epsilon_\rho}^*-
           \slas{k'} \epsilon_{\rho^\prime \nu}^*)\gamma_5  \\
& & \nonumber \\
I^4_e &=& i \frac{g_{\rho NN}\kappa_\rho}{8m_N}\epsilon_{ij3}\tau_i
            [\slas{\epsilon_{\rho^\prime}}^*\slas{\epsilon_\gamma}-
            \slas{\epsilon_\gamma}\slas{\epsilon_{\rho^\prime}}^*] \\
& & \nonumber \\
I^4_f&=& i\frac{g_{\rho NN}}{2}
        [\gamma_\mu+\frac{\kappa_\rho}{2m_N}(\gamma_\mu\slas{\tilde{k}}-
       \slas{\tilde{k}} \gamma_\mu)] \nonumber \\
     & &\times [\epsilon^{\mu *}_{\rho^\prime}(\tilde{k}+k')\cdot \epsilon_\gamma 
      -(\tilde{k}\cdot \epsilon_{\rho^\prime}^*)\epsilon^\mu_\gamma -
       (\epsilon_\gamma \cdot\epsilon_{\rho^\prime}^*)k^{'\mu}]
       \frac{\epsilon_{ij3}\tau^i}{\tilde{k}^2-m^2_\rho} \\
& & \nonumber \\
I^4_g&=&=-i\frac{f_{\pi NN}}{m_\pi}\frac{g_{\rho\pi\gamma}}{m_\pi}
     \tau^j\slas{\tilde{k}}\gamma_5\epsilon_{\alpha\beta\eta\delta}k^{'\eta}
     \epsilon_{\rho^\prime}^{\delta *} q^\alpha\epsilon^\beta_\gamma
     \frac{1}{\tilde{k}^2-m_\pi^2} \\
%& & \nonumber \\
%I^4_g&=&=g_{\sigma NN}\frac{g_{\rho\sigma\gamma}}{m_\pi}
%\delta_{j3}[(q\cdot k')(\epsilon_\rho^*\cdot \epsilon_\gamma)
%-(q\cdot \epsilon_\rho^*)(k'\cdot\epsilon_\gamma)]
%\frac{1}{\tilde{k}^2-m_\sigma^2} %\\
\end{eqnarray}

\subsection{$\gamma(q)+N(p) \rightarrow \pi(k',j)+ \Delta(p')$ }
\begin{eqnarray}
I(5)=I^5_a+I^5_b+I^5_c+I^5_d+I^5_e+I^5_f+I^5_g
\end{eqnarray}
with
\begin{eqnarray}
I^5_a&=& i\frac{f_{\pi N\Delta}}{m_\pi}\epsilon_\Delta^*\cdot k'T^j
S_N(p'+k')\Gamma_N \\
& & \nonumber \\
I^5_b&=& i\frac{f_{\pi N\Delta}}{m_\pi}\Gamma^{em,\Delta}_\nu 
\epsilon^{\nu *}_\Delta S_N(p-k')\slas{k'}\gamma_5\tau^j \\
& & \nonumber \\
I^5_c & = & -i\frac{f_{\pi \Delta\Delta}}{m_\pi}\epsilon^{*}_{\Delta \mu}
           \slas{k'}\gamma_5 T_\Delta^j
           S^{\mu\nu}_\Delta(p'+k')  \Gamma^{em,\Delta}_\nu \\
& & \nonumber \\
I^5_d&=&i\frac{f_{\pi N\Delta}}{m_\pi}\epsilon^{\eta *}_\Delta
(\frac{1}{2} + T_\Delta^3) 
[ - g_\eta^\mu \slas{\epsilon}_\gamma+ (\epsilon_\gamma)_\eta \gamma^\mu]
S^\Delta_{\mu\nu}(p-k')k^{'\nu} T^j \\
& & \nonumber \\
I^5_e&=&\frac{f_{\pi N\Delta}}{m_\pi}\epsilon_{ij3}T^i 
        \epsilon_\gamma\cdot \epsilon_\Delta^*\\
& & \nonumber \\
I^5_g&=&-\frac{f_{\pi N\Delta}}{m_\pi}\epsilon_{ij3}T^i [V^5_g + Z^5_g]
%I^5_g&=&-\frac{f_{\pi N\Delta}}{m_\pi}\epsilon_\Delta^*\cdot \tilde{k} 
%\epsilon_{ij3}T^i (\tilde{k}+k')\cdot\epsilon_\gamma\frac{1}
%{\tilde{k}^2-m^2_\pi} \label{gn-pd}
\\
& & \nonumber \\
I^5_g&=&-\frac{f_{\rho N\Delta}}{m_\rho}
\frac{g_{\rho\pi\gamma}}{m_\pi}T^j
\frac{1}{\tilde{k}^2-m^2_\rho}
[\tilde{k}\cdot\epsilon^*_\Delta\gamma^\mu
- \tilde{k}^\mu \slas{\epsilon_\Delta}^*]\gamma_5\epsilon_{\alpha\beta\eta\mu}
q^\alpha\epsilon^\beta_\gamma\tilde{k}^\eta
\end{eqnarray}
where the pion pole term $I^5_g$ consists of energy independent
interaction $V^5_g$ and energy dependent interaction
$Z^5_g$ given as
\begin{eqnarray}
V^5_g & = & \frac{1}{2E_\pi(k-k')}\frac{\epsilon_\Delta^*\cdot k_1 (k_1
 + k')\cdot\epsilon_\gamma}{E_N(q) - E_\Delta(k') - E_\pi(k-k')}
 +\epsilon_\Delta^{0 *}\epsilon_\gamma^0 \\
Z^5_g & = &  \frac{1}{2E_\pi(k-k')}\frac{\epsilon_\Delta^*\cdot k_2 (k_2
 + k')\cdot\epsilon_\gamma}{E - E_N(q) - E_\pi(k') - E_\pi(k-k')+i\epsilon}
\end{eqnarray}
with $k_1=(E_\pi(k-k'),\vec{k}'-\vec{k})$ and $k_2=(-E_\pi(k-k'),\vec{k}'-\vec{k})$.
The on-shell matrix element of $V^5_g+Z^5_g$  is given as
\begin{eqnarray}
V^5+Z^5 & = & \epsilon_\Delta^*\cdot \tilde{k}
 (\tilde{k}+k')\cdot\epsilon_\gamma\frac{1}
 {\tilde{k}^2-m^2_\pi}.
\end{eqnarray}

\section{Multipole amplitudes of $\gamma N \rightarrow \pi N$ }

For 
$\gamma N \rightarrow MB$ matrix elements, we use 
the helicity-LSJ mixed-representation defined by
Eq.(\ref{eq:mixed-rep}). It can be calculated by using 
Eq.(F1). For pseudo-scalar meson $\pi$ and $\eta$ 
production, it is
often to write the amplitudes in terms of
multipole amplitudes.
Here we want to relate our matrix element Eq.(F1) and
hence also the amplitude Eq.(\ref{eq:mixed-rep}) to this commonly
used multipole amplitude.

With the definition Eq.(\ref{eq:smatrix}) for the scattering amplitude $T$, 
we first define the  amplitude $F$  by the
on-shell T-matrix element of $\gamma N \rightarrow MB$ as
\begin{eqnarray}
<MB|F|\gamma N> & 
= & - \frac{4\pi^2}{W}\sqrt{E_N(k)E_M(k)|q_0| E_N(q)}
    \frac{1}{\sqrt{2|q_0|}}
<MB|T|\gamma N>  \label{eqft}
\end{eqnarray}
with
\begin{eqnarray}
T & = & J^\mu \epsilon_\mu= J^0 \epsilon^0 - \vec{J}\cdot\vec{\epsilon} \,,
 \label{je}
\end{eqnarray}
where $J^\mu \epsilon_\mu =\sum_{n}j^\mu(n)\epsilon_\mu$ 
is identical to that in Eq.(F2).
The most general Chew-Goldberger-Low-Nambu (CGLN) amplitudes\cite{cgln}
 $F$ 
can be written as (isospin index is suppressed)
\begin{eqnarray}
F & = &  - i\bm{\sigma}\cdot \bm{\epsilon}_\perp F_1
 -  \bm{\sigma}\cdot \hat{k} \bm{\sigma}\cdot\hat{q}\times\bm{\epsilon}_\perp
                                                   F_2
 - i\bm{\sigma}\cdot\hat{q}\hat{k}\cdot\bm{\epsilon}_\perp F_3
 - i\bm{\sigma}\cdot\hat{k}\hat{k}\cdot\bm{\epsilon}_\perp F_4
 \nonumber \\
 & &
 - i\bm{\sigma}\cdot\hat{q}\hat{q}\cdot\bm{\epsilon} F_5
 - i\bm{\sigma}\cdot\hat{k}\hat{q}\cdot\bm{\epsilon} F_6
 + i \bm{\sigma}\cdot\hat{k} \epsilon_0 F_7 
 + i \bm{\sigma}\cdot\hat{q}\epsilon_0 F_8 \,.
  \label{fvec}
\end{eqnarray}
Each coefficient in the above equation can be written in terms
of multipole amplitudes $E_{l\pm}$, $M_{l\pm}$, $L_{l\pm}$, $S_{l\pm}$
\begin{eqnarray}
F_1 & = & \sum_l[
  P_{l+1}' E_{l+}  + P_{l-1}'     E_{l-} 
+ lP_{l+1}'M_{l+} + (l+1)P_{l-1}' M_{l-}] \,,\\
F_2 & = & \sum_l[
 (l+1)P_l'M_{l+} + lP_l' M_{l-}]\,,\\
F_3 & = & \sum_l[
  P_{l+1}'' E_{l+}  + P_{l-1}''   E_{l-} 
- P_{l+1}'' M_{l+}  + P_{l-1}''   M_{l-}]\,, \\
F_4 & = & \sum_l[
- P_{l}'' E_{l+}  - P_{l}''   E_{l-} 
+ P_{l}'' M_{l+}  - P_{l}''   M_{l-}]\,, \\
F_5 & = & \sum_l[
  (l+1) P_{l+1}' L_{l+}  -  l  P_{l-1}' L_{l-}]\,, \\
F_6 & = & \sum_l[
 -(l+1) P_{l}' L_{l+}  +  l  P_{l}' L_{l-}] \,, \\
F_7 & = & \sum_l[
 -(l+1) P_{l}' S_{l+}  +  l  P_{l}' S_{l-}] \,, \\
F_8^V & = & \sum_l[
  (l+1) P_{l+1}' S_{l+}  -  l  P_{l-1}' S_{l-}] \,,
\end{eqnarray}
where $P_L(x)$ is Legendre function and $x=\hat{k}\cdot\hat{q}$.
It is of course well known that only four of the above amplitudes
are independent for photo-production and six for electro-production.

Choosing the photon direction
as $\hat{q}=\hat{z}$, Eqs.(F2) and Eqs.(G1)-(G2) clearly lead to
\begin{eqnarray}
<J|F|\lambda_\gamma s> & = &
- \frac{4\pi^2}{W}\sqrt{E_N(k)\omega_M(k)|q_0| E_N(q)}
    \frac{1}{\sqrt{2|q_0|}}
v^{JT}_{LS\pi N,\lambda_\gamma s}(k,q) \sqrt{2q_0}\,.
\end{eqnarray}
Here $s$ is the z-component of the initial nucleon spin and
we have dropped the notation $MB=\pi N$ and 
$(LS)=({\it l }=J\pm 1/2, 1/2)$ and isospin $T$
in defining the matrix element of $F$.

With the form Eq.(G3), it is easy to calculate the matrix element 
$<J|F|\lambda_\gamma s>$
in our helicity-LSJ mixed-representation. 
After some derivations, we obtain the following relations
\begin{eqnarray}
E_{l+} & =  \frac{1}{4\pi i (l+1)}[&
                           <J=l+1/2|F|\lambda=1,s=-1/2>
                                                         \nonumber \\ &&
 - \sqrt{\frac{l}{l+2}}    <J=l+1/2|F|\lambda=1,s=1/2>] \\
E_{l-} & =  \frac{1}{4\pi i l}[&
                         - <J=l-1/2|F|\lambda=1,s=-1/2>
                                                         \nonumber \\ &&
 - \sqrt{\frac{l+1}{l-1}}  <J=l-1/2|F|\lambda=1,s=1/2>] \\
M_{l+} & =  \frac{1}{4\pi i (l+1)}[& 
                           <J=l+1/2|F|\lambda=1,s=-1/2>
                                                          \nonumber \\ &&
 + \sqrt{\frac{l+2}{l}}    <J=l+1/2|F|\lambda=1,s=1/2>] \\
M_{l-} & =  \frac{1}{4\pi i l}[&
                           <J=l-1/2|F|\lambda=1,s=-1/2>
                                                           \nonumber \\ &&
 - \sqrt{\frac{l-1}{l+1}}  <J=l-1/2|F|\lambda=1,s=1/2>] \\
L_{l+} & =  - \frac{\sqrt{2}}{4\pi i (l+1)}&  <J=l+1/2|F|\lambda=0,s=1/2>\\
L_{l-} & =    \frac{\sqrt{2}}{4\pi i l} &     <J=l-1/2|F|\lambda=0,s=1/2>\\
S_{l+} & =   \frac{\sqrt{2}}{4\pi i (l+1)}&  <J=l+1/2|F|\lambda=t,s=1/2>\\
S_{l-} & =  -\frac{\sqrt{2}}{4\pi i l}&     <J=l-1/2|F|\lambda=t,s=1/2>
\end{eqnarray}
Here we used polarization vector $\epsilon^\mu(\lambda=t)= (1,\vec{0})$.
Substituting Eq.(G12) into Eqs.(G13)-(20), we can relate the usual multipole
amplitudes of $\gamma N \rightarrow \pi N$
to the matrix element Eq.(F1) in the helicity-LSJ mixed-representation.
 
\section{$\gamma N \rightarrow \pi\pi N$ amplitudes}

We consider the non-resonant 
$\gamma (q) N (p) \rightarrow \pi(k^i) \pi(k^j) N (p') $ illustrated
in Fig.\ref{fig:mbpipin}.
With the normalization defined by Eq.(\ref{eq:crst-gnpipin}), we define the 
matrix element of this amplitude with
a photon momentum $q^\mu=(\omega,\vec{q})$ as 
\begin{eqnarray}
<k^j(j),k^i(i),p^\prime \mid v_{\pi\pi N,\gamma N} \mid  p>
=\frac{1}{\sqrt{2\omega}}<k^j(j),k^i(i),p^\prime \mid
\hat{J} \mid p > \cdot \epsilon_\gamma(q) \,,
\end{eqnarray}
where $i$ and $j$ denote the isospin components of the produced $\pi\pi$, 
$\epsilon_\gamma$ is the photon polarization vector, and
\begin{eqnarray}
<k^j(j),k^i(i),p^\prime \mid \hat{J}^\mu \mid  p> 
&=& \frac{1}{(2\pi)^{9/2}}
\sqrt{\frac{m_N}{E_{N}(p^\prime)}}
\frac{1}{\sqrt{4E_{\pi}(k^i)E_{\pi}(k^j)}}
\sqrt{\frac{m_N}{E_{N}(p)}}
 \nonumber \\
& & \times \bar{u}_N({\bf k}') j^\mu u_N({\bf k})
\end{eqnarray}
with
\begin{eqnarray}
j^\mu & = & j^\mu(1) +j^\mu(2) +j^\mu(3) +j^\mu(4) +j^\mu(5)+j^\mu(6)
\end{eqnarray}
Each term of Eq.(H3) are from mechanisms 
illustrated in Fig.\ref{fig:mbpipin}. Within our formulation,
the non-resonant mechanisms are only from  diagrams 
with intermediate nucleon states. 
The exchange mesons can be $\pi, \rho$ and
$\omega$. We then have the following expressions
\begin{eqnarray}
j^\mu(1) & = & i [\frac{f_{\pi NN}}{m_\pi}]^2[
 \slas{k}^i\gamma_5 \tau^i S_N(p'+k^i)\gamma^\mu\gamma_5 \epsilon_{kj3}\tau^k
+
 \gamma^\mu\gamma_5 \epsilon_{kj3}\tau^k S_N(p - k^i)\slas{k}^i\gamma_5 \tau^i ]
 \,,  \\
j^\mu(2) & = & - [\frac{f_{\pi NN}}{m_\pi}]^2[
 \slas{k}^i\gamma_5 \tau^i S_N(p'+k^i) \slas{k}^j\gamma_5 \tau^j
 S_N(p'+k^i+k^j)J_N^\mu
 \nonumber \\
 &  &  
+  \slas{k}^i\gamma_5 \tau^i S_N(p'+k^i) J_N^\mu 
 S_N(p - k^j) \slas{k}^j\gamma_5 \tau^j
 \nonumber \\
 &  &  
+   J_N^\mu  S_N(p - k^i - k^j) \slas{k}^i\gamma_5 \tau^i
 S_N(p - k^j) \slas{k}^j\gamma_5 \tau^j] \,, \\
j^\mu(3) & = & -i [\frac{f_{\pi NN}}{m_\pi}]^2[
 \slas{k}^i\gamma_5 \tau^i S_N(p'+k^i)
(\slas{p}- \slas{p}'-\slas{k}^i)\gamma_5 \epsilon_{kj3}\tau^k
\frac{(p-p'-k^i+k^j)^\mu}{(p-p'-k^i)^2-m_\pi^2}
\nonumber \\
& & 
+(\slas{p}- \slas{p}'-\slas{k}^i)
\gamma_5 \epsilon_{kj3}\tau^k S_N(p - k^i)\slas{k}^i\gamma_5 \tau^i 
\frac{(p-p'-k^i+k^j)^\mu}{(p-p'-k^i)^2-m_\pi^2}
] \,, \\
j^\mu(4) & = & -i [\frac{f_{\pi NN}g_{\rho NN}g_{\rho\pi\gamma}}{m_\pi^2}][
 \slas{k}^i\gamma_5 \tau^i S_N(p'+k^i)
\frac{\tau^j}{2}
 [\gamma^\delta+\frac{\kappa_\rho}{4m_N}(\gamma^\delta\slas{\tilde{k}}-
\slas{\tilde{k}} \gamma^\delta)]
 \nonumber \\
& & +
\frac{\tau^j}{2}
 [\gamma^\delta+\frac{\kappa_\rho}{4m_N}(\gamma^\delta\slas{\tilde{k}}-
\slas{\tilde{k}} \gamma^\delta)
 S_N(p - k^i)\slas{k}^i\gamma_5 \tau^i ]]\frac{1}{\tilde{k}^2-m^2_\rho}
\,, \\
j^\mu(5) & = & -i [\frac{f_{\pi NN}g_{\omega NN}g_{\omega\pi\gamma}}{m_\pi^2}][
 \slas{k}^i\gamma_5 \tau^i S_N(p'+k^i)
 [\gamma^\delta+\frac{\kappa_\omega}{4m_N}(\gamma^\delta\slas{\tilde{k}}-
\slas{\tilde{k}} \gamma^\delta)]
 \nonumber \\
& & +
 [\gamma^\delta+\frac{\kappa_\omega}{4m_N}(\gamma^\delta\slas{\tilde{k}}-
\slas{\tilde{k}} \gamma^\delta)
 S_N(p - k^i)\slas{k}^i\gamma_5 \tau^i ]] \nonumber \\
& &
\times  \epsilon_{\alpha\beta\eta\delta}
{\tilde{k}}^\eta q^\alpha \epsilon^\beta_\gamma \delta_{j3}
\frac{1}{\tilde{k}^2-m^2_\omega} \,, \\
j^\mu(6) & = &
  -g_{\rho NN}g_{\rho\pi\pi}
\frac{\tau^i \delta_{j,3}+ \tau^j \delta_{i,3} - 2 \tau_3 \delta_{i,j}}{2}
 [\gamma^\mu+\frac{\kappa_\rho}{4m_N}(\gamma^\mu\slas{\tilde{k}}-
\slas{\tilde{k}} \gamma^\mu)] \frac{1}{(p-p')^2-m^2_\rho} \,.
\end{eqnarray}
In the above equations, we have defined
\begin{eqnarray}J_N^\mu & = & \frac{1 + \tau^3}{2}\gamma^\mu
  + i \frac{\kappa_s + \tau^3\kappa_V}{2m_N}\sigma^{\mu\nu}q_\nu \,,
\end{eqnarray}
and
\begin{eqnarray}
\tilde{k} = p - p^\prime - k^i \,.
\end{eqnarray}

\section{Resonant amplitudes}
In this appendix, we  give formula for using the
$N^*$ parameters listed\cite{pdg} by Particle Data Group to calculate
the resonant amplitudes defined by Eq.(\ref{eq:pdgres}).

In the rest frame of $N^*$, the amplitudes of
strong decays of a $N^*$ with a mass $M^{JT}$ and spin-iospin  $(JT)$ 
can be written as the following partial-wave form
\begin{eqnarray}
& & <\phi_{J m_J, T m_T}\mid \Gamma_{N^*\rightarrow MB} 
\mid \vec{k},  m_{j_M} m_{t_M}, m_{j_B} m_{t_B} > \nonumber \\
 & & = \sum_{LS} \sum_{all\, m_z} [
<T m_T |  t_M t_B m_{t_M} m_{t_B}> 
 <J m_J | L S m_L m_S> <S M_S |j_M j_B m_{j_M} m_{j_B}>  
\nonumber \\
& &\times  \frac{1}{(2\pi)^{3/2}}\frac{1}{\sqrt{2E_M(k)}}
\sqrt{\frac{m_B}{E_B(k)}}
\sqrt{\frac{8\pi^2M^{JT}}{m_Bk_R}}\,[G^{JT}_{LS}]\, f^{JT}_{LS}(k,k_R)
(\frac{k}{k_R})^L Y_{Lm_L}(\hat{k})] \,,
\label{eq:g1}
\end{eqnarray}
where $k_R$ is defined by $M^{JT}=E_B(k_R)+E_M(k_R)$ and the 
form factor is chosen such that $f^{JT}_{LS}(k_R,k_R)=1$.
With the normalizations
\begin{eqnarray}
<\phi_{J m_J, T m_T}|\phi_{J m_J, T m_T}> =1 \,, \nonumber  \\
<\vec{k}| \vec{k^\prime}> =\delta(\vec{k}- \vec{k^\prime}) \,,
\label{eq:g2}
\end{eqnarray}
the partial decay widths can be written as 
\begin{eqnarray}
d\Gamma_{MB}(N^*_{JT}) &=&(2\pi) \delta(M^{JT} - E_B(k) - E_M(k))
\frac{1}{2J+1}d\vec{k} \nonumber \\
& &[\sum_{m_J}\sum_{m_{j_M},m_{j_B}}
\mid <\phi_{J m_J, T m_T}\mid \Gamma_{N^*\rightarrow MB}
\mid \vec{k},  m_{j_M} m_{t_M}, m_{j_B} m_{t_B} >|^2] \,.
\label{eq:g3}
\end{eqnarray}
From Eqs.(\ref{eq:g1}) and (\ref{eq:g3}), we then have
\begin{eqnarray}
\Gamma_{MB}(N^*_{JT}) = \sum_{LS}|G^{JT}_{LS}|^2 \,.
\label{eq:g4}
\end{eqnarray}
Eq.(\ref{eq:g4}) allows us to determine
 the coupling constant $G^{JT}_{LS}$ up to its phase in terms
of the empirical partial decay widths as listed by 
Particle Data Group\cite{pdg}. Here we use the phase from the $^3P_0$ model
of Capstick and Roberts\cite{capstick-roberts}.

For $N^*\rightarrow \gamma N$ amplitudes, we use the commonly used
helicity representation to define
\begin{eqnarray}
d\Gamma_{\gamma N}(N^*_{JT}) &=&(2\pi) \delta(M^{JT} - E_N(q) - q)
\frac{1}{2J+1} d\vec{q}\nonumber \\
& &[\sum_{m_J}\sum_{\lambda_\gamma\lambda_N}
\mid <\phi_{J m_J, T m_T}\mid \Gamma_{N^*\rightarrow MB}
\mid \vec{q},  \lambda_\gamma,  \lambda_N, m_{t_N} >|^2]
\label{eq:g5}
\end{eqnarray}
With the normalizations defined by Eq.(\ref{eq:g2}), we then define
\begin{eqnarray}
& &<\phi_{Jm_J,Tm_T}|\Gamma_{N^*\rightarrow \gamma N}
|\vec{q},\lambda_\gamma\lambda_N m_{t_N} > \nonumber \\
& & =\delta_{m_T,m_{t_N}}\delta_{\lambda,(\lambda_\gamma-\lambda_N)}
\frac{1}{(2\pi)^{3/2}}\sqrt{\frac{m_N}{E_N(q)}}\frac{1}{\sqrt{2k}}
[\sqrt{2k_R} A^{JT}_\lambda] g^{JT}_\lambda(q,q_R)
d^J_{\lambda,m_J}(\theta)e^{i(\lambda-m_J)\phi} \,,
\label{eq:g6}
\end{eqnarray}
where $g^{JT}_\lambda(q,q_R)$ is a form factor
with $q_R$ defined by $M^{JT}=q_R+E_N(q_R)$ and
normalized as $g^{JT}_\lambda(q_R,q_R)=1$.
Substituting Eq.(\ref{eq:g6}) into Eq.(\ref{eq:g5}) and
noting that $|A^{JT}_{-\lambda}| = A^{JT}_{\lambda}$, we then obtain the
standard form 
\begin{eqnarray}
\Gamma_{\gamma N}(N^*_{JT}) = \frac{q_R^2}{4\pi}\frac{m_N}{M^{JT}}
\frac{8}{2J+1}[|A^{JT}_{3/2}|^2 + |A^{JT}_{1/2}|^2]
\label{eq:g7}
\end{eqnarray}

We only include 3 and 4 stars $N^*$ in our calculations of 
Eq(\ref{eq:pdgres}). We use their mean
values of $G^{JT}_{LS}$ and $A_\lambda$, as  
listed in Tables \ref{tab:nstar1}-\ref{tab:nstar3}.
\begin{table}
\caption{
The helicity amplitude $A_\lambda$ is given in unit of
$10^{-3}$ GeV$^{-1/2}$.
$G_{LS}$
is in unit of MeV$^{1/2}$.
The resonance mass $M^J_R$ and the total decay width $\Gamma^{tot}$
are  in unit of MeV.
}
\centering
\begin{tabular}{cccccccccccccccccccc}
%%%%%%%%%%%%%%%%%
   &&&&& &    & &     & &     & &     &  \\ \hline
$N^*_{TJ}(M_R)$ &&&$\Gamma^{tot}$  && & channels &  & $L,S$ & & $G_{LS}$ 
& & $A_{1/2}$ & &$A_{3/2}$ & \\ \hline \hline
$S_{11}(1535)$ &&&150&& &$\gamma N$ & & - & &- & &0.090& & 0.0& \\ \hline 
              &&&&& &$\pi N$ & & $0,1/2$ & &6.26& &-& & -& \\ \hline
              &&&&& &$\eta N$ & & $0,1/2$ & &7.55& &-& & -& \\ \hline
              &&&&& &$\pi \Delta$ & & $2,3/2$ & &1.06& &-& & -& \\ \hline
              &&&&& &$\rho N$ & & $0,1/2$ & &1.49& &-& & -& \\ \hline
              &&&&& &$\sigma N$ & & $1,1/2$ & &1.50& &-& & -& \\ 
   &&&&& &    & &     & &     & &     & \\ \hline\hline
$S_{11}(1650)$ &&&150&& &$\gamma N$ & & - & &- & &0.063& & 0.0& \\ \hline
              &&&&& &$\pi N$ & & $0,1/2$ & &12.23& &-& & -& \\ \hline
              &&&&& &$\eta N$ & & $0,1/2$ & &3.48& &-& & -& \\ \hline
              &&&&& &$\pi \Delta$ & & $2,3/2$ & &2.01& &-& & -& \\ \hline
              &&&&& &$\rho N$ & & $0,1/2$ & &1.42& &-& & -& \\ \hline
              &&&&& &        & & $2,1/2$ & &5.124& &-& & -& \\ \hline
              &&&&& &$\sigma N$ & & $1,1/2$ & &1.42& &-& & -& \\ 
   &&&&& &    & &     & &     & &     & \\ \hline \hline
$P_{11}(1440)$ &&&350&& &$\gamma N$ & & - & &- & &-0.065& & 0.0& \\ \hline
              &&&&& &$\pi N$ & & $1,1/2$ & &18.78& &-& & -& \\ \hline
              &&&&& &$\pi \Delta$ & & $1,3/2$ & &8.85& &-& & -& \\ \hline
              &&&&& &$\sigma N$ & & $0,1/2$ & &7.66& &-& & -& \\ 
   &&&&& &    & &     & &     & &     & \\ \hline\hline
$P_{11}(1710)$ &&&100&& &$\gamma N$ & & - & &- & &0.009& & 0.0& \\ \hline
              &&&&& &$\pi N$ & & $1,1/2$ & &6.22& &-& & -& \\ \hline
              &&&&& &$\eta N$ & & $1,1/2$ & &2.93& &-& & -& \\ \hline
              &&&&& &$\pi \Delta$ & & $1,3/2$ & &7.47& &-& & -& \\ \hline
              &&&&& &$\rho N$ & & $1,1/2$ & &4.93& &-& & -& \\ \hline
              &&&&& &$\sigma N$ & & $0,1/2$ & &1.19& &-& & -& \\ 
   &&&&& &    & &     & &     & &     & \\ \hline\hline
$P_{13}(1720)$ &&&150&& &$\gamma N$ & & - & &- & &0.018& & -0.019& \\ \hline
              &&&&& &$\pi N$ & & $1,1/2$ & &2.45& &-& & -& \\ \hline
              &&&&& &$\eta N$ & & $1,1/2$ & &2.20& &-& & -& \\ \hline
              &&&&& &$\rho N$ & & $1,1/2$ & &10.49& &-& & -& \\ 
\end{tabular}
\label{tab:nstar1}
\end{table}

\begin{table}
\caption{
The helicity amplitude $A_\lambda$ is given in unit of
$10^{-3}$ GeV$^{-1/2}$.
$G_{LS}$
is in unit of MeV$^{1/2}$.
The resonance mass $M^J_R$ and the total decay width $\Gamma^{tot}$
are  in unit of MeV.
}
\centering
\begin{tabular}{cccccccccccccccccccc}
%%%%%%%%%%%%%%%%%
   &&&&& &    & &     & &     & &     &  \\ \hline
$N^*_{TJ}(M_R)$ &&&$\Gamma^{tot}$  && & channels &  & $L,S$ & & $G_{LS}$
& & $A_{1/2}$ & &$A_{3/2}$ & \\ \hline \hline

   &&&&& &    & &     & &     & &     & \\ \hline \hline
$D_{13}(1520)$ &&&120&& &$\gamma N$ & & - & &- & &-0.024& & 0.166& \\ \hline
              &&&&& &$\pi N$ & & $2,1/2$ & &8.84& &-& & -& \\ \hline
              &&&&& &$\pi \Delta$ & & $0,3/2$ & &4.31& &-& & -& \\ \hline
              &&&&& &            & & $2,3/2$ & &3.69& &-& & -& \\ \hline
              &&&&& &$\rho N$ & & $0,3/2$ & &3.34& &-& & -& \\ \hline
              &&&&& &$\sigma N$ & & $1,1/2$ & &1.11& &-& & -& \\ 
   &&&&& &    & &     & &     & &     & \\ \hline\hline
$D_{13}(1700)$ &&&100&& &$\gamma N$ & & - & &- & &-0.018& & -0.002& \\ \hline
              &&&&& &$\pi N$ & & $2,1/2$ & &2.65& &-& & -& \\ \hline
              &&&&& &$\pi \Delta$ & & $0,3/2$ & &4.38& &-& & -& \\ \hline
              &&&&& &$          $ & & $2,3/2$ & &11.758& &-& & -& \\ \hline
              &&&&& &$\rho N$ & & $0,3/2$ & &3.5& &-& & -& \\ \hline
   &&&&& &    & &     & &     & &     & \\ \hline\hline
$D_{15}(1675)$ &&&150&& &$\gamma N$ & & - & &- & &0.019& & 0.015& \\ \hline
              &&&&& &$\pi N$ & & $2,1/2$ & &6.77& &-& & -& \\ \hline
              &&&&& &$\pi \Delta$ & & $2,3/2$ & &9.085& &-& & -& \\ \hline
              &&&&& &$\rho N$ & & $2,3/2$ & &1.46& &-& & -& \\ \hline
   &&&&& &    & &     & &     & &     & \\ \hline\hline
$F_{15}(1700)$ &&&130&& &$\gamma N$ & & - & &- & &-0.015& & 0.133 & \\ \hline
              &&&&& &$\pi N$ & & $3,1/2$ & &9.39& &-& & -& \\ \hline
              &&&&& &$\pi \Delta$ & & $1,3/2$ & &4.23& &-& & -& \\ \hline
              &&&&& &$          $ & & $3,3/2$ & &1.13& &-& & -& \\ \hline
              &&&&& &$\rho N$ & & $1,3/2$ & &2.52& &-& & -& \\ \hline
              &&&&& &           & & $3,3/2$ & &1.95& &-& & -& \\ \hline
              &&&&& &$\sigma N$ & & $2,1/2$ & &3.39& &-& & -& \\
   &&&&& &    & &     & &     & &     & \\ \hline\hline
$G_{17}(2190)$ &&&450&& &$\gamma N$ & & - & &- & &-0.055& & 0.081 & \\ \hline
              &&&&& &$\pi N$ & & $4,1/2$ & & 9.52 & &-& & -& \\ \hline
              &&&&& &$\rho N$ & & $2,3/2$ & &11.46& &-& & -& \\ \hline
   &&&&& &    & &     & &     & &     & \\ \hline\hline

\end{tabular}
\label{tab:nstar2}
\end{table}
\begin{table}
\caption{
The helicity amplitude $A_\lambda$ is given in unit of
$10^{-3}$ GeV$^{-1/2}$.
$G_{LS}$
is in unit of MeV$^{1/2}$.
The resonance mass $M^J_R$ and the total decay width $\Gamma^{tot}$
are  in unit of MeV.
}
\centering
\begin{tabular}{cccccccccccccccccccc}
%%%%%%%%%%%%%%%%%
   &&&&& &    & &     & &     & &     &  \\ \hline
$N^*_{TJ}(M_R)$ &&&$\Gamma^{tot}$  && & channels &  & $L,S$ & & $G_{LS}$
& & $A_{1/2}$ & &$A_{3/2}$ & \\ \hline \hline

   &&&&& &    & &     & &     & &     & \\ \hline \hline
$S_{31}(1620)$ &&&150&& &$\gamma N$ & & - & &- & &0.027& & - & \\ \hline
              &&&&& &$\pi N$ & & $0,1/2$ & &8.02& &-& & -& \\ \hline
              &&&&& &$\pi \Delta$ & & $2,3/2$ & &7.47& &-& & -& \\ \hline
              &&&&& &$\rho N$ & & $0,1/2$ & &4.57& &-& & -& \\ \hline
              &&&&& &         & & $2,3/2$ & &1.69& &-& & -& \\ \hline
   &&&&& &    & &     & &     & &     & \\ \hline\hline
$P_{31}(1910)$ &&&150&& &$\gamma N$ & & - & &- & &0.003 & & - & \\ \hline
              &&&&& &$\pi N$ & & $1,1/2$ & &14.38 & &-& & -& \\ \hline
   &&&& & $1,1/2$ & &11.5 & &-& & -& \\ \hline
   &&&&& &    & &     & &     & &     & \\ \hline\hline
$P_{33}(1600)$ &&&350&& &$\gamma N$ & & - & &- & &-0.023 & & -0.009 & \\ \hline
              &&&&& &$\pi N$ & & $1,1/2$ & &11.75 & &-& & -& \\ \hline
              &&&&& &$\pi \Delta$ & & $1,3/2$ & &17.06& &-& & -& \\ \hline
   &&&&& &    & &     & &     & &     & \\ \hline\hline
$P_{33}(1920)$ &&&200&& &$\gamma N$ & & - & &- & &0.04 & & 0.023 & \\ \hline
              &&&&& &$\pi N$ & & $1,1/2$ & &2.48 & &-& & -& \\ \hline
              &&&&& &$\pi \Delta$ & & $1,3/2$ & &7.10 & &-& & -& \\ \hline
   &&&&& &    & &     & &     & &     & \\ \hline\hline
$D_{33}(1700)$ &&&300&& &$\gamma N$ & & - & &- & &0.104& & 0.085 & \\ \hline
              &&&&& &$\pi N$ & & $2,1/2$ & &2.44& &-& & -& \\ \hline
              &&&&& &$\pi \Delta$ & & $0,3/2$ & &10.35& &-& & -& \\ \hline
              &&&&& &             & & $2,3/2$ & &2.18& &-& & -& \\ \hline
              &&&&& &$\rho N$ & & $0,3/2$ & &1.09& &-& & -& \\ \hline
   &&&&& &    & &     & &     & &     & \\ \hline\hline
$F_{35}(1905)$ &&&350&& &$\gamma N$ & & - & &- & &0.026& & -0.045 & \\ \hline
              &&&&& &$\pi N$ & & $3,1/2$ & &6.11& &-& & -& \\ \hline
              &&&&& &$\pi \Delta$ & & $1,3/2$ & &9.78& &-& & -& \\ \hline
              &&&&& &             & & $3,3/2$ & &13.53& &-& & -& \\ \hline
              &&&&& &$\rho N$ & & $1,3/2$ & &9.99& &-& & -& \\ \hline
   &&&&& &    & &     & &     & &     & \\ \hline\hline
$F_{35}(1950)$ &&&300&& &$\gamma N$ & & - & &- & &-0.076& & -0.097 & \\ \hline
              &&&&& &$\pi N$ & & $3,1/2$ & &10.38& &-& & -& \\ \hline
              &&&&& &$\pi \Delta$ & & $3,3/2$ & &9.39& &-& & -& \\ \hline
   &&&&& &    & &     & &     & &     & \\ \hline\hline
\end{tabular}
\label{tab:nstar3}
\end{table}

\end{document}